\documentclass[12pt]{article}
\usepackage{amsmath,cite,epsfig,a4,times,euscript,listings}
\usepackage[text-hat-accent]{euler}
\usepackage{graphics}
\usepackage[usenames]{color}
\renewcommand{\baselinestretch}{1.1}
\newcommand{\myTitle}[1]{\begin{center}{\bf\Huge #1}\\[5ex]\end{center}}
\newcommand{\myAuthor}[1]{\begin{center}{\Large #1}\\[2ex]\end{center}}
\newcommand{\myAffiliation}[1]{\\[1ex]{\it\large #1}}
\newcommand{\myEmail}[1]{}
\newcommand{\myDate}{\begin{center}\vspace{2.5ex}\end{center}}
\newcommand{\myAbstract}[1]{\begin{center}\renewcommand{\baselinestretch}{1}{\bf Abstract}\\[2ex]\parbox{0.8\linewidth}{\small\hspace{15pt} #1}\end{center}\vspace{\baselineskip}}
\newcommand{\myReport}[1]{\hspace{\fill} #1}
\newcommand{\myPreprint}[1]{}
\newcommand{\myKeywords}[1]{}
\newcommand{\myFigure}[1]{\begin{figure}\begin{center}#1\end{center}\end{figure}}

\newcommand{\myScript}[1]{\EuScript{#1}}

\newcommand{\slashl}{\ell\hspace{-6pt}/}
\newcommand{\slashp}{p\hspace{-6.5pt}/}

\newcommand{\slashk}{k\hspace{-6.5pt}/}
\newcommand{\slashK}{K\hspace{-7.7pt}/}

\newcommand{\slashJ}{J\hspace{-5.5pt}/}

\newcommand{\Appendix}[1]{Appendix~\ref{#1}}   
\newcommand{\Section}[1]{Section~\ref{#1}}

\newcommand{\Figure}[1]{Fig.~\ref{#1}}
\newcommand{\Equation}[1]{Eq.~(\ref{#1})}
\newcommand{\ie}{{\it i.e.}}

\newcommand{\eg}{{\it e.g.}}
\newcommand{\lOPP}{\ell}
\newcommand{\Ord}{\myScript{O}}
\newcommand{\Amp}{\myScript{A}}
\newcommand{\srac}[2]{{\textstyle\frac{#1}{#2}}}

\newcommand{\imag}{\mathrm{i}}

\newcommand{\dDell}{[d\ell]}%{\frac{d^{4-2\varepsilon}\ell}{h(\mu,\varepsilon)}}
\newcommand{\gluon}{\mathrm{g}}
\newcommand{\quark}{\mathrm{q}}
\newcommand{\antiq}{\bar{\mathrm{q}}}
\newcommand{\uQuark}{\mathrm{u}}
\newcommand{\uAntiq}{\bar{\mathrm{u}}}
\newcommand{\eleon}{\mathrm{e}}
\newcommand{\brktAA}[1]{\langle#1\rangle}
\newcommand{\brktAS}[1]{\langle#1]}

\newcommand{\brktSS}[1]{[#1]}
\newcommand{\leftA}[1]{\langle#1|}
\newcommand{\Arght}[1]{|#1\rangle}
\newcommand{\leftS}[1]{[#1|}
\newcommand{\Srght}[1]{|#1]}
\newcommand{\lop}[2]{#1\!\cdot\!#2}

\newcommand{\Den}{\EuScript{D}}
\newcommand{\GenC}{C}
\newcommand{\BoxC}{C}%^{\mathrm{Box}}}
\newcommand{\TriC}{C}%^{\mathrm{Tri}}}
\newcommand{\BubC}{C}%^{\mathrm{Bub}}}
\newcommand{\TadC}{C}%^{\mathrm{Bub}}}
\newcommand{\tGenC}{\tilde{C}}%^{\mathrm{Box}}}
\newcommand{\tBoxC}{\tilde{C}}%^{\mathrm{Box}}}
\newcommand{\tTriC}{\tilde{C}}%^{\mathrm{Tri}}}
\newcommand{\tBubC}{\tilde{C}}%^{\mathrm{Bub}}}
\newcommand{\GenPol}{\EuScript{C}}
\newcommand{\TriPol}{\GenPol}%^{\mathrm{Tri}}}
\newcommand{\BubPol}{\GenPol}%^{\mathrm{Bub}}}
\newcommand{\tGenPol}{\tilde{\GenPol}}
\newcommand{\tBoxPol}{\tilde{\GenPol}}%^{\mathrm{Box}}}
\newcommand{\tTriPol}{\tilde{\GenPol}}%^{\mathrm{Tri}}}
\newcommand{\tBubPol}{\tilde{\GenPol}}%^{\mathrm{Bub}}}
\newcommand{\tTadPol}{\tilde{\GenPol}}%^{\mathrm{Bub}}}
\newcommand{\graph}[3]{\raisebox{-#3ex}{\epsfig{file=#1.pdf,width=#2ex}}}

\newcommand{\tweakcodepar}[3]%
  {\vspace{#1ex}\newline\noindent\hspace*{4.0ex}{\small\tt #3}\vspace{#2ex}\newline\noindent}

\begin{document}

\myReport{IFJPAN-IV-2017-21}
\myPreprint{}\\[2ex]

\myTitle{%
Calculating off-shell one-loop amplitudes\\[-0.5ex] for $k_T$-dependent factorization:\\[0.2ex] a proof of concept
}

\myAuthor{%
A.~van~Hameren%
\myAffiliation{%
Institute of Nuclear Physics Polisch Academy of Sciences\\
PL-31342 Krak\'ow, Poland%
\myEmail{hameren@ifj.edu.pl}
}
}

\myDate

\myAbstract{%
A method to define and calculate one-loop amplitudes with an off-shell space-like, or $k_T$-dependent, gluon is presented.
It introduces a practical regularization to deal with the divergencies that appear due to linear denominators, and can be applied to arbitarary partonic scattering processes.
}

\myKeywords{QCD}

%\begin{document}
%

\newpage%
\setcounter{tocdepth}{2}
\tableofcontents

\section{Introduction}
The description of events with high transverse momentum ($p_T$) in the final state resulting from hadron scattering is facilitated through the factorization of the low-scale dependence from the high-scale dependence.
The latter is the partonic cross section which can be calculated using perturbative quantum chromodynamics (QCD), and the former consist of the parton distribution functions describing the scattering hadrons which cannot (yet) be calculated within QCD from first principles.
This factorization can be heuristically motivated, and is a necessity for the tractability of the computational problem.
In order for a factorization formula to be reliable it needs to be proven to be valid, or at least it needs to be shown that it admits the application of perturbation theory.
In particular, it needs to be shown that the mass singularities that are inherently present in perturbative QCD can be dealt with in a structural manner.

One of the advantages of $k_T$-factorization, or high-energy factorization~\cite{Gribov:1984tu,Catani:1990eg,Collins:1991ty}, is that it allows for a complete kinematical description at lowest order in perturbation theory by providing a momentum imbalance to the final state, and one may expect that higher-order corrections will be smaller than in factorization prescriptions that do not allow for such an imbalance.
One of the prices to pay is that it requires the momenta of the initial-state partons to be space-like rather than light-like, which complicates the calculation of the partonic cross section, in particular if one is interested in final states with more than two partons.
At tree level, this problem has been completely solved~\cite{Lipatov:1995pn,Lipatov:2000se,Antonov:2004hh,vanHameren:2012if,vanHameren:2013csa,Kotko:2014aba} up to the implementation of a fully differential parton-level Monte Carlo event generator for arbitrary processes within the Standard Model~\cite{vanHameren:2016kkz}.
%
%While the context of the cited works, and to a certain extend also the language, is different, the resulting partonic cross sections are, at least at tree-level, completely equivalent.
%
In this paper, we will only concentrate on the partonic cross section, and we will collectively refer to approaches that require partonic cross sections with off-shell initial states as $k_T$-dependent factorization.

Besides the progress in the quest for precision, also the confirmation of the reliability of $k_T$-dependent factorization requires the advancement to higher orders in perturbation theory.
For the partonic cross section, this implies the ability to go beyond tree-level and to deal with loop amplitudes.
Already at one loop, new complications arise compared to factorization prescriptions for which the initial-state partons are light-like, in the form of new fundamental one-loop integrals, with {\em linear denominators\/}, and the associated new divergencies, ``light-cone  divergencies'' or ``rapidity  divergencies'', which cannot be tackled by straightforward dimensional regularization.
Some time ago an effort was started to push $k_T$-dependent factorization beyond tree-level within the parton reggeization approach~\cite{Hentschinski:2011tz,Chachamis:2012cc,Chachamis:2013hma}, and it was pursued recently~\cite{Nefedov:2016clr,Nefedov:2017qzc}.
One of the main issues to be tackled in those works was the regularization of the mentioned divergencies, which must preferably be manifestly Lorentz covariant, respect gauge invariance, and allow for practical calculations.
This problem also occurs in calculations done in light-cone gauges (\cite{Gituliar:2014eba} and references therein), and in soft collinear effective theory \cite{Becher:2011dz,Chiu:2012ir}.
Other recent NLO calculations within $k_T$-dependent factorization are~\cite{Iancu:2016vyg,Boussarie:2016ogo,Ducloue:2017mpb,Beuf:2017bpd,Lappi:2016oup,Hanninen:2017ddy}.

In this paper we advance the approach presented in~\cite{vanHameren:2012if} to one loop.
It inherently provides a regularization of the mentioned divergencies that manifestly respects both Lorentz covariance and gauge invariance.
The regularization can perfectly consistently be applied alongside with dimensional regularization.
It will be shown that the regularization is sound, in the sense that it produces divergencies that are just logarithmic in the regularization parameter, and not both linear and logarithmic for example.
Furthermore, it will be shown that it allows for the application of the powerful so-called {\em integrand techniques\/} for the calculation of one-loop amplitudes involving several partons.

The focus will be on the validity for arbitrary numbers of partons involved in the hard scattering.
For processes involving only three or four gluons, for example, many of the issues addressed can be solved by choosing specific gauges for the polarization vectors.
Such choices will never be implied here.
What will be implied is the Feynman gauge for internal gluons.
One-loop Feynman graphs with ghost loops are not an issue because they do not involve the new regularization.

The main issue will appear to be ``high-rank'' Feynman graphs, that is Feynman graphs with a high power of the integration momentum in the numerator of the integrand compared to the number of denominators involving the integration momentum.
The essence of the approach of~\cite{vanHameren:2012if} is that the of-shell gluons are represented as auxiliary quark-antiquark pairs.
Quark lines in a loop lead to ``lower rank'' compared to gluon lines, and consequently graphs that have more than one auxiliary quark in the loop are less of an issue within this context.
Therefore, only amplitudes with a single off-shell gluon are considered in this paper, because the potentially most severe problems are already encountered in those.

The paper will continue in \Section{Sec:Tree-level} with a short repetition of the formulation of tree-level amplitudes.
Then, in \Section{Sec:One-loop}, one-loop amplitudes and their regularization will be addressed.
The main statements regarding the soundness and applicability of the regularization are given in \Section{Sec:Statements}, and the rest of the body of the paper consists of an exposition of the arguments for the statements.
This involves many details that are referred to the appendices.
One of the results of this paper will be that the necessary scalar master integrals have at most one linear denominator, and their expressions are given in \Section{scalarintegrals}.

\section{Tree-level amplitudes\label{Sec:Tree-level}}
Consider the scattering amplitude of a process involving a quark-antiquark pair.
The pair has flavor $A$, and we assume that there are no other quark-antiquark pairs of this flavor involved in the scattering process.
We take all particles out-going, and write the amplitude as
%
%%%%%%%%%%%%%%%%%%%%%%%%%%%%%%%%%%%%%%%%
\begin{equation}
\EuScript{A}\big(\emptyset\to \antiq_{A}\quark_{A}+X\big)
~.
\end{equation}
%%%%%%%%%%%%%%%%%%%%%%%%%%%%%%%%%%%%%%%%
%
The letter $X$ stands for other particles involved in the hard scattering process, \eg\ $X=\gluon\gluon$ or $X=\gluon\,\uAntiq\uQuark\,\eleon^+\eleon^-$, etc..
As mentioned in the introduction, the complications that will be encountered in the current study involve ``high-rank'' one-loop integrals, and settling them for multi-gluon amplitudes, which allow for the highest rank, will be sufficient.
Therefore, we may imagine $X$ to be just a number of gluons:
%
%%%%%%%%%%%%%%%%%%%%%%%%%%%%%%%%%%%%%%%%
\begin{equation}
\graph{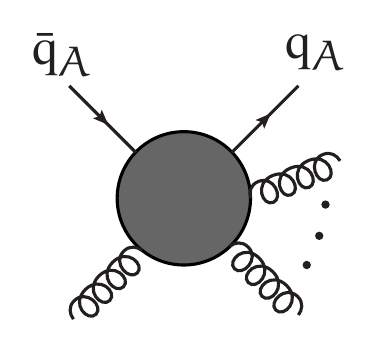}{15}{5.5}
\end{equation}
%%%%%%%%%%%%%%%%%%%%%%%%%%%%%%%%%%%%%%%%
%
The antiquark momentum $p_{A}^\mu$ and the quark momentum $p_{A'}^\mu$ are given in the following Sudakov decomposition
%
%%%%%%%%%%%%%%%%%%%%%%%%%%%%%%%%%%%%%%%%
\begin{equation}
p_{A}^\mu = \Lambda p^\mu + \alpha q^\mu + \beta k_{T}^\mu
\quad,\quad
p_{A'}^\mu = k^\mu-p_{A}^\mu %-(\Lambda-x) p^\mu - \alpha q^\mu + (1-\beta) k_{T}^\mu
\quad,\quad
k^\mu = xp^\mu+k_T^\mu
\quad,
\label{auxmom1}
\end{equation}
%%%%%%%%%%%%%%%%%%%%%%%%%%%%%%%%%%%%%%%%
%
where $p^\mu,q^\mu$ are light-like with $p\!\cdot\!q>0$, where $p\!\cdot\!k_T=q\!\cdot\!k_T=0$, and where
%
%%%%%%%%%%%%%%%%%%%%%%%%%%%%%%%%%%%%%%%%
\begin{equation}
\alpha = \frac{-\beta^2k_{T}^2}{\Lambda(p+q)^2}
\quad,\quad
\beta = \frac{1}{1+\sqrt{1-x/\Lambda}}
\quad.
\label{auxmom2}
\end{equation}
%%%%%%%%%%%%%%%%%%%%%%%%%%%%%%%%%%%%%%%%
%
With this choice, the momenta $p_{A}^\mu,p_{A'}^\mu$ satisfy the relations
%
%%%%%%%%%%%%%%%%%%%%%%%%%%%%%%%%%%%%%%%%
\begin{equation}
p_{A}^2 = p_{A'}^2 = 0
\quad,\quad
p_{A}^\mu+p_{A'}^\mu = xp^\mu+k_T^\mu
\end{equation}
%%%%%%%%%%%%%%%%%%%%%%%%%%%%%%%%%%%%%%%%
%
for any value of the parameter $\Lambda$.
The scattering amplitude depends on this parameter via its dependence on the momenta:
%%%%%%%%%%%%%%%%%%%%%%%%%%%%%%%%%%%%%%%%
\begin{equation}
\EuScript{A}(\Lambda) =
\EuScript{A}\Big(\emptyset\to \bar{q}_{A}\big(p_{A}(\Lambda)\big)\,q_{A}\big(p_{A'}(\Lambda)\big)+X\Big)
~.
\end{equation}
%%%%%%%%%%%%%%%%%%%%%%%%%%%%%%%%%%%%%%%%
%
In \cite{vanHameren:2012if}, it was shown at tree-level that in the limit of $\Lambda\to\infty$, the amplitude is directly related to the amplitude of the process in which the quark-antiquark pair is replaced by an off-shell space-like, or reggeized, gluon with momentum $k=xp+k_T$:
%%%%%%%%%%%%%%%%%%%%%%%%%%%%%%%%%%%%%%%%
\begin{equation}
\frac{|k_T|}{\Lambda}\,\EuScript{A}(\Lambda)
\;\overset{\Lambda\to\infty}{\longrightarrow}\;
\EuScript{A}\big(\emptyset\to g^*(xp+k_T)+X\big)
~,
\label{limit}
\end{equation}
%%%%%%%%%%%%%%%%%%%%%%%%%%%%%%%%%%%%%%%%
%
or graphically:
%
%%%%%%%%%%%%%%%%%%%%%%%%%%%%%%%%%%%%%%%%
\begin{equation}
\graph{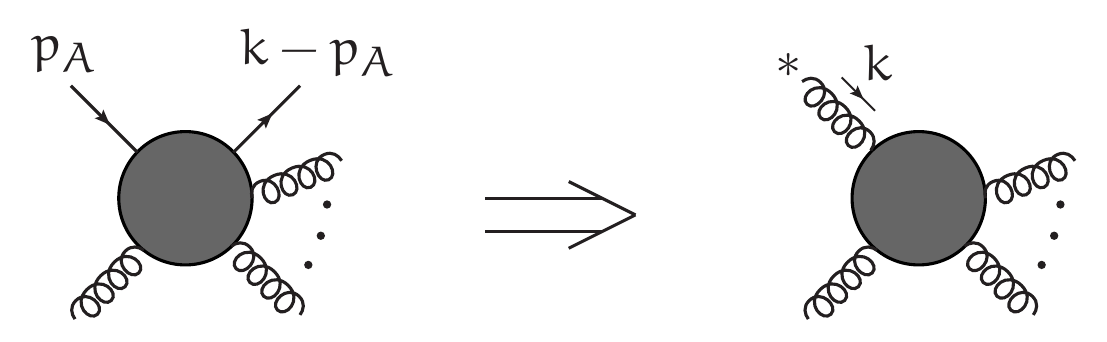}{46}{5.5}
\end{equation}
%%%%%%%%%%%%%%%%%%%%%%%%%%%%%%%%%%%%%%%%
%

The parametrization of the momenta in \cite{vanHameren:2012if} was different, and had the advantage that the quark-antiquark spinors were directly given by
$\Srght{p_{A}} = \sqrt{\Lambda}\,\Srght{p}$
and
$\Arght{p_{A'}} = \sqrt{\Lambda-x}\,\Arght{p}$,
while now these relations only hold approximately for large $\Lambda$.
The disadvantage of the choice in \cite{vanHameren:2012if} is the fact that the momenta are not real, which would lead to unnecessary complications in the current study.
Important is that with both choices, the amplitudes are completely gauge invariant for any value of $\Lambda$.
Using \Equation{auxmom1} and \Equation{auxmom2}, we have
%
%%%%%%%%%%%%%%%%%%%%%%%%%%%%%%%%%%%%%%%%
\begin{equation}
p_{A}^\mu = \Lambda p^\mu + \srac{1}{2}\,k_T^\mu + \Ord\big(\Lambda^{-1}\big)
\quad,\quad
p_{A'}^\mu = (x-\Lambda)p^\mu + \srac{1}{2}\,k_T^\mu + \Ord\big(\Lambda^{-1}\big)
\quad,
\label{auxmomentlimit}
\end{equation}
%%%%%%%%%%%%%%%%%%%%%%%%%%%%%%%%%%%%%%%%
%
here $k_T^\mu$ can be written in terms of the auxiliary momentum $q^\mu$ of the Sudakov decomposition as~\cite{vanHameren:2014iua}
%
%%%%%%%%%%%%%%%%%%%%%%%%%%%%%%%%%%%%%%%%
\begin{equation}
k_T^\mu = -\bar{\kappa}e^\mu
          -\bar{\kappa}^*e_*^\mu
\quad,\quad
e^\mu = \srac{1}{2}\brktAS{p|\gamma^\mu|q}
\quad,\quad
e_*^\mu = \srac{1}{2}\brktAS{q|\gamma^\mu|p}
\quad,
\label{kTexpansion}
\end{equation}
%%%%%%%%%%%%%%%%%%%%%%%%%%%%%%%%%%%%%%%%
%
with
%
%%%%%%%%%%%%%%%%%%%%%%%%%%%%%%%%%%%%%%%%
\begin{equation}
\bar{\kappa} = \frac{\kappa}{[pq]} = \frac{\brktAS{q|\slashk|p}}{(p+q)^2}
\quad,\quad
\bar{\kappa}^* = \frac{\kappa^*}{\brktAA{qp}} = \frac{\brktAS{p|\slashk|q}}{(p+q)^2}
\quad.
\end{equation}
%%%%%%%%%%%%%%%%%%%%%%%%%%%%%%%%%%%%%%%%
%%
%
The Weyl spinors of $p_{A}^\mu,p_{A'}^\mu$ can now, up to higher powers of $\Lambda^{-1}$, be expanded as
%
%%%%%%%%%%%%%%%%%%%%%%%%%%%%%%%%%%%%%%%%
\begin{align}
\Arght{p_{A}} = \sqrt{\Lambda}\,\Arght{p}
            - \frac{\bar{\kappa}^*\Arght{q}}{2\sqrt{\Lambda}}
\quad&,\quad
\Srght{p_{A}} = \sqrt{\Lambda}\,\Srght{p}
            - \frac{\bar{\kappa}\Srght{q}}{2\sqrt{\Lambda}}
\quad,
\label{AAprimeSpinors}\\
\Arght{p_{A'}} = \sqrt{\Lambda}\,\Arght{p}
            + \frac{\bar{\kappa}^*\Arght{q}}{2\sqrt{\Lambda}}
            - \frac{x\Arght{p}}{2\sqrt{\Lambda}}
\quad&,\quad
\Srght{p_{A'}} = -\sqrt{\Lambda}\,\Srght{p}
            - \frac{\bar{\kappa}\Srght{q}}{2\sqrt{\Lambda}}
            + \frac{x\Srght{p}}{2\sqrt{\Lambda}}
\quad.
\notag
\end{align}
%%%%%%%%%%%%%%%%%%%%%%%%%%%%%%%%%%%%%%%% 
%

One way to interpret \Equation{limit} is that it tells us how to get an expression for the right-hand-side given an expression for the left-hand-side.
In \cite{vanHameren:2012if}, however, the limit of $\Lambda\to\infty$ was analysed on a graph-by-graph basis, and Feynman rules were derived how to arrive at the right-hand-side without the need for an explicit expression for the left-hand-side.
It was found that the auxiliary quark-antiquark pair $\bar{q}_{A},q_{A}$ simply must follow eikonal Feynman rules to arrive at the limit directly.
The interaction vertex and propagator are given by
%
%%%%%%%%%%%%%%%%%%%%%%%%%%%%%%%%%%%%%%%%
\begin{equation}
\raisebox{-3.5ex}{\epsfig{figure=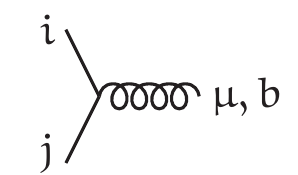,width=15ex}}
= -\mathrm{i}\,p^\mu\,T^b_{i,j}
\qquad,\qquad
\raisebox{-0.5ex}{\epsfig{figure=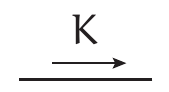,width=8ex}}
=\frac{\mathrm{i}}{p\!\cdot\!K}
\quad.
\label{Feynmanrules}
\end{equation}
%%%%%%%%%%%%%%%%%%%%%%%%%%%%%%%%%%%%%%%%
%
The momenta assigned to the eikonal quark-antiquark pair may eventually be anything that adds up to $xp^\mu+k_T^\mu$ and has vanishing invariant inner product with $p^\mu$, and need not to be light-like anymore.

\subsection{Color decomposition}
In the rest of the paper a color decomposition
%
%%%%%%%%%%%%%%%%%%%%%%%%%%%%%%%%%%%%%%%%
\begin{equation}
\EuScript{M}(\textrm{color},\textrm{spin},\textrm{momenta})
=\sum_i C_i(\textrm{color})\,\Amp_i(\textrm{spin},\textrm{momenta})
~,
\end{equation}
%%%%%%
%
of the colored amplitude $\EuScript{M}$ into color-independent {\em partial amplitudes\/} $\Amp_i$ will be assumed.
The partial amplitudes consist of planar graphs constructed using the color ordered Feynman rules of~\Figure{Fig:colorordered}.
\myFigure{%
\epsfig{figure=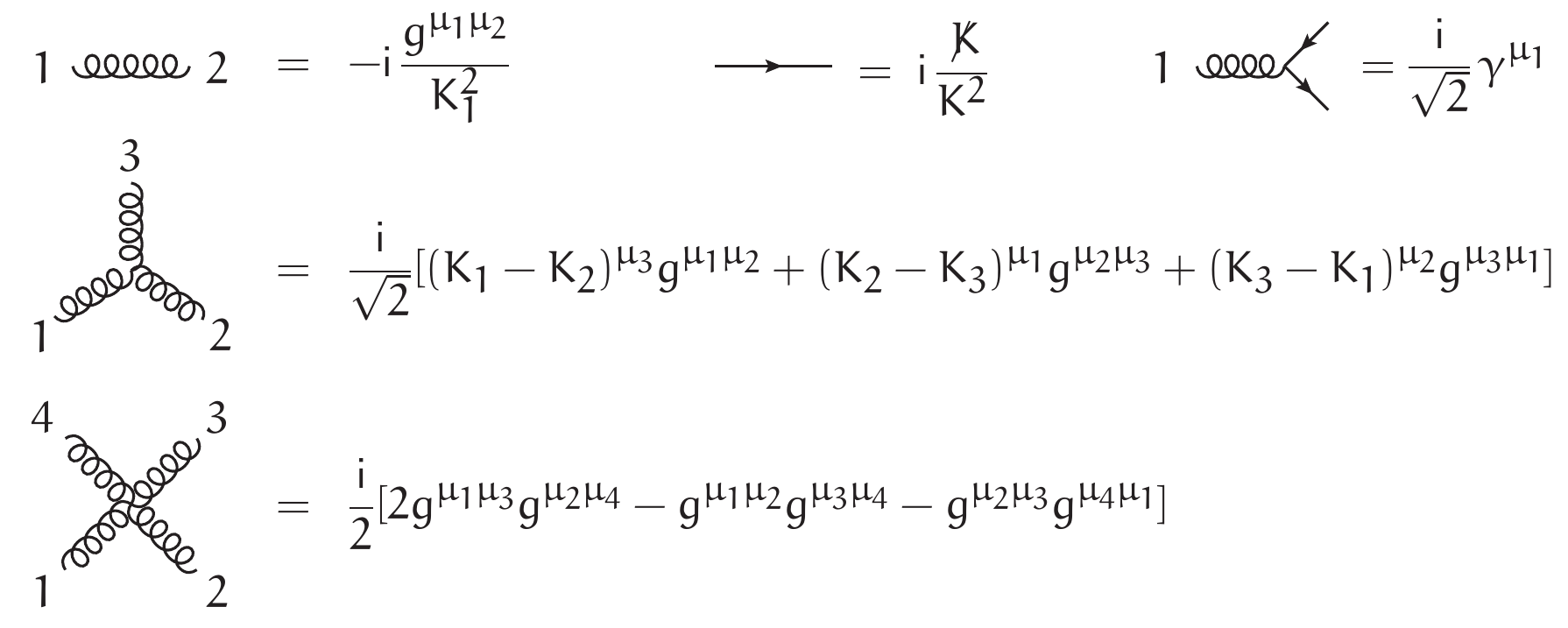,width=0.9\linewidth}
\caption{\label{Fig:colorordered}Color-ordered Feynman rules. All momenta are assumed to be incoming, and momentum conservation is understood.}
}
Also for one-loop amplitudes such decompositions exists~\cite{Bern:1994fz}.
In that case, it is enough to study the so-called {\em primitive one-loop amplitudes\/}, which are obtained by considering planar one-loop graphs and applying the color-ordered Feynman rules.
Primitive amplitudes are gauge invariant, and all necessary partial amplitudes can be obtained as linear combinations of them.
The color ordered eikonal Feynman rules are
%
%%%%%%%%%%%%%%%%%%%%%%%%%%%%%%%%%%%%%%%%
\begin{equation}
\raisebox{-3.5ex}{\epsfig{figure=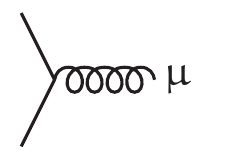,width=11ex}}
=\sqrt{2}\,\mathrm{i}\,p^\mu
\qquad,\qquad
\raisebox{-0.5ex}{\epsfig{figure=eikPropagator.pdf,width=8ex}}
=\frac{\mathrm{i}}{2p\!\cdot\!K}
\quad.
\label{orderedFeynmanrules}
\end{equation}
%%%%%%%%%%%%%%%%%%%%%%%%%%%%%%%%%%%%%%%%
%
%

\subsection{Tree-level off-shell currents}
Amplitudes with an off-shell gluon were defined as the limitting case of amplitudes for which the off-shell gluon is replaced with a quark-antiquark pair that have momenta with diverging components.
We will briefly address an approach to this limit which allows to study other types of amplitudes involving partons with diverging momenta.

We will make use of so-called tree-level off-shell currents.
An $n$-parton gluonic off-shell current $J_{1,n}^\mu$ is an $(n+1)$-gluon Green function with $n$ on-shell, amputated, legs.
The subscript indicates that it includes the $n$ on-shell gluons $1$ to $n$.
The one off-shell leg includes a propagator, and we will denote the off-shell current without this propagator with a tilde:
%
%%%%%%%%%%%%%%%%%%%%%%%%%%%%%%%%%%%%%%%%
\begin{equation}
J_{1,n}^\mu = \frac{-\imag g_{\nu}^\mu}{K_{1,n}^2}\,\tilde{J}_n^\nu
\quad,\quad
K_{1,n}^\mu = \sum_{j=1}^np_j^\mu
\quad.
\end{equation}
%%%%%%%%%%%%%%%%%%%%%%%%%%%%%%%%%%%%%%%%
%
A one-point current by definition is the polarization vector or spinor of the external parton it represent, and does not include a propagator.

Off-shell currents are defined such that if the one off-shell leg ``goes on-shell'', then it becomes an amplitude.
The procedure of going from an off-shell leg to an amputated on-shell leg can be made explicit as follows.
Suppose we have $n$ partons with light-like momenta $p_1^\mu,p_2^\mu,\ldots p_n^\mu$.
We can deform one of these momenta, say $p_i^\mu$, using a vector $e^\mu$ satisfying
%
%%%%%%%%%%%%%%%%%%%%%%%%%%%%%%%%%%%%%%%%
\begin{equation}
\lop{e}{p_i} = 0
\quad,\quad
\lop{e}{e} = 0
\quad,
\end{equation}
%%%%%%%%%%%%%%%%%%%%%%%%%%%%%%%%%%%%%%%%
%
to 
%
%%%%%%%%%%%%%%%%%%%%%%%%%%%%%%%%%%%%%%%%
\begin{equation}
p_i^\mu\to\hat{p}_i^\mu(z)=p_i^\mu+ze^\mu
~.
\end{equation}
%%%%%%%%%%%%%%%%%%%%%%%%%%%%%%%%%%%%%%%%
%
The momentum $\hat{p}_i^\mu(z)$ is still light-like for any value of $z$.
The vector $e^\mu$ can for example be constructed using an auxiliary light-like momentum $q^\mu$ as $e^\mu=\srac{1}{2}\brktAS{p_i|\gamma^\mu|q}$ or $e^\mu=\srac{1}{2}\brktAS{q|\gamma^\mu|p_i}$.
Then, the spinors of the deformed momentum become
%
%%%%%%%%%%%%%%%%%%%%%%%%%%%%%%%%%%%%%%%%
\begin{equation}
\Arght{\hat{p}_i} = \Arght{{p}_i}
\;,\;
\Srght{\hat{p}_i} = \Srght{{p}_i} + z\Srght{{q}}
\qquad\textrm{or}\qquad
\Arght{\hat{p}_i} = \Arght{{p}_i} + z\Arght{{q}}
\;,\;
\Srght{\hat{p}_i} = \Srght{{p}_i}
\quad,
\end{equation}
%%%%%%%%%%%%%%%%%%%%%%%%%%%%%%%%%%%%%%%%
%
depending on the choice for $e^\mu$.
Now we can choose $z$ such that the sum $K_{1,n}^\mu$ of all $n$ light-like momenta also becomes light-like:
%
%%%%%%%%%%%%%%%%%%%%%%%%%%%%%%%%%%%%%%%%
\begin{equation}
z_{1,n} = -\frac{K_{1,n}^2}{2\lop{e}{K_{1,n}}}
\quad\Longrightarrow\quad
\hat{K}_{1,n}^2 = 0
\quad,\quad
\hat{K}_{1,n}^\mu = K_{1,n}^\mu + z_{1,n}e^\mu
\quad.
\end{equation}
%%%%%%%%%%%%%%%%%%%%%%%%%%%%%%%%%%%%%%%%
%
An $n$-parton gluonic off-shell current $J_{1,n}^\mu$ depends on $n$ light-like momenta in such a way that
%
%%%%%%%%%%%%%%%%%%%%%%%%%%%%%%%%%%%%%%%%
\begin{equation}
\varepsilon_\mu\big(\!-\!\hat{K}_{1,n}\big)\,\tilde{J}_{1,n}^\mu\big(p_1,\ldots,p_{i-1},\hat{p}_i,p_{i+1}\ldots,p_n\big)
\end{equation}
%%%%%%%%%%%%%%%%%%%%%%%%%%%%%%%%%%%%%%%%
%
is an $(n+1)$-parton amplitude.
Here, $\varepsilon^\mu(-\hat{K}_{1,n})$ is a polarization vector for the extra on-shell gluon with momentum $-\hat{K}_{1,n}^\mu$.
Notice that the above involves the off-shell current without the propagator.

Similarly we can have (anti)quark off-shell currents $\leftS{J_{1,n}}$, $\leftS{J_{1,n}}$, $\Arght{J_{1,n}}$ and $\Srght{J_{1,n}}$.
The notation with the angular/square brackets makes sense for massless quarks within QCD where there is no interaction involving $\gamma^5$.
The type of bracket just alternates when including the propagator, \eg
%
%%%%%%%%%%%%%%%%%%%%%%%%%%%%%%%%%%%%%%%%
\begin{equation}
\leftA{J_{1,n}} = \leftS{\tilde{J}_n}\,\frac{\imag\slashK_{1,n}}{K_{1,n}^2}
\quad.
\end{equation}
%%%%%%%%%%%%%%%%%%%%%%%%%%%%%%%%%%%%%%%%
%
%
For a quark off-shell current $\leftA{J_{1,n}}$,
%
%%%%%%%%%%%%%%%%%%%%%%%%%%%%%%%%%%%%%%%%
\begin{equation}
\brktSS{\,\tilde{J}_{1,n}\big(p_1,\ldots,p_{i-1},\hat{p}_i,p_{i+1}\ldots,p_n\big)\,|-\!\hat{K}_{1,n}\,}
\end{equation}
%%%%%%%%%%%%%%%%%%%%%%%%%%%%%%%%%%%%%%%%
%
is an $(n+1)$-parton amplitude.
The equivalent of course goes through for $\leftS{J_{1,n}}$ and the anti-quark currents $\Arght{J_{1,n}}$ and $\Srght{J_{1,n}}$.

Eventually, an off-shell current consists of a sum of Feynman graphs that is complete in the sense that the above holds.
They can be defined constructively via the Berends-Giele recursive relations~\cite{Berends:1987me} depicted in \Figure{Fig:DSg} and \Figure{Fig:DSq}.
\myFigure{%
\epsfig{figure=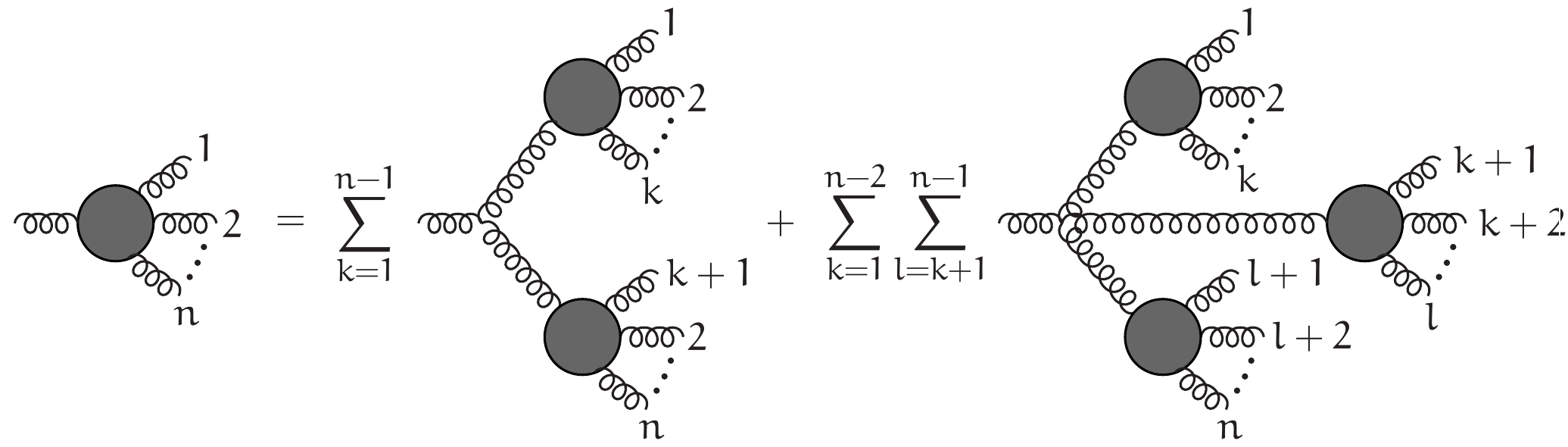,width=0.9\linewidth}
\caption{\label{Fig:DSg}Berends-Giele recursive relation for gluon currents. Enumerated partons are on-shell.}
}%
\myFigure{%
\epsfig{figure=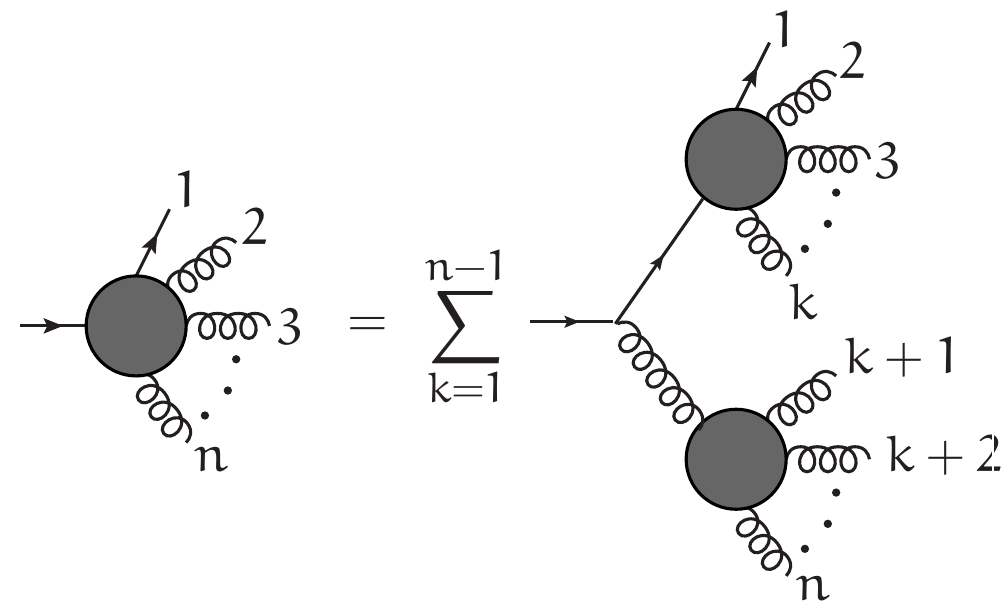,width=0.43\linewidth}
\caption{\label{Fig:DSq}Berends-Giele recursive relation for quark currents. Enumerated partons are on-shell.}
}%
These are for planar currents consisting of sums of planar graphs.
The enumerated external partons are on-shell.
The explicit vertices occuring in the figures are those from \Figure{Fig:colorordered}. %
%, and the virtual partons connecting those vertices with the blobs represent propagators, except when the blob contains only one on-shell parton.
%
The $n$ parton momenta may sum up to a light-like momentum, and then the deformation is not needed to obtain an amplitude from the ``off''-shell current.
The recursive relations thus constitute an efficient method to calculate multi-parton amplitudes.
In the following, we will use the notation
%
%%%%%%%%%%%%%%%%%%%%%%%%%%%%%%%%%%%%%%%%
\begin{equation}
K_{i,j}^\mu = \sum_{k=i}^jp_k^\mu
~,
\end{equation}
%%%%%%%%%%%%%%%%%%%%%%%%%%%%%%%%%%%%%%%%
%
and $J_{i,j}$ will refer to an off-shell current containing on-shell partons $i,i+1,\ldots,j$.

\subsection{\label{Sec:auxquarkdiv}Quark-antiquark pair with divergent momenta}
We are interested in the situation when the momenta of one or more on-shell external  partons diverge when a parameter $\Lambda$ becomes large, like in \Equation{auxmom1}.
Let us take the quark momentum $p_1^\mu=p_{A'}^\mu$ to diverge.
We will show now that to leading behavior in $p_{A'}^\mu$ the current $\leftA{J_{1,n}}$ can be written as $\leftA{p_{A'}}$ times a finite coefficient:
%
%%%%%%%%%%%%%%%%%%%%%%%%%%%%%%%%%%%%%%%%
\begin{equation}
\leftA{J_{1,n}} = E_{1,n}\,\leftA{p_{A'}}
\quad\textrm{with finite $E_{1,n}$ for diverging $p_1^\mu=p_{A'}^\mu$.}
\label{Eq383}
\end{equation}
%%%%%%%%%%%%%%%%%%%%%%%%%%%%%%%%%%%%%%%%
%
Of course, the equivalent holds for $\leftS{J_{1,n}}$, $\Arght{J_{1,n}}$, and $\Srght{J_{1,n}}$ with divergent quark/anti-quark momentum.
The recursive relation of \Figure{Fig:DSq} for quark currents is given by 
%
%%%%%%%%%%%%%%%%%%%%%%%%%%%%%%%%%%%%%%%%
\begin{equation}
\leftA{J_{1,n}} = 
\sum_{k=1,n-1}
\frac{\imag}{\sqrt{2}}\leftA{J_{1,k}}\,\slashJ_{k+1,n}\,\frac{\imag\slashK_{1,n}}{K_{1,n}^2}
~.
\end{equation}
%%%%%%%%%%%%%%%%%%%%%%%%%%%%%%%%%%%%%%%%
%
Via induction and using $\leftA{p_{A'}}\slashp_{A'}=0$ we have to leading behavior in $p_{A'}^\mu$:
%
%%%%%%%%%%%%%%%%%%%%%%%%%%%%%%%%%%%%%%%%
\begin{multline}
\leftA{J_{1,k}}\,\slashJ_{k+1,n}\,\frac{\slashK_{1,n}}{K_{1,n}^2}
\;=\;
E_{1,k}\,\leftA{p_{A'}}\,\slashJ_{k+1,n}\,\frac{\slashp_{A'}+\slashK_{2,n}}{(\slashp_{A'}+K_{2,n})^2}\\
\;=\;E_{1,k}\,\frac{2\lop{p_{A'}}{J_{k+1,n}}}{2\lop{p_{A'}}{K_{2,n}}+K_{2,n}^2}\,\leftA{p_{A'}}
\;+\;
E_{1,k}\,\leftA{p_{A'}}\,\slashJ_{k+1,n}\,\frac{\slashK_{2,n}}{2\lop{p_{A'}}{K_{2,n}}+K_{2,n}^2}
~.
\label{Eq432}
\end{multline}
%%%%%%%%%%%%%%%%%%%%%%%%%%%%%%%%%%%%%%%%
%
The second term on the right-hand-side is suppressed for diverging $p_{A'}^\mu$ compared to the first one, which has the form of \Equation{Eq383}.
Notice that the coefficient picks up the typical eikonal factor.
Closer inspection reveals that the amplitude, obtained when the (divergent) sum of momenta $K_{1,n}^\mu$ is light-like and equal to say $-p_{A}^\mu$~, is the one obtained by applying eikonal Feynman rules from the start.
The coefficients $E_{1,i}$ of the off-shell currents are finite because the number of propagators and vertices involving $p_{A'}^\mu$ match.
For the amplitude, there is one propagator less, and thus the amplitude diverges, necessitating the factor $1/\Lambda$ in \Equation{limit}.

\subsection{\label{Sec:auxglu}Gluon pair with divergent momenta}
Now consider the situation in which an on-shell external gluon, say gluon $j$, has divergent momentum $p_j^\mu=p_{A}^\mu$.
Its polarization vector $\varepsilon_j^\mu=\varepsilon_{A}^\mu$ can be chosen to be finite, \eg
%
%%%%%%%%%%%%%%%%%%%%%%%%%%%%%%%%%%%%%%%%
\begin{equation}
\varepsilon_{A}^\mu = 
\frac{\brktAS{p_{A}|\gamma^\mu|r}}{\sqrt{2}\brktSS{p_{A}|r}}
\quad\textrm{or}\quad
\varepsilon_{A'}^\mu = 
\frac{\brktAS{r|\gamma^\mu|p_{A}}}{\sqrt{2}\brktAA{r|p_{A}}}
~,
\label{Eq523}
\end{equation}
%%%%%%%%%%%%%%%%%%%%%%%%%%%%%%%%%%%%%%%%
%
depending on the helicity, for some arbitrary light-like momentum $r^\mu$.
It is straightforeward to deduce from the recursive equation that %
%to leading behavior in $p_{A}^\mu$, %
any gluon current $J_{i,k}^\mu$ with $i\leq j\leq k$, and thus containing gluon $j$, can be written as
%%%%%%%%%%%%%%%%%%%%%%%%%%%%%%%%%%%%%%%%
\begin{equation}
J_{i,k}^\mu = \varepsilon_{A}^\mu\,F_{i,k} + p_{A}^\mu\,G_{i,k}+V_{i,k}^\mu
\label{Eq454}
\end{equation}
%%%%%%%%%%%%%%%%%%%%%%%%%%%%%%%%%%%%%%%%
%
where $F_{i,k}$ and $p_{A}^\mu\,G_{i,k}$ are finite, \ie\ $G_{i,k}$ is suppressed, while $V_{i,k}^\mu$ represents the vanishing contribution.
Since four-point vertices do not contribute momentum factors in \Figure{Fig:DSg} while three-point vertices do, it is clear that the leading contribution comes from terms involving the latter.
Let us consider such a contribution to $J_{i,k}^\mu$ involving $J_{i,l}^\mu,J_{l+1,k}^\mu$.
Abbreviating the labels $(i,l)\to(1)$ and $(l+1,k)\to(2)$, and we have
%
%%%%%%%%%%%%%%%%%%%%%%%%%%%%%%%%%%%%%%%%
\begin{equation}
C_l^\mu = \frac{1}{\sqrt{2}}\frac{
\big(\lop{J_{1}}{J_{2}}\big)(K_{1}-K_{2})^\mu
+2\big(\lop{K_{2}}{J_{1}}\big)J_{2}^\mu
-2\big(\lop{K_{1}}{J_{2}}\big)J_{1}^\mu
}{(K_{1}+K_{2})^2}
\end{equation}
%%%%%%%%%%%%%%%%%%%%%%%%%%%%%%%%%%%%%%%%
%
We used current conservation already: $\lop{K_{1}}{J_{1}}=\lop{K_{2}}{J_{2}}=0$.
Let us say that $J_{1}^\mu$ contains on-shell gluon with the divergent momentum $p_{A}^\mu$, and let us write $K_{1}^\mu=p_{A}^\mu+\bar{K}_{1}^\mu$.
%
%The denominator diverges with $p_{A}^\mu$.
%
Substituting $J_{1}^\mu$ with \Equation{Eq454} reveals straightforwardly that $C_l^\mu$ conserves the structure of \Equation{Eq454}.
The coefficient of $\varepsilon_{A}^\mu$ deserves some more attention.
The contribution from $C_l^\mu$ is
%
%%%%%%%%%%%%%%%%%%%%%%%%%%%%%%%%%%%%%%%%
\begin{equation}
\frac{-1}{\sqrt{2}}\,\frac{2(\lop{p_{A}+\bar{K}_{1})}{J_{2}}}{2\lop{p_{A}}{(\bar{K}_{1}+K_{2})}+(\bar{K}_{1}+K_{2})^2}\,F_{1}\,\varepsilon_{A}^\mu
=
\frac{-1}{\sqrt{2}}\,\frac{\lop{p_{A}}{J_{2}}}{\lop{p_{A}}{(\bar{K}_{1}+K_{2})}}\,F_{1}\,\varepsilon_{A}^\mu
\end{equation}
%%%%%%%%%%%%%%%%%%%%%%%%%%%%%%%%%%%%%%%%
%
where the equality only holds to leading behavior in $p_{A}^\mu$.
So we see that the coefficent for $\varepsilon_{A}^\mu$ picks up the typical eikonal factor.
Now let $K_{1,n}^\mu=-p_{A'}^\mu$ be light-like, and remember that a tilde indicates that the propagator of the ``off-shell'' leg is not included.
Then
%
%%%%%%%%%%%%%%%%%%%%%%%%%%%%%%%%%%%%%%%%
\begin{equation}
\lop{\varepsilon_{A'}}{\tilde{J}_{1,n}}
= \tilde{F}_{1,n}\,\lop{\varepsilon_{A'}}{\varepsilon_{A}}
+ \tilde{G}_{1,n}\,\lop{\varepsilon_{A'}}{p_{A}} + \lop{\varepsilon_{A'}}{\tilde{V}_{1,n}}
\end{equation}
%%%%%%%%%%%%%%%%%%%%%%%%%%%%%%%%%%%%%%%%
%
is an $(n+1)$-gluon amplitude, where $\tilde{F}_{1,n}$ diverges with $p_{A}^\mu$, while $\tilde{G}_{1,n}$ and $\tilde{V}_{1,n}^\mu$ are finite.
Let the two polarization vectors have the opposite helicity, and insert \Equation{auxmomentlimit} and \Equation{AAprimeSpinors}.
We see that $\lop{\varepsilon_{A'}}{\varepsilon_{A}}\to1$ while %
%$\lop{p_{A'}}{\varepsilon_{A}}\to\kappa^*/\sqrt{2}$ %
$\lop{\varepsilon_{A'}}{p_{A}}$ %
is finite, so the leading contribution to the amplitude is given by $\tilde{F}_{1,n}$.
Closer inspection reveals that it is the same amplitude one would get with an auxiliary quark-antiquark pair instead of the gluons with momenta $p_{A}^\mu,p_{A'}^\mu$.
Dividing the amplitude by $\Lambda$ and taking $\Lambda\to\infty$ leads to the same amplitude as with an auxiliary quark-antiquark pair, and can be calculated with the same eikonal Feynman rules.

\subsection{\label{Sec:quarkgludiv}Quark and gluon with divergent momenta}
We will encounter tree-level amplitudes with a quark and a gluon carrying diverging momenta $p_{A'}^\mu$ and $p_{A}^\mu$ which, like before, add up to finite momentum.
Now we need to highlight that this implies that
%
%%%%%%%%%%%%%%%%%%%%%%%%%%%%%%%%%%%%%%%%
\begin{equation}
\lop{p_{A}}{p_{A'}}
\quad\textrm{and}\quad
\brktAA{p_{A}|p_{A'}}
\quad\textrm{and}\quad
\brktSS{p_{A}|p_{A'}}
\quad\textrm{are finite.}
\end{equation}
%%%%%%%%%%%%%%%%%%%%%%%%%%%%%%%%%%%%%%%%
%
The critical point regarding the behavior of the amplitude is the behavior of $\lop{p_{A'}}{J_{k+1,n}}$ in \Equation{Eq432}, in vertices where off-shell currents containing the quark and the gluon with momentum $p_{A}^\mu$ meet, so $J_{k+1,n}^\mu$ is finite and has the form of \Equation{Eq454}.
Taking into account the behavior of the momenta stated above, we see that $\lop{p_{A'}}{V_{k+1,n}}$ is finite and that $\lop{p_{A'}}{p_{A}}\,G_{k+1,n}$ is actually suppressed, because $G_{k+1,n}$ is suppressed.
The behavior of the amplitude is determined by that of $\lop{p_{A'}}{\varepsilon_{A}}$.
Considering polarization vectors of the type of \Equation{Eq523}, the behavior is eventually determined by that of $\brktAA{p_{A}|p_{A'}}$ or $\brktSS{p_{A}|p_{A'}}$, which we just saw are finite.
This of course invalidates the statements before about what gives the leading contribution, but most importantly, indicates that the amplitude itself is finite.
It is easy to see that the same holds for an antiquark instead of a quark with divergent momentum.

\subsection{\label{Sec:auxilquarkglu}Quark-antiquark pair and a gluon with divergent momenta}
We will also encounter the situation in which a quark-antiquark pair and a gluon have divergent momenta, and we need to study their behavior with these momenta.
We label them $A,A',A''$ where the gluon label is without accent.
They add up to a finite momentum, and their leading behavior may be thought of as given by 
%
%%%%%%%%%%%%%%%%%%%%%%%%%%%%%%%%%%%%%%%%
\begin{equation}
p_{A}^\mu\sim\alpha\Lambda p^\mu
\quad,\quad
p_{A'}^\mu\sim\alpha'\Lambda p^\mu
\quad,\quad
p_{A''}^\mu\sim\alpha''\Lambda p^\mu
\quad,\quad
\alpha+\alpha'+\alpha''=0
\quad,
\end{equation}
%%%%%%%%%%%%%%%%%%%%%%%%%%%%%%%%%%%%%%%%
%
with
%
%%%%%%%%%%%%%%%%%%%%%%%%%%%%%%%%%%%%%%%%
\begin{equation}
\lop{p_{A}}{p_{A'}} = \Ord\big(\Lambda\big)
\quad,\quad
\lop{p_{A'}}{p_{A''}} = \Ord\big(\Lambda\big)
\quad,\quad
\lop{p_{A''}}{p_{A}} = \Ord\big(\Lambda\big)
\quad.
\label{Eq567}
\end{equation}
%%%%%%%%%%%%%%%%%%%%%%%%%%%%%%%%%%%%%%%%
%
The latter implies that the spinor products of the momenta behave as $\sqrt{\Lambda}$.
So contrary to the previous type of amplitudes, this time the term with $\lop{p_{A'}}{\varepsilon_{A}}$ in \Equation{Eq432} {\em does\/} give the leading contribution, but behaves only as $\sqrt{\Lambda}$, leading to the amplitude to behave as $\sqrt{\Lambda}$ rather than $\Lambda$,with  which will prove to have important consequences.

\subsection{\label{Sec:qqfamdiv}Quark and antiquark of different flavor with divergent momenta}
Finally, we will also encounter amplitudes with a quark an antiquark with divergent momenta, which do not have the same flavor.
So this involves amplitudes with at least two quark-antiquark pairs.
The graphs contributing to such amplitudes must have at least one more gluon propagator than gluon three-point vertices with diverging momentum components flowing through, so they together contribute a factor $1/\Lambda$.
Quark propagators are finite, while the external quark and antiquark spinors each contribute a factor $\sqrt{\Lambda}$, so in total the amplitude is finite.

\section{One-loop amplitudes\label{Sec:One-loop}}
The derivations above and in \cite{vanHameren:2012if} are rather simple because a tree-level amplitude is a rational function of the momenta involved.
For a one-loop amplitude this is not the case anymore, because it is the integral of a rational function of momenta, including the integration momentum.
We will see in particular that this integral will lead to terms proportional to powers of $\log\Lambda$.
Before taking $\Lambda\to\infty$, all integrals are well-defined, and our approach can be interpreted as a method to regularize integrals with linear denominators: in \cite{vanHameren:2012if} we arrived at the eikonal Feynman rules for the off-shell amplitudes via the relation
%
%%%%%%%%%%%%%%%%%%%%%%%%%%%%%%%%%%%%%%%%
\begin{equation}
\imag\,\frac{\slashp_A+\slashK}{(p_A+K)^2}
\overset{\Lambda\to\infty}{\longrightarrow} \frac{\imag\,\slashp}{2p\!\cdot\!K}
~,
\end{equation}
%%%%%%%%%%%%%%%%%%%%%%%%%%%%%%%%%%%%%%%%
%
and our proposal to regularize loop graphs with linear denominators is to essentially read the above backwards:
%
%%%%%%%%%%%%%%%%%%%%%%%%%%%%%%%%%%%%%%%%
\begin{equation}
\frac{1}{2\lop{p}{K}} \to \frac{\Lambda}{(p_A+K)^2}
\label{regularization}
~.
\end{equation}
%%%%%%%%%%%%%%%%%%%%%%%%%%%%%%%%%%%%%%%%
%
In this write-up, we deal with the complication caused by the fact that counting powers of $\Lambda$ in numerators and denominators in general does not commute with the loop integral.
In order to address this issue, we must start with some formalism.

Before the limit $\Lambda\to\infty$, all existing techniques to calculate one-loop amplitudes can be applied.
In particular, the following decomposition is valid: 
%
%%%%%%%%%%%%%%%%%%%%%%%%%%%%%%%%%%%%%%%%
\begin{multline}
\int \dDell\,\frac{\EuScript{N}(\ell)}{\prod_i\Den_i(\ell)}
= \sum_{i,j,k,l} \BoxC_{ijkl}\,\mathrm{Box}_{ijkl}
             + \sum_{i,j,k} \TriC_{ijk}\,\mathrm{Tri}_{ijk}
             + \sum_{i,j} \BubC_{ij}\,\mathrm{Bub}_{ij}
             + \sum_{i} \TadC_{i}\,\mathrm{Tad}_{i}
\\
             + \EuScript{R} + \EuScript{O}(\varepsilon)
~,
\label{oneloopdecom}
\end{multline}
%%%%%%%%%%%%%%%%%%%%%%%%%%%%%%%%%%%%%%%%
%
where the {\em master integrals\/}  are defined as
%
%%%%%%%%%%%%%%%%%%%%%%%%%%%%%%%%%%%%%%%%
\begin{align}
\mathrm{Box}_{ijkl}&= \int\frac{\dDell}{\Den_i(\ell)\Den_j(\ell)\Den_k(\ell)\Den_l(\ell)}
\quad,&
\mathrm{Tri}_{ijk} &= \int\frac{\dDell}{\Den_i(\ell)\Den_j(\ell)\Den_k(\ell)}
\quad,\quad\notag\\
\mathrm{Bub}_{ij} &= \int\frac{\dDell}{\Den_i(\ell)\Den_j(\ell)}
\quad,&
\mathrm{Tad}_{i} &= \int\frac{\dDell}{\Den_i(\ell)}
~.
\end{align}
%%%%%%%%%%%%%%%%%%%%%%%%%%%%%%%%%%%%%%%%
%
The left-hand side is a one-loop integral represented by a one-loop Feynman graph, or a collection of graphs with the same loop-denominators.
The denominators are quadratic in $\ell^\mu$ and come from the propagator denominators in the one-loop Feynman graphs.
In general, they have the form
%
%%%%%%%%%%%%%%%%%%%%%%%%%%%%%%%%%%%%%%%%
\begin{equation}
\Den_i(\ell) = (\ell+K_i)^2-m_i^2+\imag\eta
~,
\end{equation}
%%%%%%%%%%%%%%%%%%%%%%%%%%%%%%%%%%%%%%%%
%
where $K_i$ is a sum of a subset of external momenta, $m_i$ is the mass of an internal particle, and $\eta$ is small and positive in order to enforce the Feynman prescription.
The ``normalization'' of the dimensionally regulated loop volume element is given by
%
%%%%%%%%%%%%%%%%%%%%%%%%%%%%%%%%%%%%%%%%
\begin{equation}
\dDell=
\frac%
{\Gamma(2-\varepsilon)\mu^{2\varepsilon}}
{\Gamma^2(1-\varepsilon)\Gamma(1+\varepsilon)\imag\pi^{2-\varepsilon}}
\,d^{4-2\varepsilon}\ell
~.
\end{equation}
%%%%%%%%%%%%%%%%%%%%%%%%%%%%%%%%%%%%%%%%
%
The sums in \Equation{oneloopdecom} are over all possible values of non-equal indices.
This may include combinations that are impossible from the point of the Feynman graphs, in which case we simply understand that the coefficients vanish.
The master integrals include (poly)-logarithms of rational functions of external momenta.
The term $\EuScript{R}$ represents the remnant rational terms caused by the divergent nature of the loop integral, and the last term reminds us that the decomposition as given above is only valid up to $\EuScript{O}(\varepsilon)$ within dimensional regularization, which however is sufficient for NLO calculations.

We will only consider loop integrals with massless denominators, so we do not need to consider one-point (tadpole) master integrals since they vanish within dimensional regularization.
The other master integrals are scalar four-point, three-point, and two-point integrals, and we will refer to them as boxes, triangles, and bubbles.
We also need to introduce some more notation regarding the denominator factors.
If we wish to highlight that a denominator comes from an auxiliary quark propagator, then we give it a subscript $A$, and if we wish to highlight that it does not, we give it a subscript $O$:
%
%%%%%%%%%%%%%%%%%%%%%%%%%%%%%%%%%%%%%%%%
\begin{align}
\Den_{Aj} \quad&\to\quad\textrm{comes from an auxiliary quark propagator.}
\notag\\
\Den_{Oj} \quad&\to\quad\textrm{comes from another propagator.}
~.
\label{dennot0}
\end{align}
%%%%%%%%%%%%%%%%%%%%%%%%%%%%%%%%%%%%%%%%
%
If we wish to highlight that a denominator depends on $\Lambda$ (or not), then we indicate this with a superscript:
%
%%%%%%%%%%%%%%%%%%%%%%%%%%%%%%%%%%%%%%%%
\begin{align}
\Den^\Lambda_{j} &= (\ell+\Lambda p+K_j)^2+\imag\eta
\notag\\
\Den^0_{j} &= (\ell+K_j)^2+\imag\eta
~.
\label{dennot}
\end{align}
%%%%%%%%%%%%%%%%%%%%%%%%%%%%%%%%%%%%%%%%
%
Realize that a $\Lambda$-dependent denominator does not necessarily have to come from an auxiliary quark propagator, since we may have shifted the loop momentum with an amount $\Lambda p$.
Consider for example the graph
%
%%%%%%%%%%%%%%%%%%%%%%%%%%%%%%%%%%%%%%%%
\begin{equation}
\graph{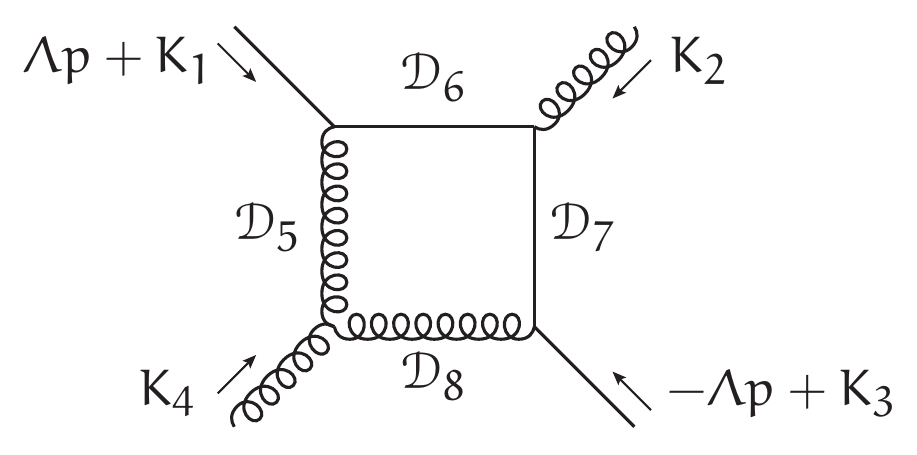}{38}{9}
\label{doubleGraph2}
~.
\end{equation}
%%%%%%%%%%%%%%%%%%%%%%%%%%%%%%%%%%%%%%%%
%
The external momenta do not have to be light-like, and tree-level blobs may be attached to the external lines.
The enumeration of the objects has no particular meaning.
We have $\Den_5=\Den_{O5}$, $\Den_8=\Den_{O8}$, and $\Den_6=\Den_{A6}$, $\Den_7=\Den_{A7}$.
One choice of momentum flow could be
%
%%%%%%%%%%%%%%%%%%%%%%%%%%%%%%%%%%%%%%%%%
\begin{align}
\Den_5&=\Den_5^0=\ell^2
\quad,\quad&&
\Den_6=\Den_6^\Lambda=(\ell+\Lambda p+K_1)^2
\nonumber\\
\Den_8&=\Den_8^0=(\ell-K_4)^2
\quad,\quad&&
\Den_7=\Den_7^\Lambda=(\ell+\Lambda p+K_1+K_2)^2
\quad,
\end{align}
%%%%%%%%%%%%%%%%%%%%%%%%%%%%%%%%%%%%%%%%
%
but also the following is possible
%
%%%%%%%%%%%%%%%%%%%%%%%%%%%%%%%%%%%%%%%%%
\begin{align}
\Den_5&=\Den_5^\Lambda=(\ell-\Lambda p - K_1)^2
\quad,\quad&&
\Den_6=\Den_6^0=\ell^2
\nonumber\\
\Den_8&=\Den_8^\Lambda=(\ell-\Lambda p - K_1-K_4)^2
\quad,\quad&&
\Den_7=\Den_7^0=(\ell+K_2)^2
\quad.
\end{align}
%%%%%%%%%%%%%%%%%%%%%%%%%%%%%%%%%%%%%%%%
%
Keeping this possibility of shifting the loop momentum in mind, we see that the only $\Lambda$-dependent master integrals we need to consider are given by
%
%%%%%%%%%%%%%%%%%%%%%%%%%%%%%%%%%%%%%%%%
\begin{align}
\int\frac{\dDell}{\Den_i^\Lambda(\ell)\Den_j^\Lambda(\ell)\Den_k^0(\ell)\Den_l^0(\ell)}
&=
\frac{  (\ln\Lambda)^2f^{(2)}_{ijkl}
      + (\ln\Lambda)  f^{(1)}_{ijkl}
      +               f^{(0)}_{ijkl} + \Ord\big(\Lambda^{-1}\big)}
     {\Lambda^2}
\label{MI42}
\\
\int\frac{\dDell}{\Den_i^\Lambda(\ell)\Den_j^0(\ell)\Den_k^0(\ell)\Den_l^0(\ell)}
&=
\frac{  (\ln\Lambda)^2f^{(2)}_{ijkl}
      + (\ln\Lambda)  f^{(1)}_{ijkl}
      +               f^{(0)}_{ijkl} + \Ord\big(\Lambda^{-1}\big)}
     {\Lambda}
\label{MI41}
\\
\int\frac{\dDell}{\Den_i^\Lambda(\ell)\Den_j^0(\ell)\Den_k^0(\ell)}
&=
\frac{  (\ln\Lambda)^2f^{(2)}_{ijk}
      + (\ln\Lambda)  f^{(1)}_{ijk}
      +               f^{(0)}_{ijk} + \Ord\big(\Lambda^{-1}\big)}
     {\Lambda}
\label{MI3}
\\
\int\frac{\dDell}{\Den_i^\Lambda(\ell)\Den_j^0(\ell)}
&=
        (\ln\Lambda)  f^{(1)}_{ij}
      +               f^{(0)}_{ij} + \Ord\big(\Lambda^{-1}\big)
\label{MI2}
\end{align}
%%%%%%%%%%%%%%%%%%%%%%%%%%%%%%%%%%%%%%%%
%
The precise form of the coefficients $f$ depends on the kinematics, and all relevant configurations with massless denominators can be found in \Section{scalarintegrals}.
Here, we only present the bubble:
%
%
%%%%%%%%%%%%%%%%%%%%%%%%%%%%%%%%%%%%%%%%
\begin{equation}
f^{(1)}_{ij} = -1 + \Ord(\varepsilon)
\quad,\quad
f^{(0)}_{ij} = \frac{1}{\varepsilon} + 2 - \ln\left(-\frac{2\lop{p}{(K_i-K_j)}+\imag\eta}{\mu^2}\right) + \Ord(\varepsilon)
\label{MI2expr}
~,
\end{equation}
%%%%%%%%%%%%%%%%%%%%%%%%%%%%%%%%%%%%%%%%
%
and the only triangle with a $\Lambda$-dependent denominator that does not follow the form of (\ref{MI3}):
%
%%%%%%%%%%%%%%%%%%%%%%%%%%%%%%%%%%%%%%%%
\begin{multline}
\mathrm{Tri}_{a1}\big(k_T^2\big)
\equiv
\graph{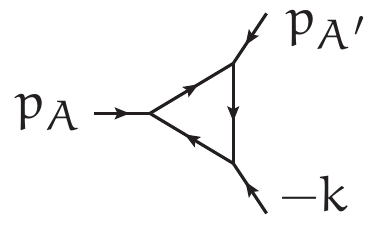}{15}{4}
=
\int\frac{\dDell}{\ell^2\,(\ell+p_A)^2\,(\ell+k)^2}
\\
=
\frac{1}{k_T^2}\left\{
  \frac{1}{\varepsilon^2}
 -\frac{1}{\varepsilon}\ln\left(\frac{k_T^2}{-\mu^2}\right)
 +\frac{1}{2}\ln^2\left(\frac{k_T^2}{-\mu^2}\right)
\right\}
+ \Ord(\varepsilon)
~.
\label{MI3viol}
\end{multline}
%%%%%%%%%%%%%%%%%%%%%%%%%%%%%%%%%%%%%%%%
%
The arrows in the graph indicate momentum flow.
The meaning of the label "$a1$" is explained in \Figure{triangles}.

\subsection{General findings\label{Sec:Statements}}
Regarding the master integrals, it is important to realize that while \Equation{MI42}, \Equation{MI41}, \Equation{MI3} are consistent with the prescription of \Equation{regularization}, bubbles \Equation{MI2} and the triangle of \Equation{MI3viol} {\em are not consistent\/}.
As a result, na\"i{}ve power counting regarding $\Lambda$ in a one-loop integrand may fail to give the correct result for the integral.
Consider for example the one-loop integral
%
%%%%%%%%%%%%%%%%%%%%%%%%%%%%%%%%%%%%%%%%
\begin{equation}
\graph{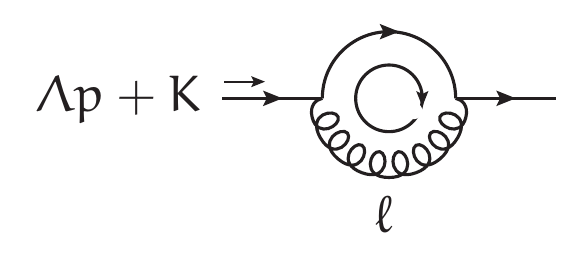}{24}{4.5}
=
\int\dDell\frac{\langle p|\gamma^\mu(\slashl + \Lambda\slashp +\slashK)\gamma_\mu|p]}
               {\ell^2(\ell+\Lambda p+K)^2}
~.
\label{vanishbubble}
\end{equation}
%%%%%%%%%%%%%%%%%%%%%%%%%%%%%%%%%%%%%%%%
%
Here, we assume the Feynman gauge, suppress color factors and coupling constants, and already took the leading $\Lambda$ contribution in the tree-level attachments at the two external lines.
The leading $\Lambda$ contribution in the numerator of the integrand above vanishes: $\langle p|\gamma^\mu\slashp\gamma_\mu|p]=0$.
A short calculation however shows that
%
%%%%%%%%%%%%%%%%%%%%%%%%%%%%%%%%%%%%%%%%
\begin{equation}
\int\dDell\frac{\langle p|\gamma^\mu(\slashl + \Lambda\slashp +\slashK)\gamma_\mu|p]}
               {\ell^2(\ell+\Lambda p+K)^2}
=
2(\varepsilon-1)\lop{p}{K}\int\dDell\frac{1}{\ell^2(\ell+\Lambda p+K)^2}
~,
\end{equation}
%%%%%%%%%%%%%%%%%%%%%%%%%%%%%%%%%%%%%%%%
%
which, according to \Equation{MI2} and \Equation{MI2expr}, does not vanish at all.
If the leading term in $\Lambda$ in the numerator of this integrand would not vanish, because for example it was terminated by other spinors, then this would cause the one-loop amplitude to diverge linearly with $\Lambda$ and severely undermine our project.

An obvious approach to calculate one-loop integrals with auxiliary quark propagators would be to take $\Lambda\to\infty$ in the integrand of left-hand side of \Equation{oneloopdecom}, that is calculate it with the eikonal Feynman rules, and apply the integrand methods of \cite{Ossola:2006us,Ellis:2007br} to arrive at the decomposition represented by the right-hand side.
The substitution (\ref{regularization}) would only be applied in the master integrals.
In light of the foregoing, this will however not lead to the correct result regarding the bubbles.
For the amplitude not to diverge worse than logarithmically with $\Lambda$, the bubble coefficients with a $\Lambda$-dependent denominator would have to vanish, but we just saw an example for which those bubbles {\em do\/} contribute, without the bad behavior.

In the rest of the paper we demonstrate that the one-loop amplitude indeed does not behave worse than linearly in $\Lambda$, that is the prescription of \Equation{limit} 
leads at most to divergencies of the type $\log^2\Lambda$.
Furthermore, for determining the coefficients for boxes and triangles, except the anomalous triangle (\ref{MI3viol}), the eikonal Feynman rules can be applied on the integrand, with the scalar integrals interpreted following \Equation{regularization}.
Finally it is shown how to calculate the coefficients for the anomalous triangle and the bubbles, as well as the rational contribution.
The exposition of these points in the following will be rather constructive, and we will derive the necessary limits for the solutions to the so-called cut equations and the master integrals.
In order to keep the argumentation coherent, some details are referred to appendices.

\section{Integrand-level reduction\label{sec:OPP}}
The integrand-level reduction methods of~\cite{Ossola:2006us,Ellis:2007br} will be essential for our argument, and we review some essential points here.
They allow for the determination of the coefficients $\GenC$ in \Equation{oneloopdecom}, and are based on the fact that the one-loop integrand, before integration, can be decomposed as
%
%%%%%%%%%%%%%%%%%%%%%%%%%%%%%%%%%%%%%%%%
\begin{multline}
\frac{\EuScript{N}(\ell)}{\prod_i\Den_i(\ell)}
= \sum_{i,j,k,l}\frac{\BoxC_{ijkl}+\tBoxPol_{ijkl}(\ell)}
                     {\Den_{i}(\ell)\Den_{j}(\ell)\Den_{k}(\ell)\Den_{l}(\ell)}
+ \sum_{i,j,k}\frac{\TriC_{ijk}+\tTriPol_{ijk}(\ell)}
                   {\Den_{i}(\ell)\Den_{j}(\ell)\Den_{k}(\ell)}
+ \sum_{i,j}\frac{\BoxC_{ij}+\tBubPol_{ij}(\ell)}
                 {\Den_{i}(\ell)\Den_{j}(\ell)}
\\
+ \sum_{i}\frac{\TadC_{i}+\tTadPol_{i}(\ell)}
                 {\Den_{i}(\ell)}
\label{integranddecom}
\end{multline}
%%%%%%%%%%%%%%%%%%%%%%%%%%%%%%%%%%%%%%%%
%
where the coefficients $\GenC$ are the same as in the integrated relation (\ref{oneloopdecom}).
Evaluating the left-hand-side and the right-hand-side for enough values of $\ell^\mu$, a linear system can be constucted to solve for the desired coefficients~$\GenC$ and, necessarily, the coefficients~$\tGenC$ hidden in the {\em spurious\/} polynomials~$\tGenPol(\ell)$.
The spurious box polynomial $\tBoxPol_{ijkl}(\ell)$ is linear in $\ell^\mu$:
%
%%%%%%%%%%%%%%%%%%%%%%%%%%%%%%%%%%%%%%%%
\begin{equation}
\tBoxPol_{ijkl}(\ell) = \tilde{C}_{ijkl}\,\lop{\ell}{e_{ijkl}}
\quad,
\end{equation}
%%%%%%%%%%%%%%%%%%%%%%%%%%%%%%%%%%%%%%%%
%
where $e_{ijkl}$ is such that
%
%%%%%%%%%%%%%%%%%%%%%%%%%%%%%%%%%%%%%%%%
\begin{equation}
\lop{(K_j-K_i)}{e_{ijkl}}=
\lop{(K_k-K_i)}{e_{ijkl}}=
\lop{(K_l-K_i)}{e_{ijkl}}=0
~.
\end{equation}
%%%%%%%%%%%%%%%%%%%%%%%%%%%%%%%%%%%%%%%%
%
In the language of~\cite{Ellis:2007br}, $e_{ijkl}^\mu$ spans the {\em trivial space} of the box.
This guarantees that the non-constant part of the polynomial integrates to zero as
%
%%%%%%%%%%%%%%%%%%%%%%%%%%%%%%%%%%%%%%%%
\begin{equation}
\int\dDell\,\frac{\tBoxPol_{ijkl}(\ell)}{\Den_i(\ell)\Den_j(\ell)\Den_k(\ell)\Den_l(\ell)} = 0
~,
\end{equation}
%%%%%%%%%%%%%%%%%%%%%%%%%%%%%%%%%%%%%%%%
%
and that the box-part of \Equation{oneloopdecom} is indeed recovered after integration.
The complement of the trivial space is called the {\em physical space\/}.
The trivial space for the triangles is two-dimensional, and the spurious triangle polynomial $\tTriPol_{ijk}(\ell)$ is qubic, and can be expanded as:
%
%%%%%%%%%%%%%%%%%%%%%%%%%%%%%%%%%%%%%%%%
\begin{equation}
\tTriPol_{ijk}(\ell)
=
 \tTriC_{ijk}^{(1)}\left(\lop{\ell}{e_{ijk}^{(1)}}\right)
+\tTriC_{ijk}^{(2)}\left(\lop{\ell}{e_{ijk}^{(2)}}\right)
+\tTriC_{ijk}^{(11)}\left(\lop{\ell}{e_{ijk}^{(1)}}\right)^2
+\tTriC_{ijk}^{(12)}\left(\lop{\ell}{e_{ijk}^{(1)}}\right)\left(\lop{\ell}{e_{ijk}^{(2)}}\right)
+\cdots
%
%= \sum_{n=1}^3\left\{
%  \tTriC_{ijk}^{1,n}\left(\lop{\ell}{e_{ijk}^{(1)}}\right)^n
%  +\tTriC_{ijk}^{2,n}\left(\lop{\ell}{e_{ijk}^{(2)}}\right)^n \right\}
~,
\end{equation}
%%%%%%%%%%%%%%%%%%%%%%%%%%%%%%%%%%%%%%%%
%
where $e_{ijk}^{(1)\mu}$ and $e_{ijk}^{(2)\mu}$ are such that
%
%%%%%%%%%%%%%%%%%%%%%%%%%%%%%%%%%%%%%%%%
\begin{equation}
\lop{(K_j-K_i)}{e_{ijk}^{(1,2)}}=
\lop{(K_k-K_i)}{e_{ijk}^{(1,2)}}=0
~.
\end{equation}
%%%%%%%%%%%%%%%%%%%%%%%%%%%%%%%%%%%%%%%%
%
This guarantees that $\tTriPol_{ijk}(\ell)$ divided by the three relevant denominators integrates to zero.
The decomposition of $\tTriPol_{ijk}(\ell)$ is not unique and depends on how $e_{ijk}^{(1,2)}$ are constructed, but has at most $6$ free coefficients $\tTriC_{ijk}$.
The trivial space of the bubbles is three-dimensional and the spurious polynomials are at most quadratic with 8 free coefficients.

The linear system to be constucted and solved for the coefficients can simplified by choosing $\ell^\mu$ such that denominators vanish.
The box coefficients can be found from
%
%%%%%%%%%%%%%%%%%%%%%%%%%%%%%%%%%%%%%%%%%
\begin{align}
\mathrm{LHS}_{ijkl}(\ell)
&= \BoxC_{ijkl}+\tBoxPol_{ijkl}(\ell)
\quad\textrm{with $\ell$ such that}\;
\Den_{i}(\ell)=\Den_{j}(\ell)=\Den_{k}(\ell)=\Den_{l}(\ell)=0
~,
\label{Eq:quadcut}
\\
\mathrm{LHS}_{ijkl}(\ell)
&\equiv
\mathrm{Res}_{ijkl}(\ell)
=
\frac{\EuScript{N}(\ell)}{\prod_{m\neq i,j,k,l}\Den_m(\ell)}
~.\label{Eq:quadLHS}
\end{align}
%%%%%%%%%%%%%%%%%%%%%%%%%%%%%%%%%%%%%%%%
%
The solution $\ell^\mu$ to the {\em cut equations\/}, setting the denominators to zero, is not unique.
In case of the box, there are $2$, exactly enough to determine both $\BoxC_{ijkl}$ and $\tBoxC_{ijkl}$.
Although eventually $\tBoxC_{ijkl}$ is not needed, it is needed to reconstuct the box polynomial in order to write the equation for the triangle coefficients:
%
%%%%%%%%%%%%%%%%%%%%%%%%%%%%%%%%%%%%%%%%
\begin{align}
\mathrm{LHS}_{ijk}(\ell)
%\EuScript{N}_4(\ell)
&=
\TriC_{ijk} + \tTriPol_{ijk}(\ell)
\quad\textrm{with $\ell$ such that}\;
\Den_{i}(\ell)=\Den_{j}(\ell)=\Den_{k}(\ell)=0
~,\label{Eq:tricut}\\
\mathrm{LHS}_{ijk}(\ell)
&\equiv
\mathrm{Res}_{ijk}(\ell)
-\sum_l\frac{\BoxC_{ijkl}+\tBoxPol_{ijkl}(\ell)}{\Den_l(\ell)}
~,\label{Eq:triLHS}\\
\mathrm{Res}_{ijk}(\ell)
&\equiv
\frac{\EuScript{N}(\ell)}{\prod_{m\neq i,j,k}\Den_m(\ell)}
~.
\label{Eq:triRes}
\end{align}
%%%%%%%%%%%%%%%%%%%%%%%%%%%%%%%%%%%%%%%%
%
Now there is an infinite number of possible choices for $\ell^\mu$, so certainly enough to determine $\TriC_{ijk}$ and the $6$ spurious coefficients.
The bubble coeffients are determined from
%
%%%%%%%%%%%%%%%%%%%%%%%%%%%%%%%%%%%%%%%%
\begin{align}
\mathrm{LHS}_{ij}(\ell)
&= \BubC_{ij} + \tBubPol_{ij}(\ell)
\quad\textrm{with $\ell$ such that}\;
\Den_{i}(\ell)=\Den_{j}(\ell)=0
~,\label{Eq:bubcut}\\
\mathrm{LHS}_{ij}(\ell)
&\equiv
\mathrm{Res}_{ij}(\ell)
-\sum_{k,l}\frac{\BoxC_{ijkl}+\tBoxPol_{ijkl}(\ell)}{\Den_k(\ell)\Den_l(\ell)}
-\sum_{k}\frac{\TriC_{ijk}+\tTriPol_{ijk}(\ell)}{\Den_k(\ell)}
~,\label{Eq:bubLHS}\\
\mathrm{Res}_{ij}(\ell)
&\equiv
\frac{\EuScript{N}(\ell)}{\prod_{m\neq i,j}\Den_m(\ell)}
~.
\end{align}
%%%%%%%%%%%%%%%%%%%%%%%%%%%%%%%%%%%%%%%%
%
Also here, there is an infinite number of choices for $\ell^\mu$, but only $9$ are needed.
Although masless tadpole master integrals vanish, their coefficients can still be necessary for the calculation of the rational contribution in \Equation{oneloopdecom}.
This is, however, not the case for tadpoles with a $\Lambda$-dependent denominator.
This is addressed in \Section{Sec:complication}.

In the following we will analyse the behavior of the coefficients for large $\Lambda$.
For a master integral that behaves as $\Lambda^{-1}$, for example, the coefficient must not behave worse than quadratically.
On the other hand, a coefficient may also turn out to behave as milder than quadratically, eliminating the contribution of the master integral altogether according to \Equation{limit}.
A particular complication, that occurs exactly with the bubbles with a $\Lambda$-dependent denominator and with the anomalous triangle (\ref{MI3viol}), is that the solutions to the cut equations may diverge with $\Lambda$.

\subsection{A more explicit example}
First, however, it is instructive to have a closer look at a slightly more explicit example.
Consider a triangle with two light-like external momenta $p_1^\mu,p_2^\mu$ and a third external momentum $k_3^\mu=-p_1^\mu-p_2^\mu$.
The three denominators are
%
%%%%%%%%%%%%%%%%%%%%%%%%%%%%%%%%%%%%%%%%
\begin{equation}
\ell^2\quad,\quad(\ell+p_1)^2\quad,\quad(\ell+p_1+p_2)^2
\quad.
\end{equation}
%%%%%%%%%%%%%%%%%%%%%%%%%%%%%%%%%%%%%%%%
%
Solutions for $\ell^\mu$ to the cut equations that put all three of these denominators to zero can conveniently be constructed with help of the vectors
%
%%%%%%%%%%%%%%%%%%%%%%%%%%%%%%%%%%%%%%%%
\begin{equation}
e_{1}^\mu = \srac{1}{2}\brktAS{p_1|\gamma^\mu|p_2}
\quad,\quad
e_{2}^\mu = \srac{1}{2}\brktAS{p_2|\gamma^\mu|p_1}
\quad,
\end{equation}
%%%%%%%%%%%%%%%%%%%%%%%%%%%%%%%%%%%%%%%%
%
that satisfy $\lop{p_{1,2}}{e_{1,2}}=0$, and span the trivial space.
Solutions to the cut equations are given by
%
%%%%%%%%%%%%%%%%%%%%%%%%%%%%%%%%%%%%%%%%
\begin{equation}
\ell_1^\mu = ze_1^\mu - p_1^\mu
\quad\textrm{and}\quad 
\ell_2^\mu = ze_2^\mu - p_1^\mu
\quad,
\end{equation}
%%%%%%%%%%%%%%%%%%%%%%%%%%%%%%%%%%%%%%%%
%
for any value of $z$.
At the solution $\ell_{1}^\mu$, any denominator other than the three from the cut equations is given by 
%
%%%%%%%%%%%%%%%%%%%%%%%%%%%%%%%%%%%%%%%%
\begin{equation}
(\ell_{1}+K)^2 = K^2 + 2z\lop{e_{1}}{K}
~,
\end{equation}
%%%%%%%%%%%%%%%%%%%%%%%%%%%%%%%%%%%%%%%%
%
and thus becomes a linear function of the variable $z$.
The same works with $\ell_2^\mu$.
So we see that \Equation{Eq:tricut} becomes an equation of rational functions in the single variable $z$.
Let us say the total number of denominators is $n$, and the highest power of $\ell$ in the numerator $\EuScript{N}(\ell)$ is $r$.
Via repeated partial fractioning
%
%%%%%%%%%%%%%%%%%%%%%%%%%%%%%%%%%%%%%%%%
\begin{equation}
\frac{1}{(a+z)(b+z)} = \frac{1}{b-a}\left(\frac{1}{a+z}-\frac{1}{b+z}\right)
\end{equation}
%%%%%%%%%%%%%%%%%%%%%%%%%%%%%%%%%%%%%%%%
%
we can turn the product of denominators into a sum of single denominators.
Then, the numerator, which is a polynomial in $z$ of order $r$, can be divided by each denominator by repeated execution of
%
%%%%%%%%%%%%%%%%%%%%%%%%%%%%%%%%%%%%%%%%
\begin{equation}
\frac{\EuScript{Q}^{(r)}(z)}{b+cz}
=
\frac{\EuScript{Q}^{(r-1)}(z)+az^r}{b+cz}
=
\frac{\EuScript{Q}^{(r-1)}(z)-(ab/c)z^{r-1}}{b+cz}
+ \frac{a}{c}\,z^{r-1}
~.
\end{equation}
%%%%%%%%%%%%%%%%%%%%%%%%%%%%%%%%%%%%%%%%
%
So finally, we find
%
%%%%%%%%%%%%%%%%%%%%%%%%%%%%%%%%%%%%%%%%
\begin{equation}
\frac{\EuScript{N}(\ell_1)}{\prod_{m\neq i,j,k}\Den_m(\ell_1)}
=
\EuScript{P}^{(r-n+3)}(z) + \sum_{m\neq i,j,k}\frac{c_m}{\Den_m(z)}
~,
\end{equation}
%%%%%%%%%%%%%%%%%%%%%%%%%%%%%%%%%%%%%%%%
%
where $\EuScript{P}^{(r-n+3)}(z)$ is a polynomial in $z$ of order $r-n+3$, and where the coefficients $c_m$ are independent of $z$.
We know that the polynomial resulting from this procedure must be of this order, because the large $z$ behavior of the original expression dictates this.
It is clear now that the task of the box subtraction terms in \Equation{Eq:triLHS} is to remove the poles in $z$ to end up with an equation of polynomials in $z$.
The coefficients of $\tTriPol(\ell)$ can then, in principle, be found by matching each term.
QCD in the Feynman gauge dictates that $r-n\leq0$, explaining the maximum cubic order of the polynomial $\tTriPol(\ell)$.

For general triangles, with denominators $(\ell+K_0)^2(\ell+K_1)^2(\ell+K_2)^2$, one can always find light-like momenta $p_1^\mu,p_2^\mu$ and coefficients $\alpha_1,\alpha_2$ such that~\cite{delAguila:2004nf}
%
%%%%%%%%%%%%%%%%%%%%%%%%%%%%%%%%%%%%%%%%
\begin{equation}
K_1^\mu-K_0^\mu = p_1^\mu + \alpha_1 p_2^\mu
\quad,\quad
K_2^\mu-K_0^\mu = p_2^\mu + \alpha_2 p_1^\mu
\quad.
\end{equation}
%%%%%%%%%%%%%%%%%%%%%%%%%%%%%%%%%%%%%%%%
%
Solutions to the cut equations then become
%
%%%%%%%%%%%%%%%%%%%%%%%%%%%%%%%%%%%%%%%%
\begin{equation}
\ell^\mu =
-K_0^\mu + z e_1^\mu + \frac{x_1x_2}{z}\,e_2^\mu
-x_1p_1^\mu - x_2p_2^\mu
~,
\end{equation}
%%%%%%%%%%%%%%%%%%%%%%%%%%%%%%%%%%%%%%%%
%
for any value of $z$, where $x_1,x_2$ are fixed by the relations
%
%%%%%%%%%%%%%%%%%%%%%%%%%%%%%%%%%%%%%%%%
\begin{equation}
x_1+\alpha_2x_2 = \alpha_2
\quad,\quad
x_2+\alpha_1x_1 = \alpha_1
\quad.
\end{equation}
%%%%%%%%%%%%%%%%%%%%%%%%%%%%%%%%%%%%%%%%
%
Now, denominator factors take the form
%
%%%%%%%%%%%%%%%%%%%%%%%%%%%%%%%%%%%%%%%%
\begin{equation}
(\ell+K)^2 = a + bz + c/z
\end{equation}
%%%%%%%%%%%%%%%%%%%%%%%%%%%%%%%%%%%%%%%%
%
and can effectively be turned into quadratic functions of $z$ by multiplying numerators and denominators in \Equation{Eq:triLHS} with $z^n$.
It needs to be stressed that, besides removing the explicit poles from the equation, the box subtraction terms further only influence terms constant in $z$.

Finally, for bubbles with denominators $(\ell+K_0)^2(\ell+K_1)^2$ the construction presented before for the specific triangle can be used by decomposing $K_1^\mu-K_0^\mu$ into a pair of light-like momenta.

\subsection{The unitarity interpretation}
The solutions to the cut equations that put denominators to zero turn internal virtual lines in one-loop graphs to on-shell lines.
In the residues the denominators are taken out, so the graphs do not diverge because of the vanishing denominators.
If we imagine that all possible graphs, contributing to a certain set of denominators made to vanish, are included, then we can understand that the residue is a product of tree-level amplitudes with the extra on-shell internal lines acting as external lines.
For example, the sum of all graphs that contain all four given denominators $\Den_i,\Den_j,\Den_k,\Den_l$ is represented by
%
%%%%%%%%%%%%%%%%%%%%%%%%%%%%%%%%%%%%%%%%
\begin{equation}
\graph{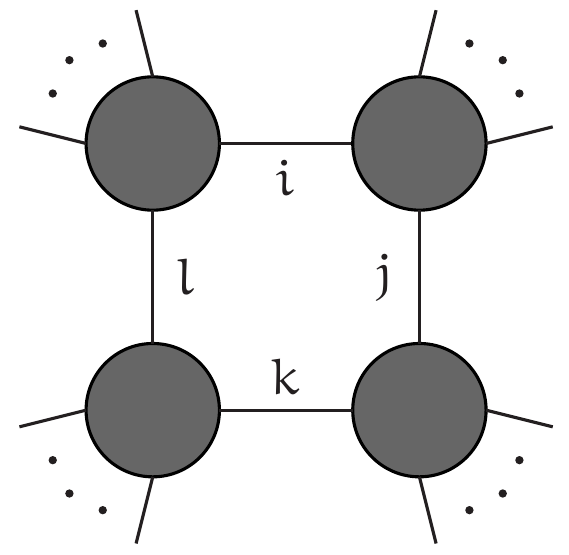}{24}{11}
\end{equation}
%%%%%%%%%%%%%%%%%%%%%%%%%%%%%%%%%%%%%%%%
%
The dots represent on-shell external lines, which are uniquely determined by the choice of the internal lines $i,j,k,l$.
In the residue of \Equation{Eq:quadLHS}, the denominators $i,j,k,l$ are excluded, and the internal momenta are put on-shell.
Consequently, the four blobs represent tree-level on-shell amplitudes, each of them with two extra on-shell lines $i,j$ and $j,k$ etc. 
Thus the residue can be calculated by sewing together four tree-level on-shell amplitudes.
The internal on-shell momenta may have complex components, but the tree-level amplitudes are still well-defined.
The same can be done for triangles with three blobs, and for bubbles with two blobs.
In the latter case, the procedure is equivalent to cutting the loop amplitude within the classic application of unitarity, hence the name.

%For us it is important that 
%%
%%%%%%%%%%%%%%%%%%%%%%%%%%%%%%%%%%%%%%%%%
%\begin{equation}
%\parbox{0.91\linewidth}{\em%
%the coefficient of the master integral vanishes if the residue vanishes at all solutions.
%\hspace{\fill}}
%\end{equation}
%%%%%%%%%%%%%%%%%%%%%%%%%%%%%%%%%%%%%%%%%
%%
%This is strictly speaking only true if the linear system is invertible, but in this respect our situation will show not to be special.
%%
%The scenario in mind here is that the residues may be proven to vanish for $\Lambda\to\infty$.

\section{The anomalous triangle contribution\label{Sec:MI3viol}}
\myFigure{%
\epsfig{figure=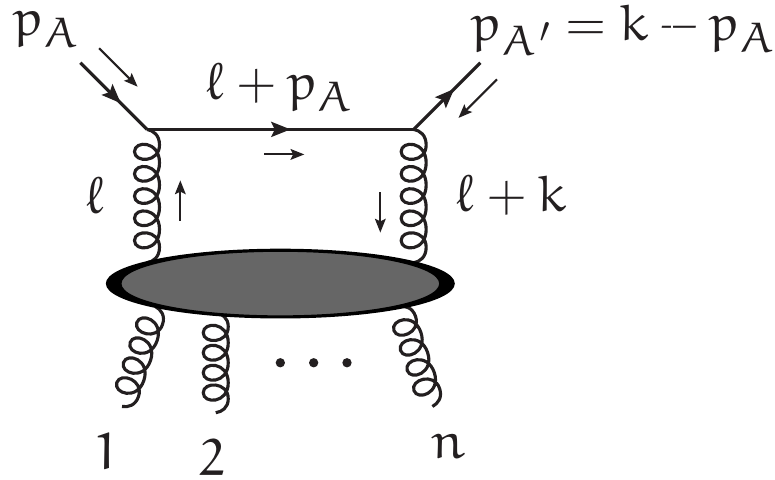,width=0.36\linewidth}\hspace{8ex}
\epsfig{figure=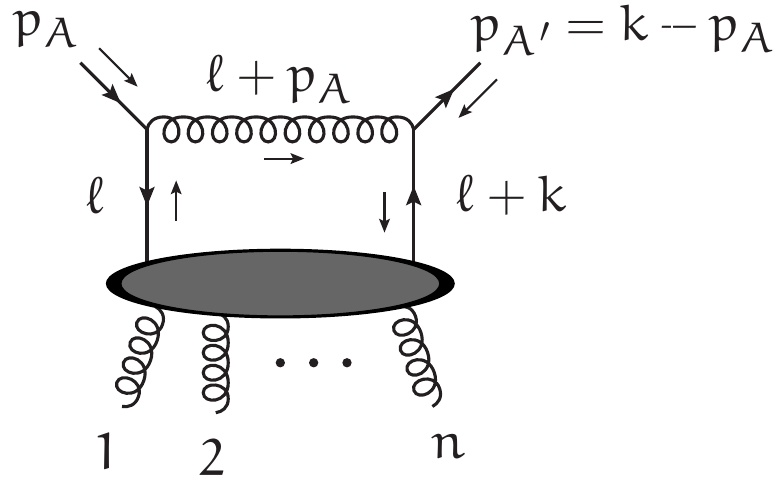,width=0.36\linewidth}
%\raisebox{2ex}{\epsfig{figure=finiteTria1.pdf,width=0.36\linewidth}}
\caption{\label{Fig:finiteTria}Graphs containing the anomalous triangle (\ref{MI3viol}). The external gluons are on-shell, and the blobs represents the sum of all possible graphs.}
%\caption{\label{Fig:finiteTria}Graphs containing triangle (\ref{MI3viol}). The external gluons $1$ to $n$ on the left are imagined to be on-shell, and the blob represents the sum of all possible graphs. The gluons on the right are not necessarily on-shell, but external currents attached to them do satisfy current conservation.}
}
We start by analysing graphs that contain the anomalous triangle (\ref{MI3viol}).
They are depicted in \Figure{Fig:finiteTria}.
First we consider the class of graphs on the left.
The class on the right will be addressed at the end of this section.

\subsection{Box coefficients for the left of \Figure{Fig:finiteTria}}
Before we consider the triangle coefficient, we consider boxes that have the three relevant denominators, plus one more, namely $(\ell+K)^2$ for some momentum $K^\mu$.
The solutions to the cut equations can be determined at the limit $\Lambda\to\infty$ directly.
We require
%
%%%%%%%%%%%%%%%%%%%%%%%%%%%%%%%%%%%%%%%%
\begin{equation}
\ell^2=\lop{p}{\ell}=(\ell+k)^2=(\ell-K)^2=0
~,
\end{equation}
%%%%%%%%%%%%%%%%%%%%%%%%%%%%%%%%%%%%%%%%
%
and the two solutions are given in terms of the quantities of \Equation{kTexpansion} by
%
%%%%%%%%%%%%%%%%%%%%%%%%%%%%%%%%%%%%%%%%
\begin{equation}
\ell_1^\mu = z_1p^\mu + \bar{\kappa}e^\mu
\;\;,\;\;
z_1 = \frac{K^2-2\bar{\kappa}\lop{e}{K}}{2\lop{p}{K}}
\quad,\quad
\ell_2^\mu = z_2p^\mu + \bar{\kappa}^*e_*^\mu
\;\;,\;\;
z_2 = \frac{K^2-2\bar{\kappa}^*\lop{e_*}{K}}{2\lop{p}{K}}
~.
\end{equation}
%%%%%%%%%%%%%%%%%%%%%%%%%%%%%%%%%%%%%%%%
%
We will need the spinors of these and the light-like momenta $\ell_{1,2}^\mu+k^\mu$:
%
%%%%%%%%%%%%%%%%%%%%%%%%%%%%%%%%%%%%%%%%
\begin{align}
\Arght{\ell_1} &= \Arght{p} & \Arght{\ell_1+k} &= x\Arght{p} - \bar{\kappa}^*\Arght{q} + z_1\Arght{p}
\notag\\
\Srght{\ell_1} &= \bar{\kappa}\Srght{q}+z_1\Srght{p} & \Srght{\ell_1+k} &= \Srght{p}
\label{spinorsbox}\\
\Arght{\ell_2} &= \bar{\kappa}^*\Arght{q}+z_2\Arght{p} & \Arght{\ell_2+k} &= \Arght{p}
\notag\\
\Srght{\ell_2} &= \Srght{p} & \Srght{\ell_2+k} &= x\Srght{p} - \bar{\kappa}\Srght{q} + z_2\Srght{p}
\notag
\end{align}
%%%%%%%%%%%%%%%%%%%%%%%%%%%%%%%%%%%%%%%%
%
The trivial space associated with the box is spanned by the vector
%
%%%%%%%%%%%%%%%%%%%%%%%%%%%%%%%%%%%%%%%%
\begin{equation}
 e_T^\mu = \alpha p^\mu - \bar{\kappa}e^\mu + \bar{\kappa}^*e_*^\mu
\quad,\quad
 \alpha = \frac{\bar{\kappa}\lop{e}{K}-\bar{\kappa}^*\lop{e_*}{K}}{\lop{p}{K}}
~,
\end{equation}
%%%%%%%%%%%%%%%%%%%%%%%%%%%%%%%%%%%%%%%%
%
defining the spurious polynomial as $\tBoxPol(\ell) = \tBoxC_K\lop{e_T}{\ell}$.
The equations for the box coefficients are given by
%
%%%%%%%%%%%%%%%%%%%%%%%%%%%%%%%%%%%%%%%%
\begin{equation}
\mathrm{Res}_{K}(\ell_1) = \BoxC_K + \tBoxC_K\frac{k_T^2}{2}
\quad,\quad
\mathrm{Res}_{K}(\ell_2) = \BoxC_K - \tBoxC_K\frac{k_T^2}{2}
~,
\end{equation}
%%%%%%%%%%%%%%%%%%%%%%%%%%%%%%%%%%%%%%%%
%
with solutions
%
%%%%%%%%%%%%%%%%%%%%%%%%%%%%%%%%%%%%%%%%
\begin{equation}
\BoxC_K = \frac{\mathrm{Res}_{K}(\ell_1)+\mathrm{Res}_{K}(\ell_2)}{2}
\quad,\quad
\tBoxC_K = \frac{\mathrm{Res}_{K}(\ell_1)-\mathrm{Res}_{K}(\ell_2)}{k_T^2}
~.
\label{Eq:319}
\end{equation}
%%%%%%%%%%%%%%%%%%%%%%%%%%%%%%%%%%%%%%%%
% 
For clearity, we label the residues and coefficients with the momentum $K$ instead of all the indices of the denominators.
The auxiliary quark line, including a factor $\imag(-\imag)^2$ from the quark propagator and two gluon propagators, becomes 
%
%%%%%%%%%%%%%%%%%%%%%%%%%%%%%%%%%%%%%%%%
\begin{equation}
-\imag\,\brktAS{p_{A'}|\frac{\imag}{\sqrt{2}}\gamma^\mu(\slashl+\slashp_A)\frac{\imag}{\sqrt{2}}\gamma^\nu\,|p_{A}}
= 2\imag\,\Lambda^2\,p^\mu p^\nu + \Ord\big(\Lambda\big)
~.
\end{equation}
%%%%%%%%%%%%%%%%%%%%%%%%%%%%%%%%%%%%%%%%
%
One factor of $\Lambda$ is absorbed by the master integral, the other one is removed following prescription (\ref{limit}).
Now we observe that the momentum $p^\mu$ can be interpreted as an un-normalized polarization vector associated with the momenta $-\ell_{1,2}^\mu$, for example:
%
%%%%%%%%%%%%%%%%%%%%%%%%%%%%%%%%%%%%%%%%
\begin{multline}
p^\mu = \srac{1}{2}\brktAS{p|\gamma^\mu|p}
      = \srac{1}{2}\brktAS{-\ell_1|\gamma^\mu|p}
\\
      = \frac{\brktSS{-\ell_1|p}}{\sqrt{2}}\,\varepsilon_-^\mu(-\ell_1,p)
      = \frac{-\bar{\kappa}\brktSS{q|p}}{\sqrt{2}}\,\varepsilon_-^\mu(-\ell_1,p)
      = \frac{\kappa}{\sqrt{2}}\,\varepsilon_-^\mu(-\ell_1,p)
~,
\end{multline}
%%%%%%%%%%%%%%%%%%%%%%%%%%%%%%%%%%%%%%%%
%
where $\varepsilon_-^\mu(-\ell_1,p)$ is the polarization vector of a negative-helicity gluon with momentum $-\ell_1^\mu$ defined using auxiliary momentum $p^\mu$.
We want the minus sign because we are interested in the momentum going into the blob.
It turns out that the auxiliary quark line can be writen as
%
%%%%%%%%%%%%%%%%%%%%%%%%%%%%%%%%%%%%%%%%
\begin{align}
-\imag\,\brktAS{p_{A'}|\frac{\imag}{\sqrt{2}}\gamma^\mu(\slashl+\slashp_A)\frac{\imag}{\sqrt{2}}\gamma^\nu\,|p_{A}}
&= \imag\,\Lambda^2\,k_T^2\,\varepsilon_+^\mu(\ell_1+k)\varepsilon_-^\nu(-\ell_1)
+ \Ord\big(\Lambda\big)
\notag\\
&= \imag\,\Lambda^2\,k_T^2\,\varepsilon_-^\mu(\ell_2+k)\varepsilon_+^\nu(-\ell_2)
+ \Ord\big(\Lambda\big)
~,
\label{Eq:379} 
\end{align}
%%%%%%%%%%%%%%%%%%%%%%%%%%%%%%%%%%%%%%%%
%
and we see that the residue of the box is given by the product of two on-shell amplitudes summed over the helicities of the internal on-shell gluon:
%
%%%%%%%%%%%%%%%%%%%%%%%%%%%%%%%%%%%%%%%%
\begin{align}
\mathrm{Res}_K(\ell_1) &= \Lambda^2\,k_T^2\sum_{h=-,+}\graph{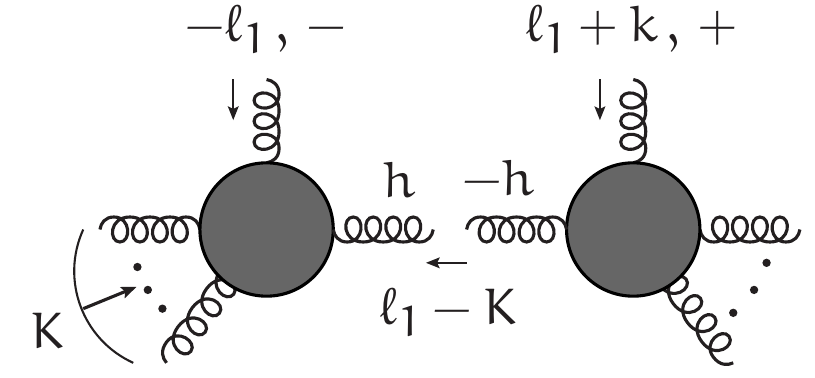}{30}{6}
\\
\mathrm{Res}_K(\ell_2) &= \Lambda^2\,k_T^2\sum_{h=-,+}\graph{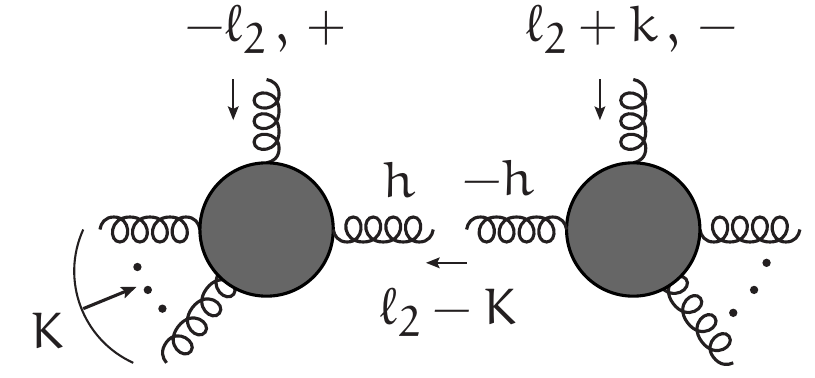}{30}{6}
\end{align}
%%%%%%%%%%%%%%%%%%%%%%%%%%%%%%%%%%%%%%%%
%
An extra factor $-\imag$ from the internal gluon propagator has been included.

%
% It is worth noting that we can interpret the momenta $-\ell_1^\mu$ and $\ell_1^\mu+k^\mu$ as ``shifted'' momenta in the sense of a BCFW recursion~\cite{Britto:2005fq}, with shift vector $p^\mu$
% %
% %%%%%%%%%%%%%%%%%%%%%%%%%%%%%%%%%%%%%%%%
% \begin{equation}
% -\ell_1^\mu = -\bar{\kappa}e^\mu - z_1p^\mu
% \quad,\quad
% \ell_1^\mu+k^\mu= xp^\mu-\bar{\kappa}^*e_*^\mu + z_1p^\mu
% ~.
% \end{equation}
% %%%%%%%%%%%%%%%%%%%%%%%%%%%%%%%%%%%%%%%%
% %
% where shift parameter $z_1$ is such that the internal gluon with momentum $\ell_1^\mu+K^\mu$ is on-shell.
% %
% In other words, we have
% %
% %%%%%%%%%%%%%%%%%%%%%%%%%%%%%%%%%%%%%%%%
% \begin{equation}
% \sum_K\frac{\mathrm{Res}_K(\ell_1)}{K^2-2\bar{\kappa}\lop{e}{K}}
% =
% \Lambda^2\,k_T^2\,\Amp_{\mathrm{tree}}\big([-\!\bar{\kappa}e]^-,[k_1]^{h_1},[k_2]^{h_2},\ldots,[k_n]^{h_n},[xp-\bar{\kappa}^*e_*]^{+}\big)
% + \Ord\big(\Lambda\big)
% ~,
% \end{equation}
% %%%%%%%%%%%%%%%%%%%%%%%%%%%%%%%%%%%%%%%%
% %
% where $\Amp_{\mathrm{tree}}$ is the tree-level $(n+2)$-gluon amplitude.
% %
% For $\ell_2$ we get the opposite helicities for the two extra gluons.

\subsection{Triangle coefficient for the left of \Figure{Fig:finiteTria}}
Now we move to the triangle.
In contrast to before, we now carefully work before the limit $\Lambda\to\infty$ because of the anomalous behavior of the triangle.
The cut equations are given by
%
%%%%%%%%%%%%%%%%%%%%%%%%%%%%%%%%%%%%%%%%
\begin{equation}
\ell^2 = (\ell+p_A)^2 = (\ell+k)^2 = 0
\end{equation}
%%%%%%%%%%%%%%%%%%%%%%%%%%%%%%%%%%%%%%%%
%
and using the fact that $k^\mu=p_A^\mu+p_{A'}^\mu$, we see that solutions are given by 
%
%%%%%%%%%%%%%%%%%%%%%%%%%%%%%%%%%%%%%%%%
\begin{equation}
\ell_1^\mu = ze_1^\mu - p_A^\mu
\quad\textrm{or}\quad
\ell_2^\mu = ze_2^\mu - p_A^\mu
\end{equation}
%%%%%%%%%%%%%%%%%%%%%%%%%%%%%%%%%%%%%%%%
%
for any value of $z$ and where
%
%%%%%%%%%%%%%%%%%%%%%%%%%%%%%%%%%%%%%%%%
\begin{equation}
e_1^\mu = \srac{1}{2}\brktAS{p_A|\gamma^\mu|p_{A'}}
\quad\textrm{and}\quad
e_2^\mu = \srac{1}{2}\brktAS{p_{A'}|\gamma^\mu|p_A}
~.
\end{equation}
%%%%%%%%%%%%%%%%%%%%%%%%%%%%%%%%%%%%%%%%
%
The vectors $e_1^\mu,e_2^\mu$ span the trivial space, and we essentially followed the construction presented in~\cite{Ossola:2006us}.
The spinors for the momenta $\ell_{1,2}^\mu$ and $\ell_{1,2}^\mu+k^\mu$ are
%
%%%%%%%%%%%%%%%%%%%%%%%%%%%%%%%%%%%%%%%%
\begin{align}
\Arght{\ell_1} &= \Arght{p_A} & \Arght{\ell_1+k} &= z\Arght{p_A} + \Arght{p_{A'}}
\notag\\
\Srght{\ell_1} &= z\Srght{p_{A'}}-\Srght{p_A} & \Srght{\ell_1+k} &= \Srght{p_{A'}}
\label{spinorstriangle}\\
\Arght{\ell_2} &= z\Arght{p_{A'}}-\Arght{p_A} & \Arght{\ell_2+k} &= \Arght{p_{A'}}
\notag\\
\Srght{\ell_2} &= \Srght{p_{A}} & \Srght{\ell_2+k} &= z\Srght{p_A} + \Srght{p_{A'}}
\notag
\end{align}
%%%%%%%%%%%%%%%%%%%%%%%%%%%%%%%%%%%%%%%%
%
Using \Equation{kTexpansion} and \Equation{AAprimeSpinors}, the solutions to the cut equations can be be found to be given by
%
%%%%%%%%%%%%%%%%%%%%%%%%%%%%%%%%%%%%%%%%
\begin{align}
\ell_1^\mu &= \left[-\Lambda(1+z)+\frac{xz}{2}\right]p^\mu
         + (1-z)\frac{\bar{\kappa}}{2}\,e^\mu
         + (1+z)\frac{\bar{\kappa}_*}{2}\,e_*^\mu
         + \Ord\big(\Lambda^{-1}\big)
\\
\ell_2^\mu &= \left[-\Lambda(1-z)-\frac{xz}{2}\right]p^\mu
          + (1-z)\frac{\bar{\kappa}}{2}\,e^\mu
          + (1+z)\frac{\bar{\kappa}_*}{2}\,e_*^\mu
          + \Ord\big(\Lambda^{-1}\big)
~.
\end{align}
%%%%%%%%%%%%%%%%%%%%%%%%%%%%%%%%%%%%%%%%
%
We see that they diverge with $\Lambda$, which could cause the residue at the solution, and therefore the triangle coefficient, to diverge with $\Lambda$.
The spurious polynomial of the triangle can be decomposed in terms of $e_1^\mu,e_2^\mu$, and inserting the solution $\ell_1^\mu$, we find, using $\lop{e_1}{e_2}=-\lop{p_A}{p_{A'}}=-k_T^2/2$, that
%
%%%%%%%%%%%%%%%%%%%%%%%%%%%%%%%%%%%%%%%%
\begin{equation}
\tTriPol(\ell_1) = 
- \frac{k_T^2}{2}\,\tTriC^{(2)}\,z
+ \frac{(k_T^2)^2}{4}\,\tTriC^{(22)}\,z^2
- \frac{(k_T^2)^3}{8}\,\tTriC^{(222)}\,z^3
~.
\end{equation}
%%%%%%%%%%%%%%%%%%%%%%%%%%%%%%%%%%%%%%%%
%
The other three coefficients $\tTriC^{(1)},\tTriC^{(11)},\tTriC^{(111)}$ can be accessed using the solution $\ell_2^\mu$.
We see that the spurious polynomial is independent of $\Lambda$ at the cuts.

As defined here, \Equation{Eq:triLHS} must not behave worse than linearly with $\Lambda$: this behavior is still canceled by the prescription of \Equation{limit}.
This triangle is anomalous in the sense that it does not absorb a factor of $\Lambda$, like the boxes in the previous section. 
Let us start with the box subtraction terms.
After applying  \Equation{limit}, the box coefficients behave as $\Ord(\Lambda)$, which comes naturally with the $\Ord\big(\Lambda^{-1}\big)$ behavior of the master integral.
This is cancelled by the similar behavior of $\Den_l(\ell_{1,2})$.
Now, $\tBoxPol_{ijkl}(\ell_{1,2})$ could still spoil the behavior, but realize that $p^\mu$ is in the physical space of the box when $\Lambda\to\infty$, so the $\Ord(\Lambda)$ behavior of $\ell_{1,2}^\mu$ is eliminated in $\tBoxPol_{ijkl}(\ell_{1,2})$.

Regarding the rest of the residue, it turns out that the auxiliary quark-line can be written as a product of polarization vectors again, like in for the box coefficients.
We find
%
%%%%%%%%%%%%%%%%%%%%%%%%%%%%%%%%%%%%%%%%
\begin{align}
-\imag\,\brktAS{p_{A'}|\frac{\imag}{\sqrt{2}}\gamma^\mu(\slashl_1+\slashp_A)\frac{\imag}{\sqrt{2}}\gamma^\nu\,|p_{A}}
%&=\frac{-z}{2}\brktAS{p_{A'}|\gamma^\mu\slashe_1\gamma^\nu|p_{A}}\\
= 2\imag\,z\,p_{A'}^\mu p_{A}^\nu
%&= -2z\frac{\brktAA{p_{A'}|\ell_1+k}}{\sqrt{2}}\varepsilon_+^\mu(\ell_1+k,p_A')
%        \frac{\brktSS{-\ell_1|p_A}}{\sqrt{2}}\varepsilon_-^\nu(-\ell_1,p_A)\\
%&= 2z^3\frac{\brktAA{p_{A'}|p_A}}{\sqrt{2}}\varepsilon_+^\mu(\ell_1+k,p_A')
%       \frac{\brktSS{p_{A'}|p_A}}{\sqrt{2}}\varepsilon_-^\nu(-\ell_1,p_A)\\
&=\imag\,z^3\,k_T^2\,\varepsilon_+^\mu(\ell_1+k)\,\varepsilon_-^\nu(-\ell_1)
\end{align}
%%%%%%%%%%%%%%%%%%%%%%%%%%%%%%%%%%%%%%%%
%%%%%%%%%%%%%%%%%%%%%%%%%%%%%%%%%%%%%%%%
\begin{align}
-\imag\brktAS{p_{A'}|\frac{\imag}{\sqrt{2}}\gamma^\mu(\slashl_2+\slashp_A)\frac{\imag}{\sqrt{2}}\gamma^\nu\,|p_{A}}
%&=\frac{-z}{2}\brktAS{p_{A'}|\gamma^\mu\slashe_2\gamma^\nu|p_{A}}\\
= 2\imag\,z\,e_2^\mu e_2^\nu
%&= -2z\frac{\brktSS{\ell_2+k|p_A}}{\sqrt{2}}\varepsilon_-^\mu(\ell_2+k,p_A)
%       \frac{\brktAA{p_{A'}|-\ell_2}}{\sqrt{2}}\varepsilon_+^\nu(-\ell_2,p_{A'})\\
%&=  2z\frac{\brktSS{p_{A'}|p_A}}{\sqrt{2}}\varepsilon_-^\mu(\ell_2+k,p_A)
%        \frac{\brktAA{p_{A'}|p_A}}{\sqrt{2}}\varepsilon_+^\nu(-\ell_2,p_{A'})\\
&= \imag\,z\,k_T^2\,\varepsilon_-^\mu(\ell_2+k)\,\varepsilon_+^\nu(-\ell_2)
\label{Eq346}
\end{align}
%%%%%%%%%%%%%%%%%%%%%%%%%%%%%%%%%%%%%%%%
%
So we see that the residues are proportional to on-shell tree-level $(n+2)$-gluon amplitudes.
The two adjacent gluons that were attached to the auxiliary quark line carry a momentum that has a contribution proportional to $\Lambda p^\mu$.
We saw in \Section{Sec:auxglu} that such amplitudes with ``auxiliary gluons'' are proportional to $\Lambda$ and we conclude that the residues are also propotional to $\Lambda$, and not a higher power of $\Lambda$.

It has to be noted that for the derivation in \Section{Sec:auxglu} to work, the auxiliary momenta with which the polarization vectors of the auxiliary gluons are defined must not be equal $p_{A'}^\mu,p_A^\mu$ themselves, because that would cause the inner product of the polarization vectors to vanish, and would invalidate the counting of powers of $\Lambda$.
This is exactly the case in \Equation{Eq346}, where $\lop{e_2}{e_2}=0$.
However, the amplitude is a gauge invariant object, and we are free to choose the auxiliary momenta for $\varepsilon_-^\mu(\ell_2+k),\varepsilon_+^\nu(-\ell_2)$, so the argument still stands.

\subsection{Coefficients for the right of \Figure{Fig:finiteTria}}
The solutions to the cut equations of course stay the same.
The box numerator can now be written as
%
%%%%%%%%%%%%%%%%%%%%%%%%%%%%%%%%%%%%%%%%
\begin{multline}
-\brktAS{p_{A'}|\frac{\imag}{\sqrt{2}}\gamma^\mu|\ell_{1,2}+k}
\,\brktAS{\ell_{1,2}+k|X|\ell_{1,2}-K}
\,\brktAS{\ell_{1,2}-K|Y|\ell_{1,2}}
\,\brktAS{\ell_{1,2}|\frac{\imag}{\sqrt{2}}\gamma_\mu|p_A}
\\
=\brktAA{p_{A'}|\ell_{1,2}}\brktSS{p_A|\ell_{1,2}+k}
\,\brktAS{\ell_{1,2}+k|X|\ell_{1,2}-K}
\,\brktAS{\ell_{1,2}-K|Y|\ell_{1,2}}
~,
\end{multline}
%%%%%%%%%%%%%%%%%%%%%%%%%%%%%%%%%%%%%%%%
%
where $\brktAS{\ell_{1,2}+k|X|\ell_{1,2}-K}$ and $\brktAS{\ell_{1,2}-K|Y|\ell_{1,2}}$ represent on-shell amplitudes with a quark-antiquark pair and a number of gluons.
Using \Equation{AAprimeSpinors} and \Equation{spinorsbox}, we see that
%
%%%%%%%%%%%%%%%%%%%%%%%%%%%%%%%%%%%%%%%%
\begin{equation}
\brktAA{p_{A'}|\ell_{1}}\brktSS{p_A|\ell_{1}+k} = \Ord\big(\Lambda^{-1}\big)
\quad,\quad
\brktAA{p_{A'}|\ell_{2}}\brktSS{p_A|\ell_{2}+k} = -\Lambda\,k_T^2 + \Ord\big(\Lambda^{0}\big)
~,
\end{equation}
%%%%%%%%%%%%%%%%%%%%%%%%%%%%%%%%%%%%%%%%
%
so the box residues carry at least one power of $\Lambda$ too few and vanish.
Notice that we chose the ``wrong'' momentum flow in the graph, that is the $\Lambda p^\mu$ contribution does not flow through the auxiliary quark line.
In the next section it is shown that also in the ``correct'' momentum flow, this contribution vanishes.

For the triangle, the numerator becomes
%
%%%%%%%%%%%%%%%%%%%%%%%%%%%%%%%%%%%%%%%%
\begin{multline}
\imag\brktAS{p_{A'}|\frac{\imag}{\sqrt{2}}\gamma^\mu|\ell_{1,2}+k}
\,\brktAS{\ell_{1,2}+k|X|\ell_{1,2}}
\,\brktAS{\ell_{1,2}|\frac{\imag}{\sqrt{2}}\gamma_\mu|p_A}
\\
=-\imag\brktAA{p_{A'}|\ell_{1,2}}\brktSS{p_A|\ell_{1,2}+k}
\,\brktAS{\ell_{1,2}+k|X|\ell_{1,2}}
~,
\end{multline}
%%%%%%%%%%%%%%%%%%%%%%%%%%%%%%%%%%%%%%%%
%
where $\brktAS{\ell_{1,2}+k|X|\ell_{1,2}}$ is a tree-level multi-gluon amplitude with one of them off-shell, constructed via an auxiliary quark-antiquark pair with momenta $\ell_{1,2}^\mu,\ell_{1,2}^\mu+k^\mu$.
So we know that $\brktAS{\ell_{1,2}+k|X|\ell_{1,2}}=\Ord(\Lambda)$.
Using \Equation{spinorstriangle} we see that $\brktAA{p_{A'}|\ell_{1,2}}\brktSS{p_A|\ell_{1,2}+k}=\pm k_T^2$, and we find that the triangle coefficient stays finite for large $\Lambda$.

\section{Box and triangle coefficients with eikonal Feynman rules}
In \Appendix{AppCutOPP} and \Appendix{AppCutEGK} we show that the formulas from both \cite{Ossola:2006us} and \cite{Ellis:2007br} for the solutions to the cut equation for boxes and remaining triangles can straightforwardly be extrapolated to $\Lambda\to\infty$ if they involve one $\Lambda$-dependent denominator, so, in the notation of \Equation{dennot}, for $\Den_i^0\Den_j^0\Den_k^\Lambda$ and $\Den_i^0\Den_j^0\Den_k^0\Den_l^\Lambda$.
The solutions are well-defined and finite.
For $\Den_i^0\Den_j^0\Den_k^\Lambda\Den_l^\Lambda$, the solutions are divergent, which can be understood from the fact that at $\Lambda\to\infty$ the equations $\Den_k^\Lambda=\Den_l^\Lambda=0$ imply that $\lop{p}{(K_k-K_l)}=0$ which is not necessarily true, so a solution does not exist.
As a consequence, the box with two $\Lambda$-dependent denominators {\em is not a master integral and decomposes into four triangles\/}:
%
%%%%%%%%%%%%%%%%%%%%%%%%%%%%%%%%%%%%%%%%
\begin{align}
%\hspace{-0.7ex}
\int\frac{\dDell\,\Lambda^2}{\Den_i^0\Den_j^0\Den_k^\Lambda\Den_l^\Lambda}
&=
 \frac{1}{2\lop{p}{(K_k-K_l)}}\int\frac{\dDell\,\Lambda}
 {\Den_i^0\Den_j^0\Den_l^\Lambda}
+\frac{1}{2\lop{p}{(K_l-K_k)}}\int\frac{\dDell\,\Lambda}
 {\Den_i^0\Den_j^0\Den_k^\Lambda}
\notag\\
&+
 \frac{1}{2\lop{p}{(K_i-K_j)}}\int\frac{\dDell\,\Lambda}
 {\Den_i^0\Den_k^\Lambda\Den_l^\Lambda}
+\frac{1}{2\lop{p}{(K_j-K_i)}}\int\frac{\dDell\,\Lambda}
 {\Den_j^0\Den_k^\Lambda\Den_l^\Lambda}
%+\Ord\big(\Lambda^{-1}\big)
+\Ord\bigg(\frac{1}{\Lambda}\bigg)
~.
\label{fourpointdecom}
\end{align}
%%%%%%%%%%%%%%%%%%%%%%%%%%%%%%%%%%%%%%%%
%
The last two triangles in the identity above are well-defined under a shift of the integration momentum $\ell^\mu\to\ell^\mu-\Lambda p^\mu$.
The identity is shown to hold by explicit calculation for a few kinematical situations in \Appendix{doubleboxes}.
Remarkably, the case for which none of the external momenta are light-like, which one would expect to be the most complicated case, turns out to be rather simple.
Notice that the identity implies that the partial fractioning
%
%%%%%%%%%%%%%%%%%%%%%%%%%%%%%%%%%%%%%%%%
\begin{equation}
\frac{\Lambda^2}{\Den_k^\Lambda\Den_l^\Lambda}
= \frac{1}{2\lop{p}{(K_l-K_k)}}
  \left(\frac{\Lambda}{\Den_k^\Lambda}-\frac{\Lambda}{\Den_l^\Lambda}\right)
+\Ord\bigg(\frac{1}{\Lambda}\bigg)
\quad,
\label{partfrac}
\end{equation}
%%%%%%%%%%%%%%%%%%%%%%%%%%%%%%%%%%%%%%%%
%
which one would naturally apply for linear denominators, {\bf is in general  not allowed} underneath an integral, and would for example lead to a decomposition into only two triangles here.
Thus we are {\it a priori\/} not allowed to apply \Equation{partfrac} repeatedly and exclude all terms with more than one $\Lambda$-depenent denominator from the decomposition of \Equation{integranddecom}.
We {\em are\/} allowed to exclude box terms with two $\Lambda$-dependent denominators, because their contribution goes to triangles.
For these, any momentum shift will still cause two out of four denominators to be $\Lambda$-dependent, thus we can conclude that, in the notation of \Equation{dennot0}, terms with the denominator combination
%
%%%%%%%%%%%%%%%%%%%%%%%%%%%%%%%%%%%%%%%%
\begin{equation}
\Den_{Oi}\Den_{Oj}\Den_{Ak}\Den_{Al}
\end{equation}
%%%%%%%%%%%%%%%%%%%%%%%%%%%%%%%%%%%%%%%%
%
do not need to be taken into account.
In the next subsection, we show that terms with the denominator combinations,
%
%%%%%%%%%%%%%%%%%%%%%%%%%%%%%%%%%%%%%%%%
\begin{equation}
\Den_{Ai}\Den_{Aj}\Den_{Ak}\Den_{Al}
\;\;,\;\;
\Den_{Oi}\Den_{Aj}\Den_{Ak}\Den_{Al}
\;\;,\;\;
\Den_{Ai}\Den_{Aj}\Den_{Ak}
\;\;,\;\;
\Den_{Oi}\Den_{Aj}\Den_{Ak}
\;\;,\;\;
\Den_{Aj}\Den_{Ak}
~,
\label{Eq:vanishingmasters}
\end{equation}
%%%%%%%%%%%%%%%%%%%%%%%%%%%%%%%%%%%%%%%%
%
do not contribute to the one-loop integral, and also do not need to be taking into account.
We conclude that we only need to take into account the combinations with at most {\em one\/} auxiliary quark propagator 
%
%%%%%%%%%%%%%%%%%%%%%%%%%%%%%%%%%%%%%%%%
\begin{equation}
\Den_{Oi}\Den_{Oj}\Den_{Ok}\Den_{Al}
\;\;,\;\;
\Den_{Oi}\Den_{Oj}\Den_{Ak}
\;\;,\;\;
\Den_{Oj}\Den_{Ak}
~.
\end{equation}
%%%%%%%%%%%%%%%%%%%%%%%%%%%%%%%%%%%%%%%%
%
As a consequence, we see {\it a posteriori\/} that if we choose the momentum routing such that every $\Lambda$-dependent denominator belongs to an auxiliary quark propagator, so every $\Den^\Lambda$ corresponds to a $\Den_A$, then we are allowed to apply \Equation{partfrac}.
%

% It has to be stated that the mechanism that makes this work is rather non-trivial.
% %
% One can see this for example by calculating the graph (\ref{doubleGraph2})
% %
% using standard Passarino-Veltman reduction~\cite{Passarino:1978jh} and keeping the full $\Lambda$ dependence.
% %
% One will find contributions from all possible master integrals, including the box.
% %
% Only with the application of \Equation{fourpointdecom} the contribution of the triangles with two denominators coming form the auxiliary quark line vanish for $\Lambda\to\infty$, and it appears that one could have applied \Equation{partfrac} from the start.
% %
% %Again, this only works for determining the contribution from the boxes and triangles, and it is essential to realize that \Equation{partfrac} cannot be applied when determining the contribution of bubbles.
% %

The above can be summarized by the statement that we are allowed to take the limit $\Lambda\to\infty$ at the integrand-level when determining the contribution from boxes and triangles, with the exception of the anomalous triangle~(\ref{MI3viol}), if we keep the ``natural'' loop momentum routing in which every $\Den_A$ corresponds to a $\Den^\Lambda$.
It turns out that there are no Feynman graphs which contain contributions from bubbles with one $\Lambda$-dependent denominator at all in this limit on the integrand, as shown in the next section.
Their contribution strictly comes from parts of the integrand that are sub-leading in $\Lambda$.
%
%In \Section{Sec:bubblesfinite}, we show that this contribution does not diverge with $\Lambda$.
%
%Another consequence of the above is that we only need to consider $\Lambda$-dependent master integrals with at most one $\Lambda$-dependent denominator.

\subsection{Vanishing denominator combinations}
\myFigure{%
\epsfig{figure=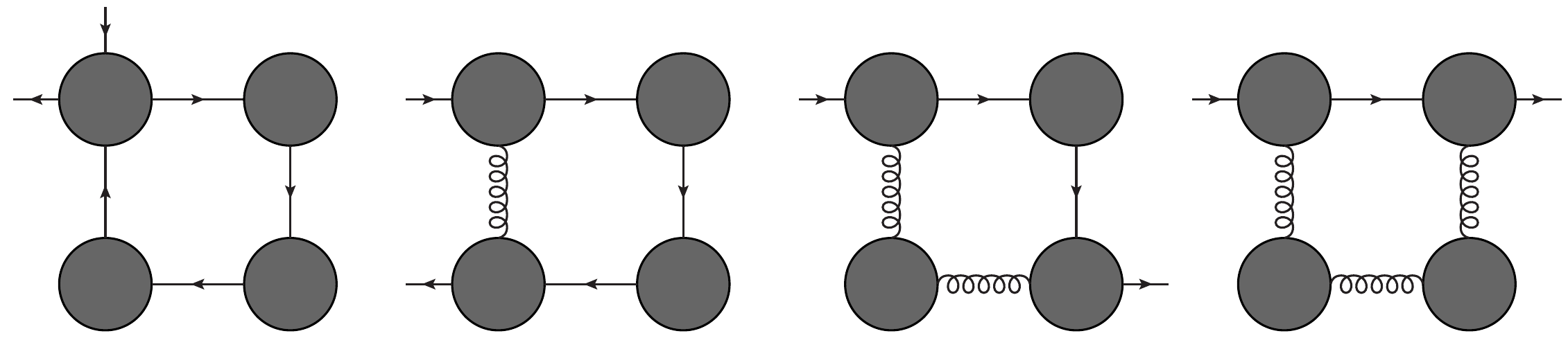,width=\linewidth}
\caption{\label{Fig:quadcuts}The four types of quadruple cuts involving the auxiliary quark line. The external quark and antiquark have divergent momenta $p_{A'}^\mu$ and $p_{A}^\mu$ and the blobs can have any number of external gluons (the ones without an external quark or antiquark have at least one).}
}%
Consider the first case in (\ref{Eq:vanishingmasters}).
The residue is depicted as the first graph in \Figure{Fig:quadcuts}.
Following the unitarity interpretation, the four blobs represent on-shell tree-level amplitudes.
Now we may choose the momentum routing such that the divergent component $\Lambda p^\mu$ does not enter the loop at all.
Then, only the upper-left blob involves divergent momenta, while the other three do not.
The upper-left blob is of the type of \Section{Sec:qqfamdiv}, which was shown to be finite, so the whole residue is finite for $\Lambda\to\infty$.
The box is also finite because it does not involve the divergent momentum components at all, and thus the whole contribution is eliminated by the prescription of \Equation{limit}.
The quark and antiquark must be considered of different flavor in the blob in order to avoid quark-loop contributions.
These do contribute to the one-loop amplitude, but are completely free from the issues addressed in this paper.
If we choose the momentum routing such that $\Lambda p^\mu$ {\em does\/} enter the loop, then it can be eliminated from the loop via shift of the loop momentum, and we see that the momenta of the cut lines at the cut solutions are still finite.

Now consider the second case in (\ref{Eq:vanishingmasters}).
The residue is represented by the second graph in \Figure{Fig:quadcuts}.
We choose the momentum routing such that the $\Lambda p^\mu$ flows through the explicitly depiced internal gluon.
It is shown in \Appendix{AppCutOPP} that the solutions to the cut equations for such situation (one $\Lambda$-dependent denominator) are finite.
Thus only the momentum of the cut gluon diverges, and the other three are finite.
The two blobs in the right represent finite tree-level amplitudes, while the two on the left are of the type of \Section{Sec:quarkgludiv}, and are also finite.
So the residue is finite, while the box has a $\Lambda$-depenent denominator and vanishes.

It is clear that the same reasoning works for the other cases in (\ref{Eq:vanishingmasters}).
We just have to realize that the subtraction terms in \eg\ \Equation{Eq:triLHS} would have to come from the boxes we just considered and thus vanish.

Contributions from graphs of the third type in \Figure{Fig:quadcuts} do not need to be considered, because the associated boxes are not master integrals and their contributions are included via triangles directly.
Let us still check if the contribution is finite.
As mentioned before, the solutions to the cut equations diverge, so all four internal lines represent divergent on-shell momenta.
The upper-left and lower-right blobs then are amplitudes of the type of \Section{Sec:auxilquarkglu} and behave as $\sqrt{\Lambda}$, while the other two are of the type of \Section{Sec:auxquarkdiv} and \Section{Sec:auxglu} and behave as $\Lambda$.
The box behaves as $\Lambda^{-2}$ so including the prescription of \Equation{limit} we find a finite contribution.

The only box contributions that need to be calculated have residues of the fourth type in \Figure{Fig:quadcuts}.
The solutions to the cut equations are finite, and in the natural momentum routing only the internal auxiliary quark line is divergent.
The upper blobs are of the type of \Section{Sec:auxquarkdiv} and diverge as $\Lambda$, while the lower blobs are finite.
The box swallows a factor $\Lambda$ and the other is eliminated by the prescription of \Equation{limit}.

\subsection{Rank of the spurious triangle polynomials}
Before we can consider the bubble master integrals, we need to have a closer look at the triangle coefficients and their behavior as function of $\Lambda$.
In the following, we will show that the spurious coefficients are suppressed and follow the behavior
%
%%%%%%%%%%%%%%%%%%%%%%%%%%%%%%%%%%%%%%%%
\begin{align}
\TriPol_{ijk}(\ell)
=
\Lambda^{r}\left[
\TriC^\mathrm{constant}_{ijk}
+\Lambda^{-1}\,\tTriPol^\mathrm{linear}_{ijk}(\ell)
+\Lambda^{-2}\,\tTriPol^\mathrm{quadratic}_{ijk}(\ell)
+\Lambda^{-3}\,\tTriPol^\mathrm{cubic}_{ijk}(\ell)
\right]
~.\label{spuriousrank}
\end{align}
%%%%%%%%%%%%%%%%%%%%%%%%%%%%%%%%%%%%%%%%
%
We saw in \Section{sec:OPP} that the behavior of the terms of $\tTriPol(\ell)$ is determined by terms with high powers of $\ell^\mu$, that is terms of high rank, in the numerator of the integrand in \Equation{Eq:triRes}.
Terms of rank $r$ equal to the total number of denominators $n$ determine the behavior of $3$-rd order terms in $\tTriPol_{ijk}(\ell)$, terms of rank $r=n-1$ the behavior of $2$-nd order terms etc.
If the rank is lower than $n-3$, the triangle does not contribute at all.

Considering the fact that numerator factors of the type $(\Lambda\slashp+\slashK+\slashl)$ coming from auxiliary quark propagators in the loop obviously contribute with lower powers of $\Lambda$ to higher powers of $\ell^\mu$, it seems clear that \Equation{spuriousrank} must hold.
Let us, however, have a more careful look at effective vertices representing all one-particle irreducible one-loop graphs with at least one internal auxiliary quark line.
They are given in \Figure{Fig:generalGraph}.
\myFigure{%
\epsfig{figure=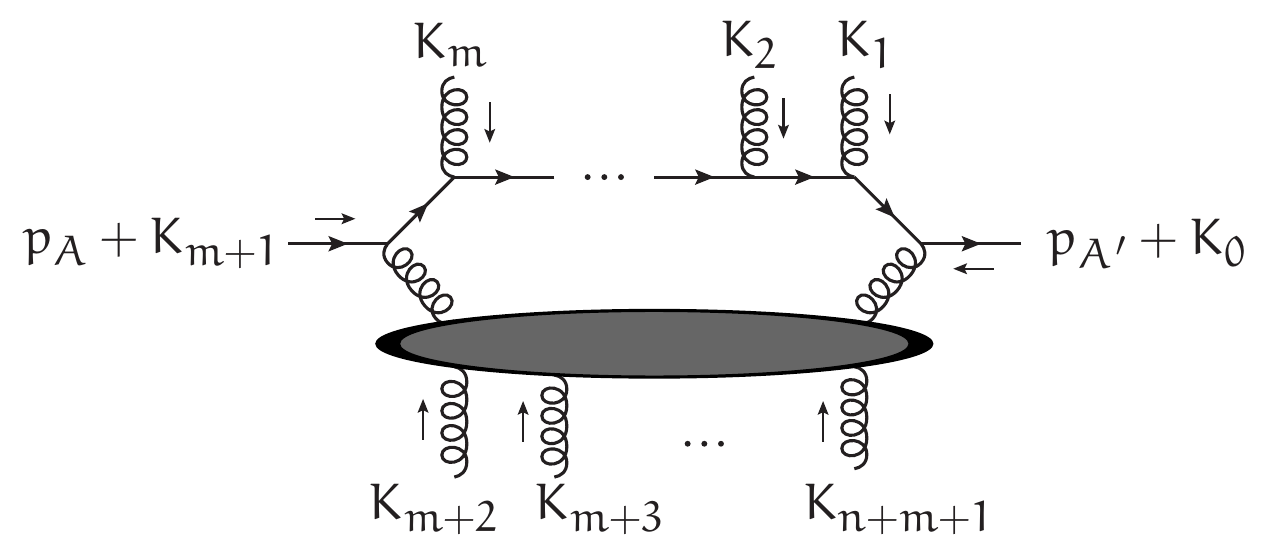,width=0.6\linewidth}
\caption{\label{Fig:generalGraph}A general effective vertex representing one-particle irreducible one-loop graphs with at least one internal auxiliary quark line . The tree-level blobs are given in \Figure{Fig:Blob}.}
}%
\myFigure{%
\epsfig{figure=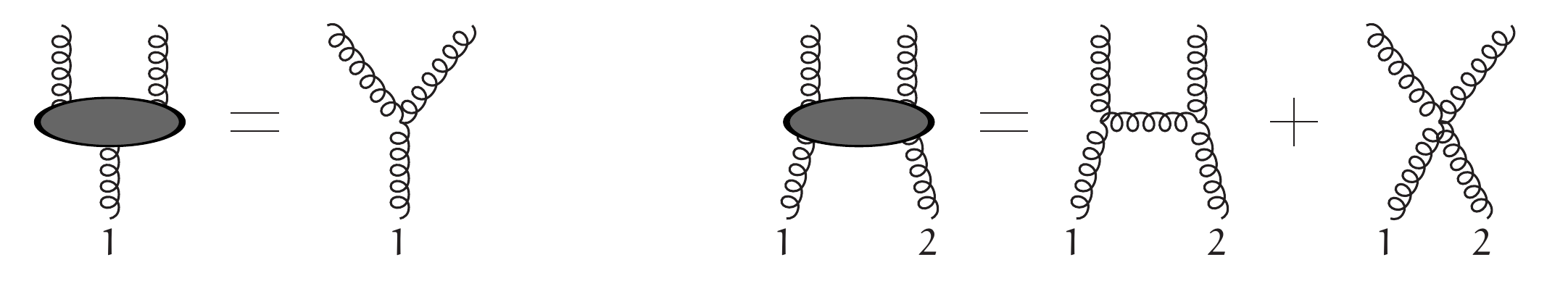,width=\linewidth}\\[1ex]
\epsfig{figure=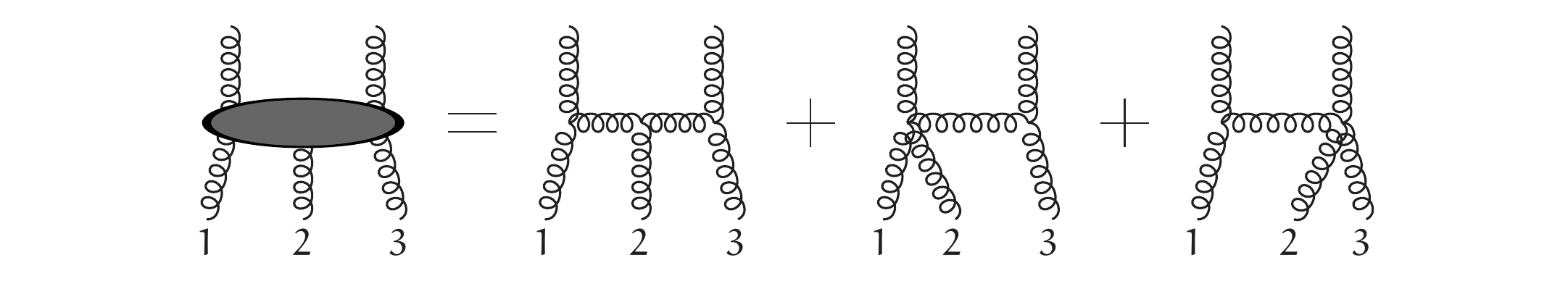,width=\linewidth}\\[1ex]
\epsfig{figure=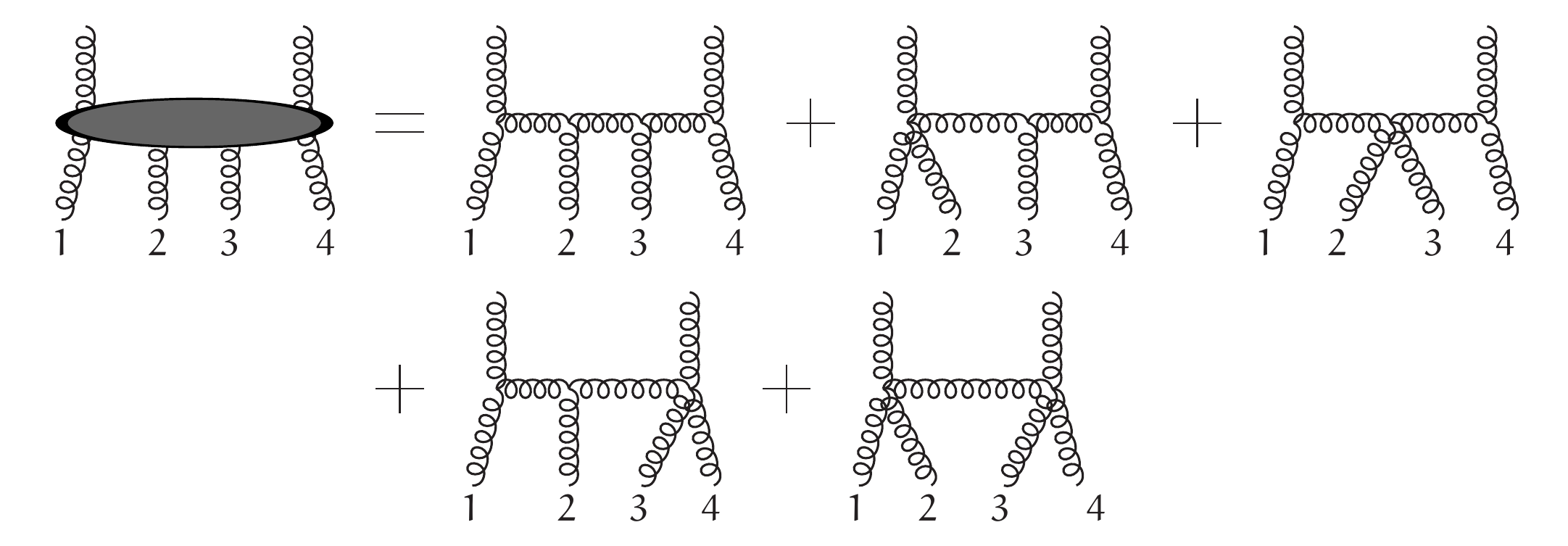,width=\linewidth}
\caption{\label{Fig:Blob}The tree-level blob in \Figure{Fig:generalGraph} for $n=1,2,3,4$.}
}%
Any of the external legs may be virtual, but we do assume that the tree-level attachments are complete and satisfy current conservation
%
%%%%%%%%%%%%%%%%%%%%%%%%%%%%%%%%%%%%%%%%
\begin{equation}
\lop{J_i}{K_i} = 0
~.
\end{equation}
%%%%%%%%%%%%%%%%%%%%%%%%%%%%%%%%%%%%%%%%
%
The tree-level attachments to the external quark and antiquark line can be expanded as
%
%%%%%%%%%%%%%%%%%%%%%%%%%%%%%%%%%%%%%%%%
\begin{equation}
\leftA{p_{A'}+K_0} = \sqrt{\Lambda}\,\leftA{p} + \frac{1}{\sqrt{\Lambda}}\,\leftA{L}
\quad,\quad
\Srght{p_{A}+K_{n+1}} = \sqrt{\Lambda}\,\Srght{p} + \frac{1}{\sqrt{\Lambda}}\,\Srght{R}
\quad.
\label{externalSpinors}
\end{equation}
%%%%%%%%%%%%%%%%%%%%%%%%%%%%%%%%%%%%%%%%
%
The exact form of $\leftA{L}$ and $\Srght{R}$ is not important, but they are necessary for bookkeeping powers of $\Lambda$.
We just need to realize that $\brktAA{p|R}\neq0$ and $\brktSS{L|p}\neq0$.

Let us say we are considering a master integral including the two denominators $\Den_i^0$ and $\Den_j^\Lambda$.
We imagine the graph of \Figure{Fig:generalGraph} to have a single common denominator, that is four-point vertices in the tree-level blob are multiplied with the necessary denominator.
It turns out to be helpful to massage the integrand a bit with the following two reduction steps.
A factor $\ell^2$ in the numerator of the integrand can be written as
%
%%%%%%%%%%%%%%%%%%%%%%%%%%%%%%%%%%%%%%%%
\begin{equation}
\ell^2 = \Den_i^0 - 2\lop{\ell}{K_i} - K_i^2
~.
\end{equation}
%%%%%%%%%%%%%%%%%%%%%%%%%%%%%%%%%%%%%%%%
%
The term $\Den_i^0$ on the right-hand-side cancels the denominator, and makes that part of the numerator irrelevant for the master integral under consideration.
The other terms have a lower rank.
A factor $2\Lambda\lop{p}{\ell}$ can be written as
%
%%%%%%%%%%%%%%%%%%%%%%%%%%%%%%%%%%%%%%%%
\begin{equation}
2\Lambda\lop{p}{\ell} = \Den_j^\Lambda-\Den_i^0 - 2\lop{(K_k-K_i)}{\ell}
- [2\Lambda\lop{p}{K_k}+K_k^2-K_i^2]
~.
\label{reduce2}
\end{equation}
%%%%%%%%%%%%%%%%%%%%%%%%%%%%%%%%%%%%%%%%
%
Now the terms $\Den_j^\Lambda$ and $\Den_i^0$ are irrelevant for the master integral, and the others carry a lower power of $\Lambda$, and/or are of lower rank.
 Let us denote general polynomial terms of rank at most $r$ by $[\ell]^r$.
There may be terms of rank lower than $r$ included in $[\ell]^r$.
With the help of the operations above, we find via explicit calculations with the program in \Appendix{formTriangle} for several values of $n,m$ that the part $\EuScript{N}_{ij}(\ell)$ of the numerator relevant for the master integral under consideration fits the pattern
%
%%%%%%%%%%%%%%%%%%%%%%%%%%%%%%%%%%%%%%%%
\begin{equation}
\EuScript{N}_{ij}(\ell)
=
\Lambda^{m+2}\left(
             [\ell]^{n-1}
+\Lambda^{-1}[\ell]^{n}
+\Lambda^{-2}[\ell]^{n+1}
+\Lambda^{-3}[\ell]^{n+2}
+\cdots
\right)
~.
\end{equation}
%%%%%%%%%%%%%%%%%%%%%%%%%%%%%%%%%%%%%%%%
%
%
The overall factor $\Lambda^{m+2}$ matches the $m+1$ denominators from the auxiliary quark lines, and the factor $\sqrt{\Lambda}\times\sqrt{\Lambda}$ from the external quark lines \Equation{externalSpinors}, which is canceled by the prescription of \Equation{limit}.
We see that the integrand indeed exibits the behavior that leads to \Equation{spuriousrank}.
Notice also that for $\Lambda\to\infty$ the rank of the numerator is too low to produce bubbles (the number of denominators is $n+m+2$), generalizing the observation regarding \Equation{vanishbubble}, that the integrand at leading power of $\Lambda$ does not produce bubbles with a $\Lambda$-dependent denominator.

\section{Bubble coefficients}
Here we show that the coefficients for bubbles with denominators 
%
%%%%%%%%%%%%%%%%%%%%%%%%%%%%%%%%%%%%%%%%
\begin{equation}
(\ell+K_0)^2
\quad,\quad
(\ell+K_1)^2
\quad,\quad
K_1^\mu = \Lambda p^\mu + K_1'^\mu
\end{equation}
%%%%%%%%%%%%%%%%%%%%%%%%%%%%%%%%%%%%%%%%
%
are finite, and we show how to calculate them.
We use labels $0,1$ rather than $i,j$ to indicate the denominators and their momenta.
We introduce the momentum
%
%%%%%%%%%%%%%%%%%%%%%%%%%%%%%%%%%%%%%%%%
\begin{equation}
k_1^\mu = K_1^\mu-K_0^\mu = \Lambda p^\mu + k_1'^\mu
\end{equation}
%%%%%%%%%%%%%%%%%%%%%%%%%%%%%%%%%%%%%%%%
%
and decompose it into two light-like momenta $k_1^\mu=p_\flat^\mu+p_\sharp^\mu$ following
%
%%%%%%%%%%%%%%%%%%%%%%%%%%%%%%%%%%%%%%%%
\begin{equation}
p_\flat^\mu = \frac{k_1^2}{2\lop{p}{k_1}}\,p^\mu
 = \left(\Lambda+\frac{k_1'^2}{2\lop{p}{k_1'}}\right)p^\mu
\quad,\quad
p_\sharp^\mu = k_1^\mu - p_\flat^\mu
             = k_1'^\mu - \frac{k_1'^2}{2\lop{p}{k_1'}}\,p^\mu
\end{equation}
%%%%%%%%%%%%%%%%%%%%%%%%%%%%%%%%%%%%%%%%
%
We construct the orthogonal vectors
%
%%%%%%%%%%%%%%%%%%%%%%%%%%%%%%%%%%%%%%%%
\begin{equation}
e_1^\mu = \srac{1}{2}\brktAS{p_\flat|\gamma^\mu|p_\sharp}
\quad,\quad
e_2^\mu = \srac{1}{2}\brktAS{p_\sharp|\gamma^\mu|p_\flat}
~,
\end{equation}
%%%%%%%%%%%%%%%%%%%%%%%%%%%%%%%%%%%%%%%%
%
and a third one to span the whole trivial space by
%%%%%%%%%%%%%%%%%%%%%%%%%%%%%%%%%%%%%%%%
\begin{equation}
e^\mu_3=e_1^\mu+e_2^\mu+\imag\big(p_\flat^\mu-p_\sharp^\mu)
= \srac{1}{2}\big(\leftA{p_\flat}-\imag\leftA{p_\sharp}\big)
   \gamma^\mu\big(\imag\Srght{p_\flat}+\Srght{p_\sharp}\big)
~.
\end{equation}
%%%%%%%%%%%%%%%%%%%%%%%%%%%%%%%%%%%%%%%%
%
These vectors satisfy $e_i^2=0$ and $-2\lop{e_i}{e_j}=k_1^2$.
Solutions to the cut equations are given by
%
%%%%%%%%%%%%%%%%%%%%%%%%%%%%%%%%%%%%%%%%
\begin{align}
\ell_1^\mu +K_0^\mu &= ze_1^\mu - p_\sharp^\mu 
                  = \srac{1}{2}\big(z\leftA{p_\flat}-\leftA{p_\sharp}\big)\gamma^\mu\Srght{p_\sharp}
\\
\ell_2^\mu +K_0^\mu &= ze_2^\mu - p_\sharp^\mu
                  = \srac{1}{2}\leftA{p_\sharp}\gamma^\mu\big(z\Srght{p_\flat}-\Srght{p_\sharp}\big)
\end{align}
%%%%%%%%%%%%%%%%%%%%%%%%%%%%%%%%%%%%%%%%
%
for any value of $z$, \ie\ they satisfy both $(\ell_{1,2}+K_0)^2=0$ and $(\ell_{1,2}+K_1)^2=0$.
Solutions constructed with both $e_1^\mu$ and $e_2^\mu$ are needed in order to access all possible terms in the spurious polynomial.
The solutions are divergent and behave as $\sqrt{\Lambda}$.
We could have chosen linear behavior by replacing the explicit $p_\sharp^\mu$ in the solutions with $p_\flat^\mu$.
Realize, however, that either choice leads to one of two cut momenta $\ell^\mu+K_0^\mu$ and $\ell^\mu+K_1^\mu$ at the solution to diverge as $\Lambda$ and the other one as $\sqrt{\Lambda}$.
More symmetrical solutions regarding the divergent behavior are given by
%
%%%%%%%%%%%%%%%%%%%%%%%%%%%%%%%%%%%%%%%%
\begin{align}
\ell_{12}^\mu+K_0^\mu &= \srac{1}{2}\Big(ze_1^\mu + \srac{1}{z}e_2^\mu - p_\flat^\mu - p_\sharp^\mu\Big)
 = \srac{1}{4}\Big(\leftA{p_\flat}-\srac{1}{z}\leftA{p_\sharp}\Big)
   \gamma^\mu\Big(\!-\Srght{p_\flat}+z\Srght{p_\sharp}\Big)
\\
\ell_{23}^\mu+K_0^\mu &= \srac{1}{2}\Big(ze_2^\mu + \srac{1}{z}e_3^\mu - p_\flat^\mu - p_\sharp^\mu\Big)
 = \srac{1}{4}\Big(\leftA{p_\flat}-(z+\imag)\leftA{p_\sharp}\Big)
   \gamma^\mu\Big(\!-\srac{z-\imag}{z}\Srght{p_\flat}+\srac{1}{z}\Srght{p_\sharp}\Big)
\\
\ell_{13}^\mu+K_0^\mu &= \srac{1}{2}\Big(ze_1^\mu + \srac{1}{z}e_3^\mu - p_\flat^\mu - p_\sharp^\mu\Big)
 = \srac{1}{4}\Big(\leftA{p_\flat}-\srac{1}{z-\imag}\leftA{p_\sharp}\Big)
   \gamma^\mu\Big(\!-\srac{z-\imag}{z}\Srght{p_\flat}+\srac{z^2+1}{z}\Srght{p_\sharp}\Big)
\label{bubblesimsol}
\end{align}
%%%%%%%%%%%%%%%%%%%%%%%%%%%%%%%%%%%%%%%%
%
for any value of $z$.
\myFigure{%
\epsfig{figure=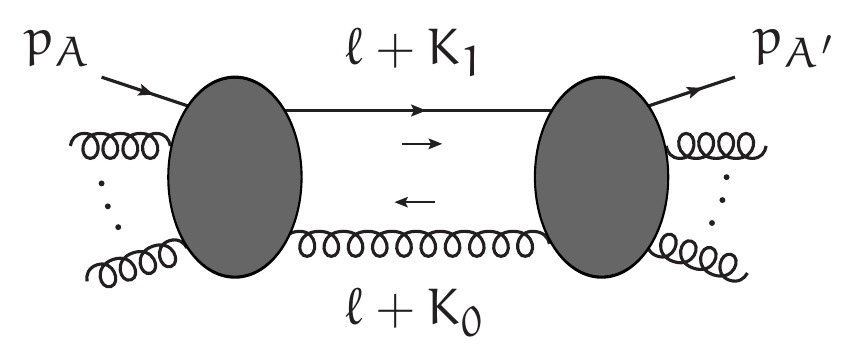,width=0.42\linewidth}
\caption{\label{Fig:bubblecut}Momentum flow in the residue of the bubble cut.}
}%
In the limit of $\Lambda\to\infty$, the cut momenta at the solutions become
%
%%%%%%%%%%%%%%%%%%%%%%%%%%%%%%%%%%%%%%%%
\begin{align}
&\ell^\mu_{12}+K_0^\mu = -\srac{1}{2}\Lambda p^\mu + \Ord\big(\sqrt{\Lambda}\big)&
&,&
&\ell^\mu_{12}+K_1^\mu =  \srac{1}{2}\Lambda p^\mu + \Ord\big(\sqrt{\Lambda}\big)&
,\\
&\ell^\mu_{23}+K_0^\mu = -\srac{z-\imag}{2z}\Lambda p^\mu + \Ord\big(\sqrt{\Lambda}\big)&
&,&
&\ell^\mu_{23}+K_1^\mu =  \srac{z+\imag}{2z}\Lambda p^\mu + \Ord\big(\sqrt{\Lambda}\big)&
,\\
&\ell^\mu_{13}+K_0^\mu = -\srac{z-\imag}{2z}\Lambda p^\mu + \Ord\big(\sqrt{\Lambda}\big)&
&,&
&\ell^\mu_{13}+K_1^\mu =  \srac{z+\imag}{2z}\Lambda p^\mu + \Ord\big(\sqrt{\Lambda}\big)&
,
\end{align}
%%%%%%%%%%%%%%%%%%%%%%%%%%%%%%%%%%%%%%%%
%
and the spinors behave as
%
%%%%%%%%%%%%%%%%%%%%%%%%%%%%%%%%%%%%%%%%
\begin{align}
&\Arght{\ell_{ij}+K_0} \propto \sqrt{\Lambda}\,\Arght{p} + \Ord\big(\Lambda^0\big)
\quad,\quad
 \Srght{\ell_{ij}+K_0} \propto \sqrt{\Lambda}\,\Srght{p} + \Ord\big(\Lambda^0\big)
\quad,
\\
&\Arght{\ell_{ij}+K_1} \propto \sqrt{\Lambda}\,\Arght{p} + \Ord\big(\Lambda^0\big)
\quad,\quad
 \Srght{\ell_{ij}+K_1} \propto \sqrt{\Lambda}\,\Srght{p} + \Ord\big(\Lambda^0\big)
\quad.
\end{align}
%%%%%%%%%%%%%%%%%%%%%%%%%%%%%%%%%%%%%%%%
%
We choose these ``symmetric'' solutions for the following.
They are also equivalent to the ones in \Appendix{bubblecutsolEGK}.

\subsection{Residue}
\Figure{Fig:bubblecut} depicts the momentum flow in the residue of the integrand at the cut.
We recognize that each blob is a tree-level amplitude of the type described in \Section{Sec:auxilquarkglu}, and behaves as $\sqrt{\Lambda}$.
% 
% % with the association
% %
% %%%%%%%%%%%%%%%%%%%%%%%%%%%%%%%%%%%%%%%%
% \begin{equation}
%  p_B^\mu = p_A^\mu
% \quad,\quad
% p_{B'}^\mu = \ell^\mu + K_0^\mu
% \quad,\quad
% p_{B''}^\mu = -\ell^\mu-K_1^\mu
% \end{equation}
% %%%%%%%%%%%%%%%%%%%%%%%%%%%%%%%%%%%%%%%%
% %
% for the left blob, and
% %
% %%%%%%%%%%%%%%%%%%%%%%%%%%%%%%%%%%%%%%%%
% \begin{equation}
%  p_B^\mu = -\ell^\mu - K_0^\mu
% \quad,\quad
%  p_{B'}^\mu = p_{A'}^\mu
% \quad,\quad
%  p_{B''}^\mu = \ell^\mu+K_1^\mu
% \end{equation}
% %%%%%%%%%%%%%%%%%%%%%%%%%%%%%%%%%%%%%%%%
% %
% for the right blob.
% %
% We see that the ratios of \Equation{gaugecondition} indeed become independent of the arbitrary auxiliary momentum in the limit of $\Lambda\to\infty$, and that the behavior of each blob is determined by the factors of \Equation{Eq393}.
% %
% The ratios are finite in the limit $\Lambda\to\infty$, so the behavior of the blobs is determined by the factors $\brktAA{p_{B'}|p_{B''}}$ and $\brktSS{p_{B''}|p_{B}}$.
% %
% The leading behavior in the spinors of all three momenta $p_B^\mu,p_{B'}^\mu,p_{B''}^\mu$ proportional to $\sqrt{\Lambda}\Arght{p}$ and $\sqrt{\Lambda}\Srght{p}$, so the leading behavior in $\brktAA{p_{B'}|p_{B''}}$ and $\brktSS{p_{B''}|p_{B}}$ is eliminated, 
% and each of them behaves at most as $\sqrt{\Lambda}$.
% %
Thus the residue behaves at most as $\Lambda$, and this is eventually eleminated by the prescription of \Equation{limit}, leading to a finite contribution to the amplitude.

\subsection{Subtraction terms}
The above was only regarding the first term on the right-hand-side of \Equation{Eq:bubLHS}, and we still need to check the subraction terms.
The box subtraction terms are clearly safe.
The spurious polynomial could give a contribution proportional to $\Lambda$, but does not because $p^\mu$ is in the physical space for $\Lambda\to\infty$.
The triangle terms are also safe thanks to \Equation{spuriousrank}.
The constant coefficient gives a contribution proportional to $\Lambda$, but the spurious polynomial vanishes again because $p^\mu$ is in the physical space for $\Lambda\to\infty$.

\subsection{The remnant rational contribution}
The rank of one-loop graphs with at least one quark propagator in the loop is only high enough to lead to non-vanishing linear terms in the spurious bubble polynomial.
It can be expanded in terms of the vectors spanning the trivial space following
% %
% %%%%%%%%%%%%%%%%%%%%%%%%%%%%%%%%%%%%%%%%
% \begin{align}
% \BubPol(\ell-K_0) = \BubC 
%               &+ \tBubC_1(\lop{e_1}{\ell})
%               + \tBubC_2(\lop{e_2}{\ell})
%               + \tBubC_3(\lop{e_3}{\ell})
% \notag\\
%              &+ \tBubC_{11}(\lop{e_1}{\ell})(\lop{e_1}{\ell})
%               + \tBubC_{22}(\lop{e_2}{\ell})(\lop{e_2}{\ell})
%               + \tBubC_{33}(\lop{e_3}{\ell})(\lop{e_3}{\ell})
% \notag\\
%              &+ \tBubC_{12}(\lop{e_1}{\ell})(\lop{e_2}{\ell})
%               + \tBubC_{23}(\lop{e_2}{\ell})(\lop{e_3}{\ell})
%               + \tBubC_{31}(\lop{e_3}{\ell})(\lop{e_1}{\ell})
% ~.
% \end{align}
% %%%%%%%%%%%%%%%%%%%%%%%%%%%%%%%%%%%%%%%%
% %
%%%%%%%%%%%%%%%%%%%%%%%%%%%%%%%%%%%%%%%%
\begin{align}
\BubPol(\ell) = \BubC 
              + \sum_{i=1}^3\tBubC_i\,\lop{e_i}{(\ell+K_0)}
%  + \sum_{i,j=1}^3\tBubC_{ij}\,\lop{e_i}{(\ell+K_0)}\,\lop{e_j}{(\ell+K_0)}
~.
\end{align}
%%%%%%%%%%%%%%%%%%%%%%%%%%%%%%%%%%%%%%%%
%
%Only $9$ of the coefficients are independent, but this is not important for the following.
%
For bubbles with a $\Lambda$-dependent denominator, the polynomial evaluates to
%
%
%%%%%%%%%%%%%%%%%%%%%%%%%%%%%%%%%%%%%%%%%
%\begin{align}
%\BubPol(\ell_{kl}(z)) = \BubC 
%  + \sum_{i=1}^3\tBubC_i'\,\EuScript{F}_i(z)
%  + \sum_{i,j=1}^3\tBubC_{ij}'\,\EuScript{F}_{ij}(z)
%~,
%\end{align}
%%%%%%%%%%%%%%%%%%%%%%%%%%%%%%%%%%%%%%%%
%
%%%%%%%%%%%%%%%%%%%%%%%%%%%%%%%%%%%%%%%%
\begin{align}
\BubPol(\ell_{12}(z)) &= \BubC 
  -\frac{\Lambda\lop{p}{k_1'}}{2}\big(\tBubC_1+\tBubC_3\big)\frac{1}{z}
  -\frac{\Lambda\lop{p}{k_1'}}{2}\big(\tBubC_2+\tBubC_3\big)z
\notag\\
\BubPol(\ell_{23}(z)) &= \BubC 
 -\frac{\Lambda\lop{p}{k_1'}}{2}\big(\tBubC_1+\tBubC_2\big)\frac{1}{z}
 -\frac{\Lambda\lop{p}{k_1'}}{2}\big(\tBubC_1+\tBubC_3\big)z
\label{Eq:bubblespurious}\\
\BubPol(\ell_{31}(z)) &= \BubC 
 -\frac{\Lambda\lop{p}{k_1'}}{2}\big(\tBubC_1+\tBubC_2\big)\frac{1}{z}
 -\frac{\Lambda\lop{p}{k_1'}}{2}\big(\tBubC_2+\tBubC_3\big)z
\notag
\end{align}
%%%%%%%%%%%%%%%%%%%%%%%%%%%%%%%%%%%%%%%%
%
at the solutions given before.
%where we define
%
%%%%%%%%%%%%%%%%%%%%%%%%%%%%%%%%%%%%%%%%%
%\begin{equation}
%%\tBubC_i' = \frac{\Lambda\lop{p}{k_1'}}{-2}\,\tBubC_i
%\tBubC_i' = -\Lambda\lop{p}{k_1'}\,\tBubC_i
%\quad,\quad
%%\tBubC_{ij}' = \left(\frac{\Lambda\lop{p}{k_1'}}{-2}\right)^2\,\tBubC_{ij}
%\tBubC_{ij}' = \big(-\Lambda\lop{p}{k_1'}\big)^2\,\tBubC_{ij}
%\label{spuriousbubblesuppression}
%\end{equation}
%%%%%%%%%%%%%%%%%%%%%%%%%%%%%%%%%%%%%%%%%
%,
%and where $\EuScript{F}_i(z),\EuScript{F}_{ij}(z)$ are simple polynomials in both $z$ and $1/z$ that are finite for $\Lambda\to\infty$.
%
%The coefficients $\BubC,\tBubC_i',\tBubC_{ij}'$ are obtained by solving \Equation{Eq:bubcut} (remember that the indices refer to something else there).
%
It appears that the spurious coefficients are suppressed compared to the constant coefficient $\BubC$.

There are no tadpole contributions, so the spurious coefficients are not needed to construct subtraction terms.
They are needed, however, to obtain the rational contribution $\EuScript{R}$  in \Equation{oneloopdecom}.
Within the integrand-based methods of one-loop calculations, it appears as the error one makes by treating the loop momentum as $4$-dimensional in the integrand, while it is not, and in the approach of \cite{Ossola:2006us} $\EuScript{R}$ is separated into two parts.
The first part $\EuScript{R}_1$ arises because of a mismatch between $4$-dimensional and $(4-2\varepsilon)$-dimensional denominators.
The second $\EuScript{R}_2$ arises because the loop momentum in the numerator is treated as if it were $4$-dimensional.
The latter is calculated with tree-level graphs containing extra renormalization-type vertices~\cite{Draggiotis:2009yb}.
They follow the tree-level power-counting and are finite with $\Lambda\to\infty$. 

$\EuScript{R}_1$ can for example be calculated following \cite{Ossola:2007bb}.
There it is explained that it can be obtained by multiplying spurious coefficients with so-called higher-dimensional integrals of the type
%
%%%%%%%%%%%%%%%%%%%%%%%%%%%%%%%%%%%%%%%%
\begin{equation}
I_{i_1i_2\cdots i_n}^{\mu_1\mu_2\cdots\mu_r} =
\int\dDell\big(\ell_{-2\varepsilon}^2\big)^{n-1}
\frac{\ell_4^{\mu_1}\ell_4^{\mu_2}\cdots\ell_4^{\mu_r}}
               {\Den_{i_1}(\ell)\Den_{i_2}(\ell)\cdots\Den_{i_n}(\ell)}
\end{equation}
%%%%%%%%%%%%%%%%%%%%%%%%%%%%%%%%%%%%%%%%
%
where $\ell_4^\mu$ represents the $4$-dimensional components of the integration momentum and $\ell_{-2\varepsilon}^\mu$ represents the higher-dimensional components, and where the set of denominators must include the two from the bubble under consideration.
The constant $\BubC$ is multiplied with scalar integrals
%
%%%%%%%%%%%%%%%%%%%%%%%%%%%%%%%%%%%%%%%%
\begin{equation}
I_{i_1i_2\cdots i_n} = \frac{-1}{n(n-1)} + \Ord(\varepsilon)
\end{equation}
%%%%%%%%%%%%%%%%%%%%%%%%%%%%%%%%%%%%%%%%
%
which, to lowest order in $\varepsilon$, are independent of the momenta involved.
Contributions from the linear coefficients are given by
%
%%%%%%%%%%%%%%%%%%%%%%%%%%%%%%%%%%%%%%%%
\begin{multline}
\sum_{j=1}^3\tBubC_i\,e_{i,\mu}I_{i_1i_2\cdots i_n}^\mu
=
\frac{1}{(n-1)n(n+1)}\sum_{j=1}^3\tBubC_i\Big[m\Lambda\lop{e_i}{p}+\Ord\Big(\sqrt{\Lambda}\Big)\Big]
 + \Ord(\varepsilon)
\\
=\frac{\imag m}{(n-1)n(n+1)}(-\Lambda\lop{p}{k_1'})\tBubC_3+\Ord\Big(1/\sqrt{\Lambda}\Big) + \Ord(\varepsilon)
~,
\end{multline}
%%%%%%%%%%%%%%%%%%%%%%%%%%%%%%%%%%%%%%%%
%
where $m$ is the number of $\Lambda$-dependent denominators.
This quantity is finite, because we saw in \Equation{Eq:bubblespurious} that the spurious coefficients are suppressed.

\subsubsection{\label{Sec:complication}Complication with higher-dimensional integrals
}
Coefficients for master integrals without a $\Lambda$-dependent denominator may be calculated at $\Lambda\to\infty$.
Only graphs of \Figure{Fig:generalGraph} with $m=0$, so with only one auxiliary quark propagator, and with only three-gluon vertices in the blob (\Figure{Fig:Blob}) have a rank that is high enough at this limit to contribute non-vanishing constant bubble coefficients.
The rank is not high enough for non-vanishing spurious bubble coefficients. 
However, the rational terms are calculated with higher-dimensional integrals including {\em all\/} denominators, and we saw before that they diverge with $\Lambda$.
Consequently, for these graphs, the $\Lambda$-suppressed part of the numerator must be reduced separately, in order to find the suppressed spurious bubble coefficients of bubbles without $\Lambda$-dependent denominators, because they give a finite contribution together with the divergent higher-dimensional integrals.
For the same reason, also the suppressed constant tadpole coefficients must extracted from these graphs.

%Now we remember from \Equation{spuriousbubblesuppression} that the linear coefficients are suppressed by $\Lambda$.
%%
%Furthermore, we have
%%
%%%%%%%%%%%%%%%%%%%%%%%%%%%%%%%%%%%%%%%%%
%\begin{equation}
%\lop{e_1}{p} = \lop{e_2}{p} = 0
%\quad,\quad
%\lop{e_3}{p} =-\imag\lop{k_1'}{p}
%~,
%\end{equation}
%%%%%%%%%%%%%%%%%%%%%%%%%%%%%%%%%%%%%%%%%
%%
%so we find to leading power in $\Lambda$ and $\varepsilon$
%%
%%%%%%%%%%%%%%%%%%%%%%%%%%%%%%%%%%%%%%%%%
%\begin{equation}
%\sum_{j=1}^3\tBubC_j\,e_{j,\mu}I_{i_1i_2\cdots i_n}^\mu
%=
%\frac{\imag m\tBubC_3'}{(n-1)n(n+1)}
%~.
%\end{equation}
%%%%%%%%%%%%%%%%%%%%%%%%%%%%%%%%%%%%%%%%%
%%
%For the contributions from the quadratic coefficients we find
%%
%%%%%%%%%%%%%%%%%%%%%%%%%%%%%%%%%%%%%%%%%
%\begin{equation}
%\sum_{j,k=1}^3\tBubC_{jk}\,e_{j,\mu}e_{k,\nu}I_{i_1i_2\cdots i_n}^{\mu\nu}
%=
%\frac{m(m+1)\tBubC_{33}'}{(n-1)n(n+1)(n+2)}
%~.
%\end{equation}
%%%%%%%%%%%%%%%%%%%%%%%%%%%%%%%%%%%%%%%%%

\section{Expressions for the scalar master integrals\label{scalarintegrals}}
\begin{figure}
\begin{center}
\epsfig{figure=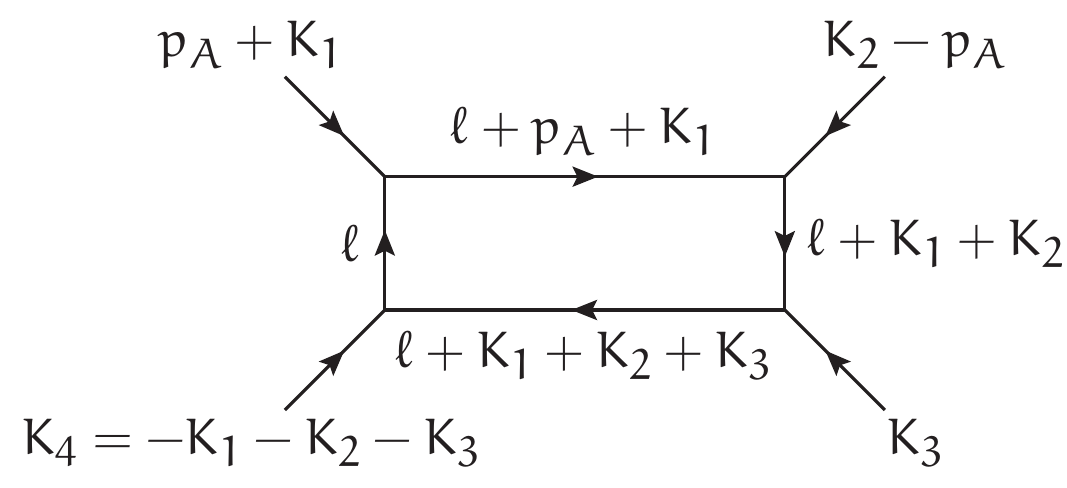,width=0.48\linewidth}
\caption{\label{Fig:box00}Momentum flow for the box integral (\ref{Eq:box00}).}
\end{center}
\end{figure}

We saw before that the necessary master integrals have at most one linear denominator.
Consider the general box with one such denominator
%
%%%%%%%%%%%%%%%%%%%%%%%%%%%%%%%%%%%%%%%%
\begin{align}
I &= 
    \int \dDell\,\frac{1}
    {(\ell+Q_0)^2\;2\lop{p}{(\ell+Q_1)}\;(\ell+Q_2)^2\;(\ell+Q_3)^2}
\notag\\
  &= 
    \int \dDell\,\frac{1}
    {\ell^2\;2\lop{p}{(\ell+K_1)}\;(\ell+K_1+K_2)^2\;(\ell+K_1+K_2+K_3)^2}
\notag\\
\end{align}
%%%%%%%%%%%%%%%%%%%%%%%%%%%%%%%%%%%%%%%%
%
where
%
%%%%%%%%%%%%%%%%%%%%%%%%%%%%%%%%%%%%%%%%
\begin{equation}
K_1=Q_1-Q_0 \quad,\quad K_2=Q_2-Q_1 \quad,\quad K_3=Q_3-Q_2
~.
\end{equation}
%%%%%%%%%%%%%%%%%%%%%%%%%%%%%%%%%%%%%%%%
%
According to our prescription, this integral becomes
%
%%%%%%%%%%%%%%%%%%%%%%%%%%%%%%%%%%%%%%%%
\begin{equation}
I \to 
    \int \dDell\,\frac{\Lambda}
    {\ell^2\;(\ell+p_A+K_1)^2\;(\ell+K_1+K_2)^2\;(\ell+K_1+K_2+K_3)^2}
~,
\label{Eq:box00}
\end{equation}
%%%%%%%%%%%%%%%%%%%%%%%%%%%%%%%%%%%%%%%%
%
which corresponds to the graph in \Figure{Fig:box00}.
We suppress the causal $\imag\eta$ here, and want to stress that there is no ambiguity in the sign for the linear denominator, because it came from the quadratic denominator in the first place.
In order to find expressions for all non-equivalent kinematical situations, $p_A^\mu=\Lambda p^\mu+\alpha q^\mu+\beta k_T^\mu$ from \Equation{auxmom1} can be replaced with $\Lambda p^\mu$ in most cases.
Only in Feynman graphs for which the off-shell gluon is directly attached to the loop there are contributions from master integrals with $K_1^\mu=0$ and $K_2^\mu=k^\mu=xp^\mu+k_T^\mu$ for which this replacement is inaccurate, because it would lead to non-vanishing invariant $s_2=(K_2-p_A)^2$ while it does vanish for $K_2^\mu=k^\mu$.
For the purpose of finding expressions of the master integrals, this is only an issue when $K_1^\mu=0$, because the expression for $K_1^\mu\neq0$ and $K_2^\mu=k^\mu$ is equivalent to the one for $K_1^\mu=0$ and $K_2^\mu\neq k^\mu$.

In order to find the expressions for $\Lambda\to\infty$, we can simply take the expressions for the integrals of the type of \Equation{Eq:box00} from literature, and take the limit in these.
For example, for the kinematical situation in which $K_3$ and $K_4$ are both light-like while $p_A+K_1$ and $K_2-p_A$ are not, we recognize that
%
%%%%%%%%%%%%%%%%%%%%%%%%%%%%%%%%%%%%%%%%
\begin{equation}
I = \Lambda\,\mathrm{Box}_4(s_3,s_4,s_{12},s_{23})
~,
\end{equation}
%%%%%%%%%%%%%%%%%%%%%%%%%%%%%%%%%%%%%%%%
%
with
%
%%%%%%%%%%%%%%%%%%%%%%%%%%%%%%%%%%%%%%%%
\begin{gather}
 s_3=(p_A+K_1)^2\to2\Lambda\lop{p}{K_1}
\quad,\quad
s_4=(K_2-p_A)\to-2\Lambda\lop{p}{K_2}
\notag\\
s_{12}=(K_1+K_2)^2
\quad,\quad
s_{23}=(K_2+K_3-p_A)^2\to-2\Lambda\lop{p}{(K_2+K_3)}
\quad,
\end{gather}
%%%%%%%%%%%%%%%%%%%%%%%%%%%%%%%%%%%%%%%%
%
and where $\mathrm{Box_4}$ refers to box integral number $4$ in the classification of \cite{Ellis:2007qk}, \ie\ formula (4.23) in that publication:
%
%%%%%%%%%%%%%%%%%%%%%%%%%%%%%%%%%%%%%%%%
\begin{multline}
\mathrm{Box}_4(s_3,s_4,s_{12},s_{23})
=1/(s_{12}s_{23})\\
\times\Bigg\{
\frac{2}{\varepsilon^2}\Bigg[
\bigg(\frac{-s_{12}}{\mu^2}\bigg)^{\!-\varepsilon}
+\bigg(\frac{-s_{23}}{\mu^2}\bigg)^{\!-\varepsilon}
-\bigg(\frac{-s_3}{\mu^2}\bigg)^{\!-\varepsilon}
-\bigg(\frac{-s_4}{\mu^2}\bigg)^{\!-\varepsilon}
%\\\hspace{\fill}
+\frac{1}{2}\bigg(\frac{-s_3}{\mu^2}\bigg)^{\!-\varepsilon}\!
            \bigg(\frac{-s_4}{\mu^2}\bigg)^{\!-\varepsilon}\!
            \bigg(\frac{\mu^2}{-s_{12}}\bigg)^{\!-\varepsilon}
\Bigg]
\\
-2\mathrm{Li}_2\bigg(1-\frac{s_3}{s_{23}}\bigg)
-2\mathrm{Li}_2\bigg(1-\frac{s_4}{s_{23}}\bigg)
-\ln^2\bigg(\frac{-s_{12}}{-s_{23}}\bigg)
\Bigg\}
+\Ord(\varepsilon)
~.
\end{multline}
%%%%%%%%%%%%%%%%%%%%%%%%%%%%%%%%%%%%%%%%
%
This procedure has been performed for all non-equivalent boxes and triangles, and the results are given below.

\subsection{Triangles}
Consider the triangle integral
%
%%%%%%%%%%%%%%%%%%%%%%%%%%%%%%%%%%%%%%%%
\begin{equation}
\graph{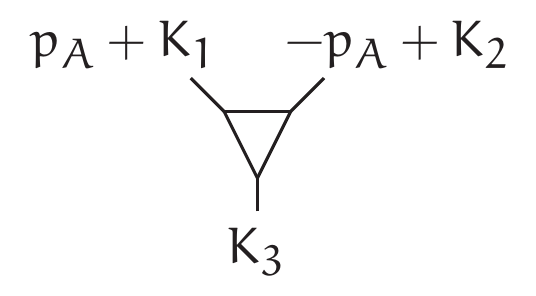}{22}{6}
=\;
\int\dDell\,\frac{\Lambda}{\ell^2\,(\ell+p_A+K_1)^2\,(\ell-K_3)^2}
~,
\end{equation}
%%%%%%%%%%%%%%%%%%%%%%%%%%%%%%%%%%%%%%%%
%
The external momenta are incoming, and we have
%
%%%%%%%%%%%%%%%%%%%%%%%%%%%%%%%%%%%%%%%%
\begin{equation}
K_1^\mu+K_2^\mu+K_3^\mu= 0~.
\end{equation}
%%%%%%%%%%%%%%%%%%%%%%%%%%%%%%%%%%%%%%%%
%
We introduce the notation
%
%%%%%%%%%%%%%%%%%%%%%%%%%%%%%%%%%%%%%%%%
\begin{equation}
s_3=K_3^2
\quad,\quad
\sigma_1=2\lop{p}{K_1}
\quad,\quad
\sigma_2=-2\lop{p}{K_2}
\quad.
\end{equation}
%%%%%%%%%%%%%%%%%%%%%%%%%%%%%%%%%%%%%%%%
%
The variables $\sigma_i$ give the leading term in $\Lambda$ of the external invariants, for example\ $(-\Lambda p+K_2)^2=\Lambda\sigma_2+\Ord\big(\Lambda^0\big)$.
By convention, momenta indicated with a capital $K$ are not specified to be light-like or not.
All non-equivalent possibilities are depicted in \Figure{triangles}, and their expressions are given below, up to and including $\Ord(\varepsilon^0)$ and $\Ord\big(\Lambda^0\big)$.
\begin{figure}
\begin{center}
\epsfig{figure=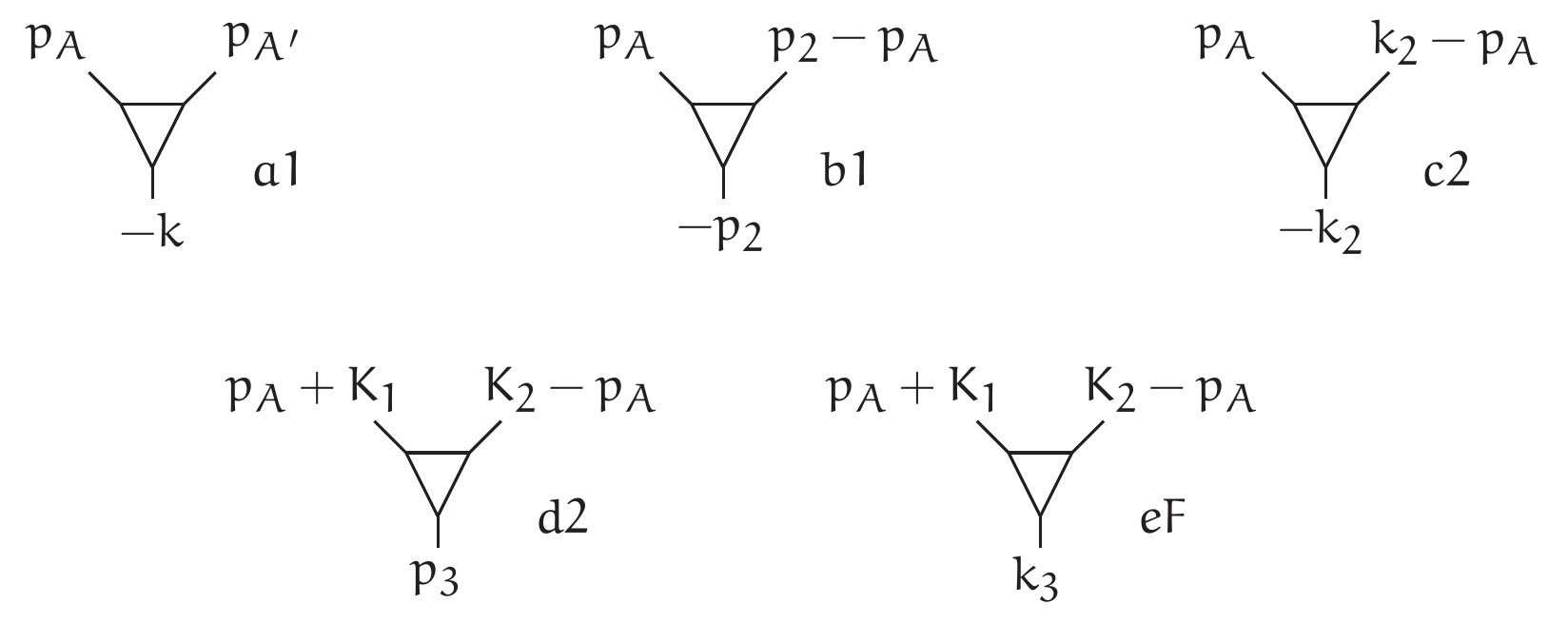,width=0.8\linewidth}
\caption{\label{triangles}All possible triangle master integrals with one linear denominator. Momenta with a ``$p$'' are light-like, momenta with a ``$k$'' are not light-like, while the momenta with a ``$K$'' can be either. We write $p_{A'}$ instead of $k-p_A$ to highlight that this momentum is light-like. The number in the labels corresponds to the classification of \cite{Ellis:2007qk}, and $F$ stands for ``finite''.}
\end{center}
\end{figure}
We replace $p_A$ with $\Lambda p$, which is correct up to sub-leading powers of $\Lambda$.
The first one with $K_1=0$ and $(K_2-P_A)^2=0$ is the anomalous triangle of \Equation{MI3viol}.
Abreviating
%
%%%%%%%%%%%%%%%%%%%%%%%%%%%%%%%%%%%%%%%%
\begin{equation}
\lambda = \ln\Lambda
~,
\end{equation}
%%%%%%%%%%%%%%%%%%%%%%%%%%%%%%%%%%%%%%%%
%
we find for the others
%
%%%%%%%%%%%%%%%%%%%%%%%%%%%%%%%%%%%%%%%%
\begin{align}
\mathrm{Tri_{b1}}(\sigma_2)
&=\frac{1}{\sigma_2}\left(\frac{-\mu^2}{\Lambda\sigma_2}\right)^\varepsilon\frac{1}{\varepsilon^2}
\\&\hspace{0ex}=
\frac{1}{\sigma_2}\left\{
  \frac{\lambda^2}{2}
 +\lambda\ln\left(\frac{\sigma_2}{-\mu^2}\right)
 -\frac{\lambda}{\varepsilon}
 +\frac{1}{\varepsilon^2}
 -\frac{1}{\varepsilon}\ln\left(\frac{\sigma_2}{-\mu^2}\right)
 +\frac{1}{2}\ln^2\left(\frac{\sigma_2}{-\mu^2}\right)
\right\}
\notag
\end{align}
%%%%%%%%%%%%%%%%%%%%%%%%%%%%%%%%%%%%%%%%
%
%%%%%%%%%%%%%%%%%%%%%%%%%%%%%%%%%%%%%%%%
\begin{align}
\mathrm{Tri}_{c2}(\sigma_2,s_3)
&=\frac{1}{\sigma_2}\left\{
     \left(\frac{-\mu^2}{\Lambda\sigma_2}\right)^\varepsilon\frac{1}{\varepsilon^2}
    -\left(\frac{-\mu^2}{s_3}\right)^\varepsilon\frac{1}{\varepsilon^2}
   \right\}
\\&\hspace{0ex}=
\frac{1}{\sigma_2}\left\{
  \frac{\lambda^2}{2}
 +\lambda\ln\left(\frac{\sigma_2}{-\mu^2}\right)
 -\frac{\lambda}{\varepsilon}
 -\frac{1}{\varepsilon}\ln\left(\frac{\sigma_2}{s_3}\right)
 +\frac{1}{2}\ln\left(\frac{\sigma_2}{s_3}\right)
             \ln\left(\frac{\sigma_2s_3}{(-\mu^2)^2}\right)
\right\}
\notag
\end{align}
%%%%%%%%%%%%%%%%%%%%%%%%%%%%%%%%%%%%%%%%
%
%%%%%%%%%%%%%%%%%%%%%%%%%%%%%%%%%%%%%%%%
\begin{equation}
\mathrm{Tri}_{d2}(\sigma_1,\sigma_2)
=
\frac{1}{\sigma_1-\sigma_2}\left\{
  \left(\ln\Lambda-\frac{1}{\varepsilon}\right)\ln\left(\frac{\sigma_1}{\sigma_2}\right)
 +\frac{1}{2}\ln\left(\frac{\sigma_1}{\sigma_2}\right)
             \ln\left(\frac{\sigma_1\sigma_2}{(-\mu^2)^2}\right)
\right\}
\end{equation}
%%%%%%%%%%%%%%%%%%%%%%%%%%%%%%%%%%%%%%%%
%
%%%%%%%%%%%%%%%%%%%%%%%%%%%%%%%%%%%%%%%%
\begin{equation}
\mathrm{Tri}_{eF}(\sigma_1,\sigma_2,s_3)
=
\frac{1}{\sigma_1-\sigma_2}\left\{
  \ln\Lambda\ln\left(\frac{\sigma_1}{\sigma_2}\right)
 +2\mathrm{Li}_2\left(1-\frac{\sigma_2}{\sigma_1}\right)
 +\ln\left(\frac{\sigma_1}{\sigma_2}\right)\ln\left(\frac{\sigma_1}{s_3}\right)
\right\}
\end{equation}
%%%%%%%%%%%%%%%%%%%%%%%%%%%%%%%%%%%%%%%%
%
The last one can for example be found starting from (40) in~\cite{Denner:1991qq} with $m_1\to\infty$.
Here, and in the following, we assume the analytical continuation as in~\cite{vanHameren:2009dr} with the interpretation  
%
%%%%%%%%%%%%%%%%%%%%%%%%%%%%%%%%%%%%%%%%
\begin{equation}
s_i\to-s_i-\imag\eta
\quad,\quad
\sigma_i\to-\sigma_i-\imag\eta
\quad,\quad
-\mu^2\to\mu^2
~,
\end{equation}
%%%%%%%%%%%%%%%%%%%%%%%%%%%%%%%%%%%%%%%%
%
with positive $\eta$ and positive $\mu^2$.
For a logarithm, this means for example
%
%%%%%%%%%%%%%%%%%%%%%%%%%%%%%%%%%%%%%%%%
\begin{equation}
\ln\left(\frac{\sigma_1\sigma_2}{-\mu^2s_3}\right)
\to
 \ln\left(-\frac{\sigma_1+\imag\eta}{\mu^2}\right)
+\ln\left(-\frac{\sigma_2+\imag\eta}{\mu^2}\right)
-\ln\left(-\frac{s_3+\imag\eta}{\mu^2}\right)
~.
\end{equation}
%%%%%%%%%%%%%%%%%%%%%%%%%%%%%%%%%%%%%%%%
%
Notice that $\mathrm{Tri}_{eF}$ is symmetric in $\sigma_1,\sigma_2$, which can be seen by applying the relation 
%%%%%%%%%%%%%%%%%%%%%%%%%%%%%%%%%%%%%%%%
\begin{equation}
\mathrm{Li}_2(1-x) = -\mathrm{Li}_2\left(1-\frac{1}{x}\right)
                     -\frac{1}{2}\ln^2(x)
~.
\label{IdLi2}
\end{equation}
%%%%%%%%%%%%%%%%%%%%%%%%%%%%%%%%%%%%%%%%
%
This relation, together with $\mathrm{Li}_2(1)=\frac{\pi^2}{6}$, is also applied when $x\propto\Lambda$.

\subsection{Boxes}
\begin{figure}
\begin{center}
\epsfig{figure=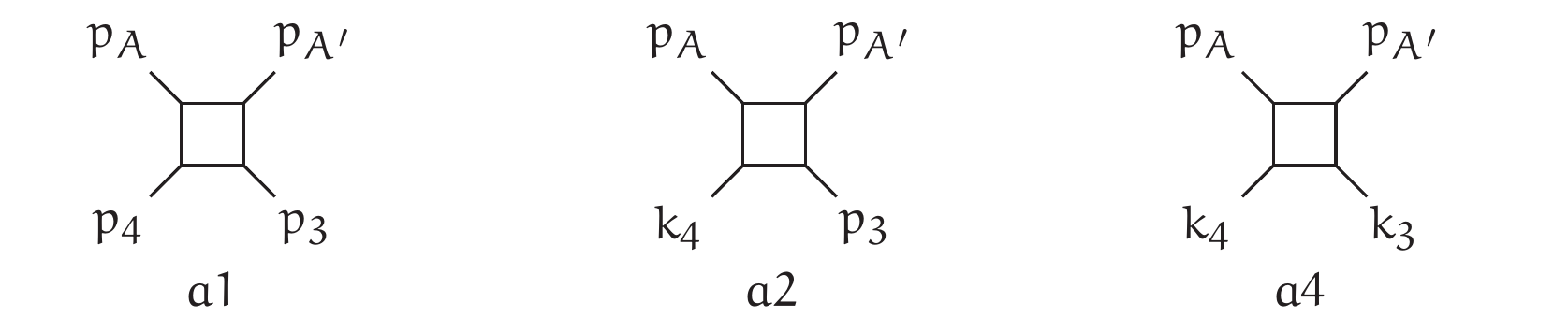,width=0.8\linewidth}\\[3ex]
\epsfig{figure=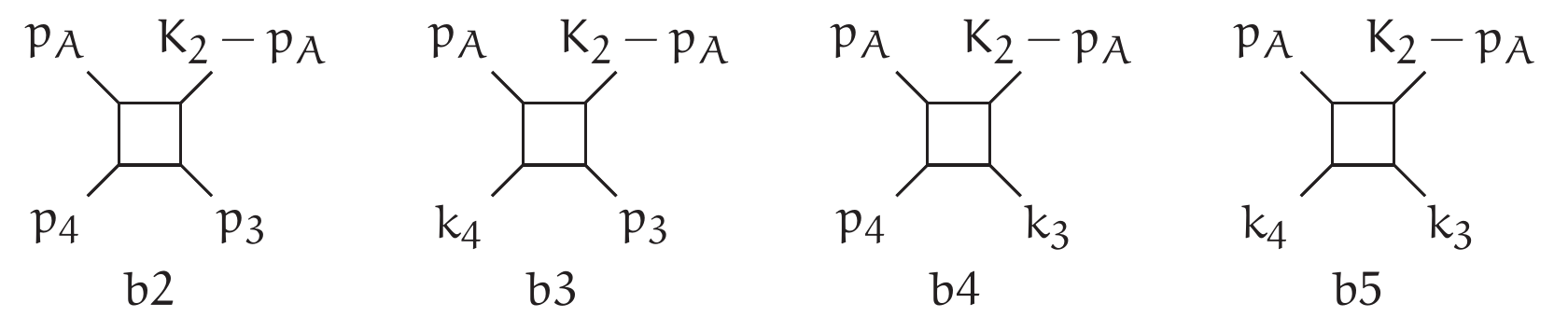,width=0.8\linewidth}\\[3ex]
\epsfig{figure=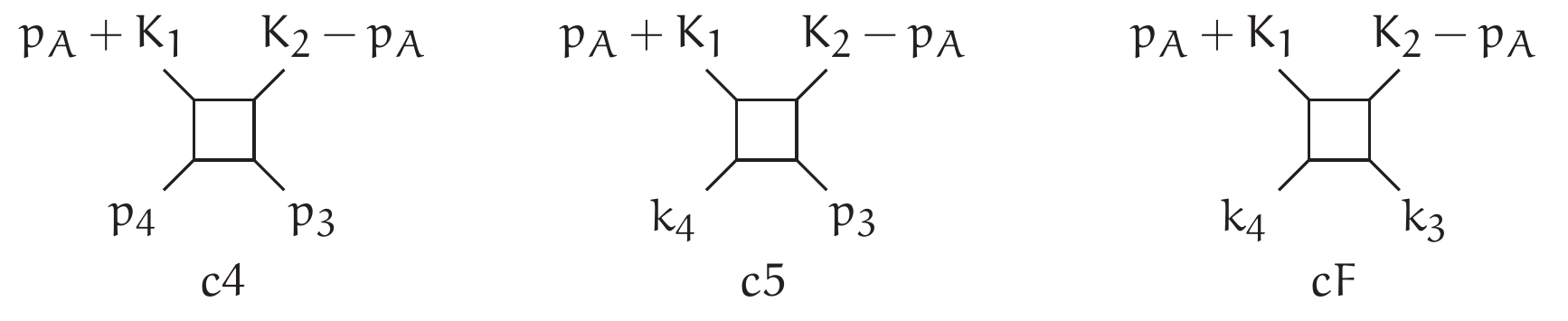,width=0.8\linewidth}
\caption{\label{boxes}All possible box master integrals with one linear denominator. Momenta with a ``$p$'' are light-like, momenta with a ``$k$'' are not light-like, while the momenta with a ``$K$'' can be either. We write $p_{A'}$ instead of $k-p_A$ in the first three boxes to highlight that this momentum is light-like. The number in the labels corresponds to the classification of \cite{Ellis:2007qk}, and $F$ stands for ``finite''.}
\end{center}
\end{figure}

Consider the box integral
%%%%%%%%%%%%%%%%%%%%%%%%%%%%%%%%%%%%%%%%
\begin{equation}
\graph{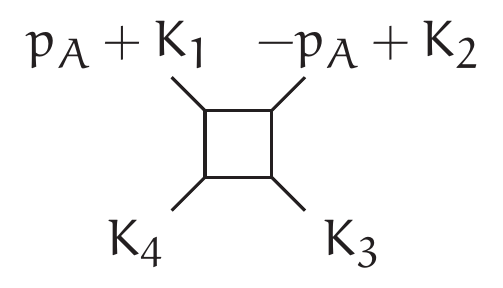}{20}{5.5}
=\;
\int\dDell\,\frac{\Lambda}{\ell^2\,(\ell+p_A+K_1)^2
                           \,(\ell-K_3-K_4)^2\,(\ell-K_4)^2}
~,
\end{equation}
%%%%%%%%%%%%%%%%%%%%%%%%%%%%%%%%%%%%%%%%
%
The external momenta are incoming, and we have
%
%%%%%%%%%%%%%%%%%%%%%%%%%%%%%%%%%%%%%%%%
\begin{equation}
K_1^\mu+K_2^\mu+K_3^\mu+K_4^\mu = 0~.
\end{equation}
%%%%%%%%%%%%%%%%%%%%%%%%%%%%%%%%%%%%%%%%
%
We introduce the notation
%
%%%%%%%%%%%%%%%%%%%%%%%%%%%%%%%%%%%%%%%%
\begin{gather}
s_3=K_3^2
\quad,\quad
s_4=K_4^2
\quad,\quad
s_{34} = (K_3+K_4)^2 = (K_1+K_2)^2
\quad,
\notag\\
\sigma_1=2\lop{p}{K_1}
\quad,\quad
\sigma_2=-2\lop{p}{K_2}
\quad,\quad
\sigma_{14}=2\lop{p}{(K_1+K_4)}
\quad.
\end{gather}
%%%%%%%%%%%%%%%%%%%%%%%%%%%%%%%%%%%%%%%%
%
Here we present the expression for all non-equivalent kinematical situations given in \Figure{boxes}.
We replace $p_A$ with $\Lambda p$, which is correct up to subleading powers of $\Lambda$, except in the first three boxes ($a1,a2,a4$).
All expressions are derived from those in \cite{Ellis:2007qk}, except the one for $\mathrm{Box}_{cF}$, which was derived from the relevant expression in \cite{Denner:1991qq}.
Notice that only the first three boxes depend on $\log\Lambda$.
\begin{multline}
\mathrm{Box}_{a1}(k_T^2,\sigma_{14})
=
\frac{1}{\sigma_{14}k_T^2}\bigg\{
-\frac{2\lambda}{\varepsilon}
+2\lambda\ln\left(\frac{k_T^2}{-\mu^2}\right)
\;+\;
\frac{4}{\varepsilon^2}
-\frac{2}{\varepsilon}\ln\left(\frac{k_T^2}{-\mu^2}\right)
-\frac{2}{\varepsilon}\ln\left(\frac{\sigma_{14}}{-\mu^2}\right)
\\
-\pi^2
+2\ln\left(\frac{k_T^2}{-\mu^2}\right)\ln\left(\frac{\sigma_{14}}{-\mu^2}\right)
\bigg\}
\\
=\frac{1}{\sigma_{14}k_T^2}\bigg\{
\left(\frac{-\mu^2}{k_T^2}\right)^\varepsilon
  \left[\frac{4}{\varepsilon^2}
       -\frac{2}{\varepsilon}\ln\frac{\Lambda\sigma_{14}}{k_T^2}
  \right]
  -\pi^2
\bigg\}
\label{onshellbox1}
\end{multline}
\begin{multline}
\mathrm{Box}_{a2}(s_4,k_T^2,\sigma_{14})
=
\frac{1}{\sigma_{14}k_T^2}\bigg\{
-\frac{2\lambda}{\varepsilon}
+2\lambda\ln\left(\frac{k_T^2}{-\mu^2}\right)
\;+\;
\frac{2}{\varepsilon^2}
-\frac{2}{\varepsilon}\ln\left(\frac{k_T^2}{-\mu^2}\right)
-\frac{2}{\varepsilon}\ln\left(\frac{\sigma_{14}}{s_4}\right)
\\
-\frac{2\pi^2}{3}
+2\mathrm{Li}_2\left(1-\frac{k_T^2}{s_4}\right)
+\ln^2\left(\frac{k_T^2}{-\mu^2}\right)
+2\ln\left(\frac{k_T^2}{-\mu^2}\right)\ln\left(\frac{\sigma_{14}}{s_4}\right)
\bigg\}
\\
=\frac{1}{\sigma_{14}k_T^2}\bigg\{
\left(\frac{-\mu^2}{k_T^2}\right)^\varepsilon
  \left[\frac{2}{\varepsilon^2}
       -\frac{2}{\varepsilon}\ln\frac{\Lambda\sigma_{14}}{s_4}
  \right]
  +2\mathrm{Li}_2\left(1-\frac{k_T^2}{s_4}\right)-\frac{2\pi^2}{3}
\bigg\}
\label{onshellbox2}
\end{multline}
\begin{multline}
\mathrm{Box}_{a4}(s_3,s_4,k_T^2,\sigma_{14})
=
\frac{1}{\sigma_{14}k_T^2}\bigg\{
-\frac{2\lambda}{\varepsilon}
+2\lambda\ln\left(\frac{k_T^2}{-\mu^2}\right)
\;+\;
\frac{1}{\varepsilon^2}
 - \frac{1}{\varepsilon}\ln\left(\frac{k_T^2}{-\mu^2}\right)
 - \frac{1}{\varepsilon}\ln\left(\frac{\sigma_{14}^2}{s_3s_4}\right)
\\
+\frac{1}{2}\ln^2\left(\frac{k_T^2}{-\mu^2}\right)
+\ln\left(\frac{k_T^2}{-\mu^2}\right)\ln\left(\frac{\sigma_{14}^2}{s_3s_4}\right)
-\frac{1}{2}\ln^2\left(\frac{s_3}{s_4}\right)
-\frac{2\pi^2}{3}
\bigg\}
\\
=\frac{1}{\sigma_{14}k_T^2}\bigg\{
\left(\frac{-\mu^2}{k_T^2}\right)^\varepsilon
  \left[\frac{1}{\varepsilon^2}
       -\frac{1}{\varepsilon}\ln\frac{\Lambda^2\sigma_{14}^2}{s_3s_4}
  \right]
  -\frac{1}{2}\ln^2\frac{s_3}{s_4}
  -\frac{2\pi^2}{3}
\bigg\}
\label{onshellbox4}
\end{multline}
\begin{multline}
\mathrm{Box}_{b2}(\sigma_2,s_{34},\sigma_{14})
=
\frac{1}{\sigma_{14}s_{34}}\bigg\{
\frac{2}{\varepsilon^2}
- \frac{2}{\varepsilon}\ln\left(\frac{s_{34}\sigma_{14}}{-\mu^2\sigma_2}\right)
\\
-2\mathrm{Li}_2\left(1-\frac{\sigma_2}{\sigma_{14}}\right)
+\ln\left(\frac{s_{34}}{-\mu^2}\right)
 \ln\left(\frac{s_{34}\sigma_{14}^2}{-\mu^2\sigma_{2}^2}\right)
%+\ln\left(\frac{\sigma_2\sigma_{14}}{s_{34}^2}\right)
% \ln\left(\frac{\sigma_2}{\sigma_{14}}\right)
%+\ln\left(\frac{s_{34}}{-\mu^2}\right)
%+\ln\left(\frac{\sigma_{14}}{-\mu^2}\right)
%-\ln\left(\frac{\sigma_2}{-\mu^2}\right)
\bigg\}
\end{multline}
\begin{multline}
\mathrm{Box}_{b3}(\sigma_2,s_4,s_{34},\sigma_{14})
=
\\
\frac{1}{\sigma_{14}s_{34}-\sigma_2s_{34}}\bigg\{
\frac{1}{\varepsilon}\ln\left(\frac{\sigma_2s_4}{\sigma_{14}s_{34}}\right)
-2\mathrm{Li}_2\left(1-\frac{\sigma_2}{\sigma_{14}}\right)
-2\mathrm{Li}_2\left(1-\frac{s_4}{s_{34}}\right)
+2\mathrm{Li}_2\left(1-\frac{\sigma_2s_4}{\sigma_{14}s_{34}}\right)
\\
-\ln\left(\frac{\sigma_2}{\sigma_{14}}\right)
 \ln\left(\frac{s_{34}^2}{(-\mu^2)^2}\right)
+\ln^2\left(\frac{s_{34}}{-\mu^2}\right)
-\ln^2\left(\frac{s_4}{-\mu^2}\right)
\bigg\}
\end{multline}
\begin{multline}
\mathrm{Box}_{b4}(\sigma_2,s_3,s_{34},\sigma_{14})
=
\frac{1}{\sigma_{14}s_{34}}\bigg\{
\frac{1}{\varepsilon^2}
+\frac{1}{\varepsilon}\ln\left(\frac{-\mu^2\sigma_2s_3}{\sigma_{14}s_{34}^2}\right)
+\frac{\pi^2}{3}
+2\mathrm{Li}_2\left(1-\frac{s_{34}}{s_{3}}\right)
\\
+2\ln\left(\frac{\sigma_{14}s_{34}}{\sigma_2s_3}\right)
  \ln\left(\frac{s_{34}}{-\mu^2}\right)
+\frac{1}{2}\ln^2\left(\frac{\sigma_2s_3}{-\mu^2\sigma_{14}}\right)
%+\ln\left(\frac{\sigma_2\sigma_{14}}{s_{34}^2}\right)
% \ln\left(\frac{\sigma_2}{\sigma_{14}}\right)
%\\
%+\ln^2\left(\frac{s_{34}}{-\mu^2}\right)
%+\ln^2\left(\frac{\sigma_{14}}{-\mu^2}\right)
%-\ln^2\left(\frac{\sigma_2}{-\mu^2}\right)
%-\ln^2\left(\frac{s_3}{-\mu^2}\right)
%+\frac{1}{2}\ln^2\left(\frac{\sigma_2s_3}{-\mu^2\sigma_{14}}\right)
\bigg\}
\end{multline}
\begin{multline}
\mathrm{Box}_{b5}(\sigma_2,s_3,s_4,s_{34},\sigma_{14}) =
\frac{1}{\sigma_{14}s_{34}-\sigma_2s_4}\bigg\{
\frac{1}{\varepsilon}\ln\left(\frac{\sigma_2s_4}{\sigma_{14}s_{34}}\right)
+2\mathrm{Li}_2\left(1-\frac{\sigma_2s_4}{\sigma_{14}s_{34}}\right)
\\
-\ln\left(\frac{\sigma_2s_4}{\sigma_{14}s_{34}}\right)
\left[
\ln\left(\frac{s_{34}}{-\mu^2}\right)
+\frac{1}{2}\ln\left(\frac{\sigma_{14}s_4}{\sigma_2s_{34}}\right)
\right]
\bigg\}
\end{multline}
\begin{multline}
\mathrm{Box}_{c4}(\sigma_1,\sigma_2,s_{34},\sigma_{14})
=\\
\frac{1}{\sigma_{14}s_{34}}\bigg\{
\frac{1}{\varepsilon^2} - \frac{1}{\varepsilon}\ln\left(\frac{s_{34}\sigma_{14}^2}{-\mu^2\sigma_1\sigma_2}\right)
-2\mathrm{Li}_2\left(1-\frac{\sigma_1}{\sigma_{14}}\right)
-2\mathrm{Li}_2\left(1-\frac{\sigma_2}{\sigma_{14}}\right)
\\
+\frac{1}{2}\ln^2\left(\frac{s_{34}}{-\mu^2}\right)
-\frac{1}{2}\ln^2\left(\frac{\sigma_1}{\sigma_2}\right)
+\ln\left(\frac{s_{34}}{-\mu^2}\right)
 \ln\left(\frac{\sigma_{14}^2}{\sigma_1\sigma_2}\right)
\bigg\}
\end{multline}
\begin{multline}
\mathrm{Box}_{c5}(\sigma_1,\sigma_2,s_4,s_{34},\sigma_{14})
=\\
\frac{1}{s_{34}\sigma_{14}-s_4\sigma_2}
\bigg\{
\frac{1}{\varepsilon}\ln\left(\frac{s_4\sigma_2}{s_{34}\sigma_{14}}\right)
+2\mathrm{Li}_2\left(1-\frac{\sigma_{14}}{\sigma_2}\right)
-2\mathrm{Li}_2\left(1-\frac{s_4}{s_{34}}\right)
\\
+2\mathrm{Li}_2\left(1-\frac{s_4\sigma_2}{s_{34}\sigma_{14}}\right)
+\ln\left(\frac{s_4\sigma_2}{s_{34}\sigma_{14}}\right)
            \left[
 \frac{1}{2}\ln\left(\frac{s_4s_{34}}{\sigma_2\sigma_{14}}\right)
-\ln\left(\frac{s_{34}\sigma_{14}s_4}{-\mu^2\sigma_1\sigma_2}\right)
\right]
\bigg\}
\end{multline}
\begin{multline}
\mathrm{Box}_{cF}(\sigma_1,\sigma_2,s_3,s_4,s_{34},\sigma_{14}) = 
\frac{-1}{a(x_1-x_2)}\bigg\{
\ln\left(\frac{x_1}{x_2}\right)\left[
\frac{1}{2}\ln(x_1x_2) - \ln\left(\frac{\sigma_1s_{34}}{\sigma_2s_4}\right)
\right]
\\
+\mathrm{Li}_2\left(1-x_1\frac{s_3}{s_{34}}\right)
-\mathrm{Li}_2\left(1-x_2\frac{s_3}{s_{34}}\right)
+\mathrm{Li}_2\left(1-x_1\frac{\sigma_{14}}{\sigma_1}\right)
-\mathrm{Li}_2\left(1-x_2\frac{\sigma_{14}}{\sigma_1}\right)
\bigg\}
\end{multline}
with
%
%%%%%%%%%%%%%%%%%%%%%%%%%%%%%%%%%%%%%%%%
\begin{equation}
\{-x_1,-x_2\} \quad\textrm{solutions to}\quad
ax^2 + bx + c = 0
%ax^2 + bx + c-\imag\eta\sigma_2 = 0
\quad,
\end{equation}
%%%%%%%%%%%%%%%%%%%%%%%%%%%%%%%%%%%%%%%%
%
where
%
%%%%%%%%%%%%%%%%%%%%%%%%%%%%%%%%%%%%%%%%
\begin{equation}
a = \sigma_{14}s_3
\quad,\quad
b = \sigma_{14}s_{34} + \sigma_1s_3 - \sigma_2s_4
\quad,\quad
c = \sigma_2s_{34}
\quad,
\end{equation}
%%%%%%%%%%%%%%%%%%%%%%%%%%%%%%%%%%%%%%%%

\section{\label{Sec:summary}Summary}
Factorization prescriptions that allow for perturbative QCD calculations for hadron scattering should preferably do so beyond leading order.
This implies the possibility to calculate one-loop amplitudes for the partonic cross section.
This issue was addressed regarding factorization prescriptions that assign a non-vanishing transverse momentum to the initial-state partons.
In particular, a regularization was studied that deals with the divergencies occurring due to linear propagator denominators in the loop integrals.
It respects gauge invariance, is manifestly Lorentz covariant, and allows for practical calculations for arbitrary partonic processes.

\subsection*{Acknowledgments}
The author would like to thank K.\ Kutak, P.\ Kotko, and C.\ Papadopoulos for valuable discussions.
All graphs were drawn with {\sc JaxoDraw}.
This work was supported by grant of National Science Center, Poland, No. 2015/17/B/ST2/01838.
%

%\bibliography{/home/user0/latex/bibliography}{}\bibliographystyle{JHEP}
\providecommand{\href}[2]{#2}\begingroup\raggedright\endgroup

\begin{appendix}
\addtocontents{toc}{\protect\setcounter{tocdepth}{1}}
\section{Solutions to the cut equations following OPP\label{AppCutOPP}}
In this appendix, we follow the notation of \cite{Ossola:2006us}.
We consider boxes and triangles
%
%%%%%%%%%%%%%%%%%%%%%%%%%%%%%%%%%%%%%%%%
\begin{equation}
I_4
=
\int\frac{d^{4-2\varepsilon}q}{(q+p_0)^2(q+p_1)^2(q+p_2)^2(q+p_3)^2}
\;,\;\;
I_3
=
\int\frac{d^{4-2\varepsilon}q}{(q+p_0)^2(q+p_1)^2(q+p_2)^2}
~.
\end{equation}
%%%%%%%%%%%%%%%%%%%%%%%%%%%%%%%%%%%%%%%%
%
The letter $k$ is now reserved for the momenta $k_i=p_i-p_0$.
We determine the coefficients $\bar{x}_1^0,\bar{x}_2^0,\bar{x}_3^\pm,\bar{x}_4^\pm$ and vectors $\bar{\lOPP}_1,\bar{\lOPP}_2,\bar{\lOPP}_3,\bar{\lOPP}_4$ in the solutions
%
%%%%%%%%%%%%%%%%%%%%%%%%%%%%%%%%%%%%%%%%
\begin{equation}
q^\mu
=
-p_0^\mu + \ell^\mu
\quad,\quad
\ell^\mu =
\bar{x}_1^0\bar{\lOPP}_1^\mu + \bar{x}_2^0\bar{\lOPP}_2^\mu 
                 + \bar{x}^\pm_3\bar{\lOPP}_3^\mu + \bar{x}^\pm_4\bar{\lOPP}_4^\mu
\label{cutOPP}
\end{equation}
%%%%%%%%%%%%%%%%%%%%%%%%%%%%%%%%%%%%%%%%
%
to the cut equations 
%
%%%%%%%%%%%%%%%%%%%%%%%%%%%%%%%%%%%%%%%%
\begin{gather}
(q+p_0)^2=(q+p_1)^2=(q+p_2)^2=(q+p_3)^2=0
\;,\notag\\
(q+p_0)^2=(q+p_1)^2=(q+p_2)^2=0
\;,
\end{gather}
%%%%%%%%%%%%%%%%%%%%%%%%%%%%%%%%%%%%%%%%
%
for the case that one $p_i$ is proportional to $\Lambda$ for $\Lambda\to\infty$, or two of them in case of the box.
The bar over the quantities indicates that they are the finite result in this limit.
The momentum $p$ without subscript still refers to the direction of the off-shell gluon.

\subsection{Box with $p_3=\Lambda p+p_3'$}
One is of course free to choose how to number the denominators.
This choice for the divergent momentum is particularly convenient.
It does not influence the formulas for the light-like vectors $\lOPP_i$, nor the coefficients $x_1^0=\bar{x}_1^0,x_2^0=\bar{x}_2^0$ in formula (3.7) of \cite{Ossola:2006us}.
Only the expressions for $A,B$ in formula (3.9) change.
Writing $k_3=\Lambda p+k_3'$ and taking $\Lambda\to\infty$, we find
%
%%%%%%%%%%%%%%%%%%%%%%%%%%%%%%%%%%%%%%%%
\begin{equation}
A\to-\frac{\lop{p}{\lOPP_3}}{\lop{p}{\lOPP_4}}
\quad,\quad
B\to\frac{\lop{p}{k_3}-x_1^0(\lop{p}{\lOPP_1})-x_2^0(\lop{p}{\lOPP_2})}{\lop{p}{\lOPP_4}}
~.
\end{equation}
%%%%%%%%%%%%%%%%%%%%%%%%%%%%%%%%%%%%%%%%
%
The coefficients $\bar{x}_3^\pm,\bar{x}_4^\pm$ then follow from the quadratic equation with the value for $A,B$ above.

\subsection{Box or triangle with $p_1=\Lambda p+p_1'$}
We need to determine the light-like vectors $\lOPP_i$ keeping relevant orders of $\Lambda$.
We get
%
%%%%%%%%%%%%%%%%%%%%%%%%%%%%%%%%%%%%%%%%
\begin{gather}
\gamma\to2\Lambda\lop{p}{k_2}
\quad,\quad
\alpha_1\to\frac{\lop{p}{k_1}}{\lop{p}{k_2}}
\quad,\quad
\alpha_2\to\frac{k_2^2}{2\Lambda\lop{p}{k_2}}
\quad,\quad
\beta\to1
\notag\\
\lOPP_1^\mu\to\Lambda\bar{\lOPP}_1^\mu=\Lambda p^\mu
\quad,\quad
\lOPP_2^\mu\to \bar{\lOPP}_2^\mu=k_2^\mu - \frac{k_2^2}{2\lop{p}{k_2}}\,p^\mu
\notag\\
\lOPP_{3}^\mu\to\sqrt{\Lambda}\,\bar{\lOPP}_{3}=\sqrt{\Lambda}\,\langle p|\gamma^\mu|\bar{\lOPP}_2]
\quad,\quad
\lOPP_{4}^\mu\to\sqrt{\Lambda}\,\bar{\lOPP}_{4}=\sqrt{\Lambda}\,\langle\bar{\lOPP}_2|\gamma^\mu|p]
\end{gather}
The coefficients $\bar{x}_{1,2}^0$ follow from
\begin{gather}
x_1^0\to\frac{\bar{x}_1^0}{\Lambda}=\frac{d_2-d_0+k_2^2\,\lop{p}{k_1}/\lop{p}{k_2}}{2\Lambda\lop{p}{k_2}}
\quad,\quad
x_2^0\to\bar{x}_2^0=\frac{\lop{p}{k_1}}{\lop{p}{k_2}}
\end{gather}
while the quadratic equation for the others has parameters
%
%%%%%%%%%%%%%%%%%%%%%%%%%%%%%%%%%%%%%%%%
\begin{equation}
A \to -\frac{\lop{k_3}{\bar{\lOPP}_3}}{\lop{k_3}{\bar{\lOPP}_4}}
\quad,\quad
B \to \frac{d_3-d_0-2\bar{x}_1^0(\lop{k_3}{\bar{\lOPP}_1})
                   -2\bar{x}_2^0(\lop{k_3}{\bar{\lOPP}_2})}
           {\sqrt{\Lambda}(\lop{k_3}{\bar{\lOPP}_4})}
\quad,\quad
C\to \frac{\bar{x}_1^0\bar{x}_2^0}{4\Lambda}
~,
\label{ABCk1OrdLambda}
\end{equation}
%%%%%%%%%%%%%%%%%%%%%%%%%%%%%%%%%%%%%%%%
%
so that the solutions behave as
%
%%%%%%%%%%%%%%%%%%%%%%%%%%%%%%%%%%%%%%%%
\begin{equation}
x_{3,4}^{\pm}\to\bar{x}_{3,4}^\pm/\sqrt{\Lambda}
~.
\end{equation}
%%%%%%%%%%%%%%%%%%%%%%%%%%%%%%%%%%%%%%%%
%
For the three-point coefficients, we find
%
%%%%%%%%%%%%%%%%%%%%%%%%%%%%%%%%%%%%%%%%
\begin{equation}
x_{3k}^\pm\to\frac{\bar{x}_{3k}^\pm}{\sqrt{\Lambda}} = \frac{\pm\sqrt{\bar{x}_1^0\bar{x}_2^0}\,e^{-\imag\pi/k}}{2\sqrt{\Lambda}}
\quad,\quad
x_{4k}^\pm\to\frac{\bar{x}_{4k}^\pm}{\sqrt{\Lambda}} = \frac{\pm\sqrt{\bar{x}_1^0\bar{x}_2^0}\,e^{\imag\pi/k}}{2\sqrt{\Lambda}}
~.
\end{equation}

\subsection{Box with $p_1=\Lambda p+p_1'$ and $p_3=\Lambda p+p_3'$}
For this case, the scalar integral is well-defined, but the solutions to the cut equations, however, diverge.
Putting the divergent $k_3^\mu$ into (\ref{ABCk1OrdLambda}), we see that one of the solutions $x_{3,4}^\pm$ vanishes while the other, say $x_3^+$ and $x_4^-$, behaves as $\Ord\big(\sqrt{\Lambda}\big)$, so
%
%%%%%%%%%%%%%%%%%%%%%%%%%%%%%%%%%%%%%%%%
\begin{align}
(q_0^+)^\mu
&=
-p_0^\mu + \bar{x}_1^0p^\mu + \bar{x}_2^0\bar{\lOPP}_2^\mu 
                 + \Lambda\bar{x}^+_3\bar{\lOPP}_3^\mu
\notag\\
(q_0^-)^\mu
&=
-p_0^\mu + \bar{x}_1^0p^\mu + \bar{x}_2^0\bar{\lOPP}_2^\mu 
                 + \Lambda\bar{x}^-_4\bar{\lOPP}_4^\mu
~.
\end{align}
%%%%%%%%%%%%%%%%%%%%%%%%%%%%%%%%%%%%%%%%
%
This reflects the fact that for the linear denominators at $\Lambda\to\infty$, there is no common solution with $\lop{p}{(q+p_1')}=\lop{p}{(q+p_3')}=0$, because this implies $\lop{p}{(p_1'-p_3')}=0$ which is not necessarily true.
Indeed, as already stated in the main text with \Equation{fourpointdecom}, the box with two $\Lambda$-dependent denominators is not a master integral, and can be written as a linear combination of triangles.

\section{Solutions to the cut equations following EGK\label{AppCutEGK}}
In this appendix, we follow the notation of~\cite{Ellis:2007br}.
We consider the box
%
%%%%%%%%%%%%%%%%%%%%%%%%%%%%%%%%%%%%%%%%
\begin{equation}
I_4
=
\int\frac{d^{4-2\varepsilon}\ell}{(\ell+K_0)^2\,(\ell+K_1)^2\,(\ell+K_2)^2\,(\ell+K_3)^2}
\end{equation}
%%%%%%%%%%%%%%%%%%%%%%%%%%%%%%%%%%%%%%%%
%
and corresponding triangles and bubbles with fewer denominators.
We determine the solutions to the cut equations if one of the denominators depends on $\Lambda$ for $\Lambda\to\infty$.
We will need the following combinations of denominator momenta:
%
%%%%%%%%%%%%%%%%%%%%%%%%%%%%%%%%%%%%%%%%
\begin{equation}
k_1^\mu = K_1^\mu-K_0^\mu
\quad,\quad
k_2^\mu = K_2^\mu-K_1^\mu
\quad,\quad
k_3^\mu = K_3^\mu-K_2^\mu
\quad.\quad
\end{equation}
%%%%%%%%%%%%%%%%%%%%%%%%%%%%%%%%%%%%%%%%
%
Also, we will need the generalized Kronecker delta
%
%%%%%%%%%%%%%%%%%%%%%%%%%%%%%%%%%%%%%%%%
\begin{equation}
\delta^{q_1q_2}_{k_1k_2}
= (\lop{q_1}{k_1})(\lop{q_2}{k_2})
- (\lop{q_1}{k_2})(\lop{q_2}{k_1})
\end{equation}
%%%%%%%%%%%%%%%%%%%%%%%%%%%%%%%%%%%%%%%%
%
and
%
%%%%%%%%%%%%%%%%%%%%%%%%%%%%%%%%%%%%%%%%
\begin{align}
\delta^{q_1q_2q_3}_{k_1k_2k_3}
&= (\lop{q_1}{k_1})(\lop{q_2}{k_2})(\lop{q_3}{k_3})
-  (\lop{q_1}{k_1})(\lop{q_2}{k_3})(\lop{q_3}{k_2})
\notag\\
&+ (\lop{q_1}{k_2})(\lop{q_2}{k_3})(\lop{q_3}{k_1})
-  (\lop{q_1}{k_2})(\lop{q_2}{k_1})(\lop{q_3}{k_3})
\notag\\
&+ (\lop{q_1}{k_3})(\lop{q_2}{k_1})(\lop{q_3}{k_2})
-  (\lop{q_1}{k_3})(\lop{q_2}{k_2})(\lop{q_3}{k_1})
\end{align}
%%%%%%%%%%%%%%%%%%%%%%%%%%%%%%%%%%%%%%%%
%
with the understanding that replacing $q_i$ with $\mu$ in the delta-symbol means replacing $\lop{q_i}{k_j}$ with $k_j^\mu$ everywhere in the expression.

\subsection{Box with $K_3 = \Lambda p+K_3'$}
The solutions to the cut equations
%
%%%%%%%%%%%%%%%%%%%%%%%%%%%%%%%%%%%%%%%%
\begin{equation}
(\ell+K_0)^2=(\ell+K_1)^2=(\ell+K_2)^2=(\ell+K_3)^2=0
\end{equation}
%%%%%%%%%%%%%%%%%%%%%%%%%%%%%%%%%%%%%%%%
%
are written in the form
%
%%%%%%%%%%%%%%%%%%%%%%%%%%%%%%%%%%%%%%%%
\begin{equation}
\ell^\mu = -K_0 + V_4^\mu + \alpha_1n_1^\mu
\end{equation}
%%%%%%%%%%%%%%%%%%%%%%%%%%%%%%%%%%%%%%%%
%
where
%
%%%%%%%%%%%%%%%%%%%%%%%%%%%%%%%%%%%%%%%%
\begin{equation}
V_4^\mu = -\frac{k_1^2}{2}\,v_1^\mu
          -\frac{k_2^2+2\lop{k_2}{k_1}}{2}\,v_2^\mu
          -\frac{k_3^2+2\lop{k_3}{(k_2+k_1)}}{2}\,v_3^\mu
\end{equation}
%%%%%%%%%%%%%%%%%%%%%%%%%%%%%%%%%%%%%%%%
%
and
%
%%%%%%%%%%%%%%%%%%%%%%%%%%%%%%%%%%%%%%%%
\begin{equation}
v_1^\mu = \frac{1}{\Delta}\,\delta^{\mu k_2k_3}_{k_1k_2k_3}
\quad,\quad
v_2^\mu = \frac{1}{\Delta}\,\delta^{k_1\mu k_3}_{k_1k_2k_3}
\quad,\quad
v_3^\mu = \frac{1}{\Delta}\,\delta^{k_1k_2\mu}_{k_1k_2k_3}
\quad,\quad
n_1^\mu = \frac{1}{\sqrt{\Delta}}\,\epsilon^{\mu k_1k_2k_3}
\quad,\quad
\end{equation}
%%%%%%%%%%%%%%%%%%%%%%%%%%%%%%%%%%%%%%%%
%
where $\epsilon$ denotes the Levi-Cevita symbol and 
%
%%%%%%%%%%%%%%%%%%%%%%%%%%%%%%%%%%%%%%%%
\begin{equation}
\Delta =  \delta^{k_1k_2k_3}_{k_1k_2k_3}
~.
\end{equation}
%%%%%%%%%%%%%%%%%%%%%%%%%%%%%%%%%%%%%%%%
%
With this parametrization, the cut equations reduce to
%
%%%%%%%%%%%%%%%%%%%%%%%%%%%%%%%%%%%%%%%%
\begin{equation}
\alpha_1^2 = -\lop{V_4}{V_4}
~,
\end{equation}
%%%%%%%%%%%%%%%%%%%%%%%%%%%%%%%%%%%%%%%%
%
which has two solutions for $\alpha_1$.
For $k_3=\Lambda p+k_3'$ we find to leading power in $\Lambda$
%
%%%%%%%%%%%%%%%%%%%%%%%%%%%%%%%%%%%%%%%%
\begin{align}
\delta^{\mu k_2k_3}_{k_1k_2k_3}
&= \frac{\Lambda^2}{4}\big[-k_1^\mu\sigma_2^2
            +k_2^\mu\sigma_1\sigma_2
            +2p^\mu((\lop{k_1}{k_2})\sigma_2-s_2\sigma_1)\big]
\notag\\
\delta^{k_1\mu k_3}_{k_1k_2k_3}
&=\frac{\Lambda^2}{4}\big[k_1^\mu\sigma_1\sigma_2
           -k_2^\mu\sigma_1^2
           +2p^\mu((\lop{k_1}{k_2})\sigma_1-s_1\sigma_2)\big]
\notag\\
\delta^{k_1k_2\mu}_{k_1k_2k_3}
&=\frac{\Lambda}{2}\big[k_1^\mu((\lop{k_1}{k_2})\sigma_2-s_2\sigma_1)
         +k_2^\mu((\lop{k_1}{k_2})\sigma_1-s_1\sigma_2)
         +2p^\mu\big(s_1s_2-(\lop{k_1}{k_2})^2\big)\big]
\notag\\
\epsilon^{\mu k_1k_2k_3}
&=\Lambda\epsilon^{\mu k_1k_2p}
\notag\\
\Delta &= -\frac{\Lambda^2}{4}\lop{(\sigma_2k_1-\sigma_1k_2)}{(\sigma_2k_1-\sigma_1k_2)} 
\end{align}
%%%%%%%%%%%%%%%%%%%%%%%%%%%%%%%%%%%%%%%%
%
where we denote
%
%%%%%%%%%%%%%%%%%%%%%%%%%%%%%%%%%%%%%%%%
\begin{equation}
s_i=k_i^2
\quad,\quad
\sigma_i=2\lop{p}{k_i}
~.
\end{equation}
%%%%%%%%%%%%%%%%%%%%%%%%%%%%%%%%%%%%%%%%
%
Taking into account also that, to leading order in $\Lambda$,
%
%%%%%%%%%%%%%%%%%%%%%%%%%%%%%%%%%%%%%%%%
\begin{equation}
k_3^2+2\lop{k_3}{(k_2+k_1)} = \Lambda(\sigma_3+\sigma_2+\sigma_1)
~,
\end{equation}
%%%%%%%%%%%%%%%%%%%%%%%%%%%%%%%%%%%%%%%%
%
we see that $V_4^\mu,n_1^\mu$, and thus the solutions to the cut equations, are finite and well-defined for $\Lambda\to\infty$.

\subsubsection{Box with $K_3=\Lambda p+K_3'$ and $K_2=\Lambda p+K_2'$}
If two denominator momenta depend on $\Lambda$, we see that
%
%%%%%%%%%%%%%%%%%%%%%%%%%%%%%%%%%%%%%%%%
\begin{gather}
\delta^{\mu k_2k_3}_{k_1k_2k_3} = \Ord\big(\Lambda^2\big)
\;\;,\;\;
\delta^{k_1\mu k_3}_{k_1k_2k_3} = \Ord\big(\Lambda^2\big)
\;\;,\;\;
\delta^{k_1k_2 \mu}_{k_1k_2k_3} = \Ord\big(\Lambda^2\big)
\;\;,\;\;
\epsilon^{\mu k_1k_2k_3} = \Ord\big(\Lambda\big)
\;\;,\;\;
\notag\\
\Delta  = \Ord\big(\Lambda^2\big)
\;\;,\;\;
k_2^2+2\lop{k_2}{k_1} = \Ord\big(\Lambda\big)
\;\;,\;\;
k_3^2+2\lop{k_3}{(k_k+k_1)} = \Ord\big(\Lambda\big)
~,
\end{gather}
%%%%%%%%%%%%%%%%%%%%%%%%%%%%%%%%%%%%%%%%
%
and we see that now $V_4^\mu$ diverges with $\Lambda$.

\subsection{Triangle with $K_2=\Lambda p+K_2'$}
The solutions to the cut equations
%
%%%%%%%%%%%%%%%%%%%%%%%%%%%%%%%%%%%%%%%%
\begin{equation}
(\ell+K_0)^2=(\ell+K_1)^2=(\ell+K_2)^2=0
\end{equation}
%%%%%%%%%%%%%%%%%%%%%%%%%%%%%%%%%%%%%%%%
%
are written in the form
%
%%%%%%%%%%%%%%%%%%%%%%%%%%%%%%%%%%%%%%%%
\begin{equation}
\ell^\mu = -K_0^\mu + V_3^\mu + \alpha_1n_1^\mu + \alpha_2n_2^\mu
\end{equation}
%%%%%%%%%%%%%%%%%%%%%%%%%%%%%%%%%%%%%%%%
%
where
%
%%%%%%%%%%%%%%%%%%%%%%%%%%%%%%%%%%%%%%%%
\begin{equation}
V_3^\mu = -\frac{k_1^2}{2}\,v_1^\mu
          -\frac{k_2^2+2\lop{k_2}{k_1}}{2}\,v_2^\mu
\end{equation}
%%%%%%%%%%%%%%%%%%%%%%%%%%%%%%%%%%%%%%%%
%
and
%
%%%%%%%%%%%%%%%%%%%%%%%%%%%%%%%%%%%%%%%%
\begin{equation}
v_1^\mu = \frac{1}{\Delta}\,\delta^{\mu k_2}_{k_1k_2}
\quad,\quad
v_2^\mu = \frac{1}{\Delta}\,\delta^{k_1\mu}_{k_1k_2}
\quad,\quad
\Delta =  \delta^{k_1k_2}_{k_1k_2}
~.
\end{equation}
%%%%%%%%%%%%%%%%%%%%%%%%%%%%%%%%%%%%%%%%
%
The vectors $n_1^\mu,n_2^\mu$ must be constructed such that
%
%%%%%%%%%%%%%%%%%%%%%%%%%%%%%%%%%%%%%%%%
\begin{equation}
\lop{n_i}{v_j}=0
\quad,\quad
\lop{n_i}{n_j} = \delta_{ij}
~.
\end{equation}
%%%%%%%%%%%%%%%%%%%%%%%%%%%%%%%%%%%%%%%%
%
With this parametrization, the cut equations reduce to
%
%%%%%%%%%%%%%%%%%%%%%%%%%%%%%%%%%%%%%%%%
\begin{equation}
\alpha_1^2 + \alpha_2^2 = -\lop{V_3}{V_3}
~,
\end{equation}
%%%%%%%%%%%%%%%%%%%%%%%%%%%%%%%%%%%%%%%%
%
which has an infinite set of solutions for $\alpha_1,\alpha_2$.
Setting $k_2=\Lambda p+k_2'$, we find to leading power in $\Lambda$
%
%%%%%%%%%%%%%%%%%%%%%%%%%%%%%%%%%%%%%%%%
\begin{gather}
\delta^{\mu k_2}_{k_1k_2}
= \frac{\Lambda^2}{2}\big[-p^\mu\sigma_1\big]
\quad,\quad
\delta^{k_1\mu}_{k_1k_2}
=\frac{\Lambda}{2}\big[-k_1^\mu\sigma_1+2p^\mu s_1\big]
\notag\\
\Delta = -\frac{\Lambda^2}{4}\,\sigma_1^2 
\quad,\quad
k_2^2+2\lop{k_2}{k_1} = \Lambda(\sigma_2+\sigma_1)
~.
\end{gather}
%%%%%%%%%%%%%%%%%%%%%%%%%%%%%%%%%%%%%%%%
%
So the vector $V_3^\mu$ is finite and well-defined in the limit $\Lambda\to\infty$.
The vectors $n_1^\mu,n_2^\mu$ can be constructed without complications such that they are orthogonal to both $k_1^\mu$ and $p^\mu$.

\subsection{\label{bubblecutsolEGK}Bubble with $K_1=\Lambda p+K_1'$\label{EGKbubble}}
The solutions to the cut equations
%
%%%%%%%%%%%%%%%%%%%%%%%%%%%%%%%%%%%%%%%%
\begin{equation}
(\ell+K_0)^2=(\ell+K_1)^2=0
\end{equation}
%%%%%%%%%%%%%%%%%%%%%%%%%%%%%%%%%%%%%%%%
%
are written in the form
%
%%%%%%%%%%%%%%%%%%%%%%%%%%%%%%%%%%%%%%%%
\begin{equation}
\ell^\mu = -K_0^\mu - \frac{1}{2}\,k_1^\mu
         + \alpha_1n_1^\mu + \alpha_2n_2^\mu + \alpha_3n_3^\mu
\end{equation}
%%%%%%%%%%%%%%%%%%%%%%%%%%%%%%%%%%%%%%%%
%
where the vectors $n_1^\mu,n_2^\mu,n_3^\mu$ must be constructed such that
%
%%%%%%%%%%%%%%%%%%%%%%%%%%%%%%%%%%%%%%%%
\begin{equation}
\lop{n_i}{k_1}=0
\quad,\quad
\lop{n_i}{n_j} = \delta_{ij}
~.
\end{equation}
%%%%%%%%%%%%%%%%%%%%%%%%%%%%%%%%%%%%%%%%
%
With this parametrization, the cut equations reduce to
%
%%%%%%%%%%%%%%%%%%%%%%%%%%%%%%%%%%%%%%%%
\begin{equation}
\alpha_1^2 + \alpha_2^2 + \alpha_3^2 = -\frac{\lop{k_1}{k_1}}{4}
~,
\end{equation}
%%%%%%%%%%%%%%%%%%%%%%%%%%%%%%%%%%%%%%%%
%
which has and infinite set of solutions for $\alpha_1,\alpha_2,\alpha_3$.
Setting $k_1=\Lambda p+k_1'=\Lambda p+K_1'-K_0$, we find
%
%%%%%%%%%%%%%%%%%%%%%%%%%%%%%%%%%%%%%%%%
\begin{equation}
\ell^\mu = - \frac{\Lambda}{2}\,p^\mu
  + \sqrt{\Lambda}\big[\bar{\alpha}_1n_1^\mu 
                     + \bar{\alpha}_2n_2^\mu
                     + \bar{\alpha}_3n_3^\mu\big]
  - \frac{1}{2}(K_0+K_1')^{\mu}
~,
\label{bubblecut}
\end{equation}
%%%%%%%%%%%%%%%%%%%%%%%%%%%%%%%%%%%%%%%%
%
where $\bar{\alpha}_1,\bar{\alpha}_2,\bar{\alpha}_3$ are solutions to the equation
%
%%%%%%%%%%%%%%%%%%%%%%%%%%%%%%%%%%%%%%%%
\begin{equation}
\bar{\alpha}_1^2+\bar{\alpha}_2^2+\bar{\alpha}_3^2
= -\frac{\lop{p}{k_1'}}{2}
  -\frac{\lop{k_1'}{k_1'}}{4\Lambda}
~.
\end{equation}
%%%%%%%%%%%%%%%%%%%%%%%%%%%%%%%%%%%%%%%%
%
So we see that for this case, the solutions $\ell$ to the cut equations diverge for $\Lambda\to\infty$.

\section{Decompostion of boxes with two $\Lambda$-dependent denominators\label{doubleboxes}}
In the following we calculate a few examples of box integrals with two $\Lambda$-dependent denominators, and show that they can be written as a linear combination of triangle integrals.
The general box looks like
%%%%%%%%%%%%%%%%%%%%%%%%%%%%%%%%%%%%%%%%
\begin{equation}
\graph{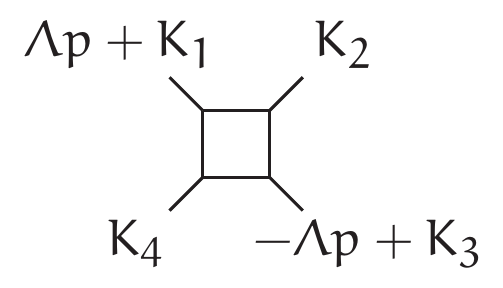}{20}{5.5}
=\;
\int\dDell\,\frac{\Lambda^2}{\ell^2\,(\ell+\Lambda p+K_1)^2
                           \,(\ell+\Lambda p+K_1+K_2)^2\,(\ell-K_4)^2}
~.
\end{equation}
%%%%%%%%%%%%%%%%%%%%%%%%%%%%%%%%%%%%%%%%
%
The external momenta are incoming, and we have
%
%%%%%%%%%%%%%%%%%%%%%%%%%%%%%%%%%%%%%%%%
\begin{equation}
K_1^\mu+K_2^\mu+K_3^\mu+K_4^\mu = 0~.
\end{equation}
%%%%%%%%%%%%%%%%%%%%%%%%%%%%%%%%%%%%%%%%
%
We introduce the notation
%
%%%%%%%%%%%%%%%%%%%%%%%%%%%%%%%%%%%%%%%%
\begin{gather}
s_2=K_2^2
\quad,\quad
s_4=K_4^2
\quad,
\notag\\
\sigma_1=2\lop{p}{K_1}
\quad,\quad
\sigma_2=2\lop{p}{K_2}
\quad,\quad
\sigma_3=-2\lop{p}{K_3}
\quad,\quad
\sigma_4=2\lop{p}{K_4}
\quad,
\\
\sigma_{12}=\sigma_1+\sigma_2
\quad,\quad
\sigma_{14}=\sigma_1+\sigma_4
\quad.
\notag
\end{gather}
%%%%%%%%%%%%%%%%%%%%%%%%%%%%%%%%%%%%%%%%
%
Although $\sigma_2,\sigma_4$ do not correspond to the leading behavior in $\Lambda$ of external invariants, they will proof to be useful, for example in the following identities implied by momentum conservation: 
%
%%%%%%%%%%%%%%%%%%%%%%%%%%%%%%%%%%%%%%%%
\begin{equation}
\sigma_1+\sigma_2-\sigma_3+\sigma_4=0
\quad,\quad
\sigma_{12}\sigma_{14}-\sigma_1\sigma_3 = \sigma_2\sigma_4
~.
\end{equation}
%%%%%%%%%%%%%%%%%%%%%%%%%%%%%%%%%%%%%%%%
%

\subsubsection*{No $K_i$ light-like}
For the case in which none of the momenta $K_i$ are light-like, we find from (41) in~\cite{Denner:1991qq}
%
%%%%%%%%%%%%%%%%%%%%%%%%%%%%%%%%%%%%%%%%
\begin{equation}
\mathrm{Box}_{dF}%(\sigma_1,s_2,\sigma_3,s_4;\sigma_2,\sigma_4)
=
\frac{1}{\sigma_2\sigma_4}\left\{
  2\mathrm{Li}_2\left(1-\frac{\sigma_1\sigma_3}{\sigma_{12}\sigma_{14}}\right)
 -\ln\left(\frac{\sigma_1\sigma_3}{\sigma_{12}\sigma_{14}}\right)
  \ln\left(\frac{\Lambda^2\sigma_{14}\sigma_{12}}{s_2s_4}\right)
\right\}
~.
\label{BoxdF}
\end{equation}
%%%%%%%%%%%%%%%%%%%%%%%%%%%%%%%%%%%%%%%%
%
Consider relation (2.7) from~\cite{Ellis:2007qk} with $P^2=\sigma_{1}$, $Q^2=\sigma_{3}$, $s=\sigma_{12}$, $t=\sigma_{14}$, so that
%
%%%%%%%%%%%%%%%%%%%%%%%%%%%%%%%%%%%%%%%%
\begin{equation}
a = \frac{\sigma_{1}+\sigma_{3}-\sigma_{12}-\sigma_{14}}{\sigma_1\sigma_3-\sigma_{12}\sigma_{14}}
 = -\frac{\sigma_1+\sigma_2-\sigma_3+\sigma_4}{\sigma_1\sigma_3-\sigma_{12}\sigma_{14}}
=0
~,
\end{equation}
%%%%%%%%%%%%%%%%%%%%%%%%%%%%%%%%%%%%%%%%
%
and the relation reduces to
%
%%%%%%%%%%%%%%%%%%%%%%%%%%%%%%%%%%%%%%%%
\begin{align}
\mathrm{Li}\left(1-\frac{\sigma_1\sigma_3}{\sigma_{12}\sigma_{14}}\right)
&=
\mathrm{Li}\left(1-\frac{\sigma_1}{\sigma_{12}}\right)
+\mathrm{Li}\left(1-\frac{\sigma_1}{\sigma_{14}}\right)
\notag\\
&+\mathrm{Li}\left(1-\frac{\sigma_3}{\sigma_{12}}\right)
+\mathrm{Li}\left(1-\frac{\sigma_3}{\sigma_{14}}\right)
+\frac{1}{2}\ln^2\left(\frac{\sigma_{12}}{\sigma_{14}}\right)
~.
\label{LiRelation}
\end{align}
%%%%%%%%%%%%%%%%%%%%%%%%%%%%%%%%%%%%%%%%
%
Substituting this into \Equation{BoxdF}, we find
%
%%%%%%%%%%%%%%%%%%%%%%%%%%%%%%%%%%%%%%%%
\begin{equation}
\mathrm{Box}_{dF}%(\sigma_1,s_2,\sigma_3,s_4;\sigma_2,\sigma_4)
=
 \frac{\mathrm{Tri}_{eF}(\sigma_{12},\sigma_{3},s_4)}{-\sigma_2}
+\frac{\mathrm{Tri}_{eF}(\sigma_{14},\sigma_{1},s_4)}{\sigma_2}
+\frac{\mathrm{Tri}_{eF}(\sigma_{12},\sigma_{1},s_2)}{\sigma_4}
+\frac{\mathrm{Tri}_{eF}(\sigma_{14},\sigma_{3},s_2)}{-\sigma_4}
~.
\end{equation}
%%%%%%%%%%%%%%%%%%%%%%%%%%%%%%%%%%%%%%%%
%

\subsubsection*{Only $K_2$ light-like}
For the case that $K_2^2=0$, we find, with application of \Equation{LiRelation} on (4.24) of ~\cite{Ellis:2007qk},
%
%%%%%%%%%%%%%%%%%%%%%%%%%%%%%%%%%%%%%%%%%
%\begin{multline}
%\mathrm{Box}_{5d}
%=
%\frac{1}{\sigma_2\sigma_4}\bigg\{
%\frac{1}{\varepsilon}\ln\left(\frac{\sigma_1\sigma_3}{\sigma_{12}\sigma_{14}}\right)
%+2\mathrm{Li}_2\left(1-\frac{\sigma_1\sigma_{3}}{\sigma_{12}\sigma_{14}}\right)
%\\
%+2\mathrm{Li}_2\left(1-\frac{\sigma_{14}}{\sigma_{3}}\right)
%-2\mathrm{Li}_2\left(1-\frac{\sigma_{1}}{\sigma_{12}}\right)
%+\frac{1}{2}\ln\left(\frac{\sigma_1\sigma_3}{\sigma_{12}\sigma_{14}}\right)
%            \ln\left(\frac{\sigma_3s_4^2\mu^4}{\Lambda^4\sigma_{14}^3\sigma_1\sigma_{12}}\right)
%\bigg\}
%\end{multline}
%%%%%%%%%%%%%%%%%%%%%%%%%%%%%%%%%%%%%%%%%
%
%%%%%%%%%%%%%%%%%%%%%%%%%%%%%%%%%%%%%%%%
\begin{multline}
\mathrm{Box}_{d5}
=
\frac{1}{\sigma_2\sigma_4}\bigg\{
\frac{1}{\varepsilon}\ln\left(\frac{\sigma_1\sigma_3}{\sigma_{12}\sigma_{14}}\right)
+2\mathrm{Li}_2\left(1-\frac{\sigma_{1}}{\sigma_{14}}\right)
+2\mathrm{Li}_2\left(1-\frac{\sigma_{3}}{\sigma_{12}}\right)
\\
+\ln^2\left(\frac{\sigma_{12}}{\sigma_{14}}\right)
-\frac{1}{2}\ln^2\left(\frac{\sigma_{1}}{\sigma_{12}}\right)
-\frac{1}{2}\ln^2\left(\frac{\sigma_{3}}{\sigma_{14}}\right)
%
%+\frac{1}{2}\ln\left(\frac{\sigma_{1}}{\sigma_{14}}\right)
%            \ln\left(\frac{\sigma_{12}^2}{\sigma_{1}\sigma_{14}}\right)
%+\frac{1}{2}\ln\left(\frac{\sigma_{3}}{\sigma_{12}}\right)
%            \ln\left(\frac{\sigma_{14}^2}{\sigma_{3}\sigma_{12}}\right)
%
-\ln\left(\frac{\sigma_1\sigma_3}{\sigma_{12}\sigma_{14}}\right)
  \ln\left(\frac{\Lambda^2\sigma_{12}\sigma_{14}}{-\mu^2s_4}\right)
\bigg\}
~,
\end{multline}
%%%%%%%%%%%%%%%%%%%%%%%%%%%%%%%%%%%%%%%%
%
which decomposes as
%
%%%%%%%%%%%%%%%%%%%%%%%%%%%%%%%%%%%%%%%%
\begin{equation}
\mathrm{Box}_{d5}%(\sigma_1,s_2,\sigma_3,s_4;\sigma_2,\sigma_4)
=
 \frac{\mathrm{Tri}_{eF}(\sigma_{12},\sigma_{3},s_4)}{-\sigma_2}
+\frac{\mathrm{Tri}_{eF}(\sigma_{14},\sigma_{1},s_4)}{\sigma_2}
+\frac{\mathrm{Tri}_{d2}(\sigma_{12},\sigma_{1})}{\sigma_4}
+\frac{\mathrm{Tri}_{d2}(\sigma_{14},\sigma_{3})}{-\sigma_4}
~.
\end{equation}
%%%%%%%%%%%%%%%%%%%%%%%%%%%%%%%%%%%%%%%%
%

\subsubsection*{Vanishing $K_1^\mu$}
For the case that $K_1^\mu=0$, we find starting from (4.24) of ~\cite{Ellis:2007qk}
%
%%%%%%%%%%%%%%%%%%%%%%%%%%%%%%%%%%%%%%%%
\begin{multline}
\mathrm{Box}_{d5'}
=
\frac{1}{\sigma_{12}\sigma_{14}}\bigg\{
\frac{1}{\varepsilon}\ln\left(\frac{s_2s_4}{\Lambda^2\sigma_{12}\sigma_{14}}\right)
-\frac{1}{2}\ln^2\left(\frac{s_2}{\Lambda\sigma_{12}}\right)
-\frac{1}{2}\ln^2\left(\frac{s_4}{\Lambda\sigma_{14}}\right)
\\
-\ln\left(\frac{s_2s_4}{\Lambda^2\sigma_{12}\sigma_{14}}\right)
 \ln\left(\frac{\Lambda\sigma_{12}\sigma_{14}}{-\mu^2\sigma_3}\right)
-\frac{\pi^2}{3}
\bigg\}
~.
\end{multline}
%%%%%%%%%%%%%%%%%%%%%%%%%%%%%%%%%%%%%%%%
%
Here, we have $\sigma_1=0$, so $\sigma_{12}=-\sigma_2$ and $\sigma_{14}=-\sigma_4$.
Now, \Equation{LiRelation} reduces to
%
%%%%%%%%%%%%%%%%%%%%%%%%%%%%%%%%%%%%%%%%
\begin{align}
 \mathrm{Li}\left(1-\frac{\sigma_3}{\sigma_{12}}\right)
+\mathrm{Li}\left(1-\frac{\sigma_3}{\sigma_{14}}\right)
=
-\frac{\pi^2}{6}-\frac{1}{2}\ln^2\left(\frac{\sigma_{12}}{\sigma_{14}}\right)
~,
\end{align}
%%%%%%%%%%%%%%%%%%%%%%%%%%%%%%%%%%%%%%%%
%
and using this, we find
%
%%%%%%%%%%%%%%%%%%%%%%%%%%%%%%%%%%%%%%%%
\begin{equation}
\mathrm{Box}_{d5'}%(\sigma_1,s_2,\sigma_3,s_4;\sigma_2,\sigma_4)
=
 \frac{\mathrm{Tri}_{eF}(\sigma_{12},\sigma_{3},s_4)}{-\sigma_2}
+\frac{\mathrm{Tri}_{c2}(\sigma_{14},s_4)           }{\sigma_2}
+\frac{\mathrm{Tri}_{c2}(\sigma_{12},s_2)           }{\sigma_4}
+\frac{\mathrm{Tri}_{eF}(\sigma_{14},\sigma_{3},s_2)}{-\sigma_4}
~.
\end{equation}
%%%%%%%%%%%%%%%%%%%%%%%%%%%%%%%%%%%%%%%%
%

\section{Program for numerator behavior\label{formTriangle}}
Here follows a FORM~\cite{Ruijl:2017dtg} program for the evaluation of the numerator of graphs of the type in~\Figure{Fig:generalGraph}.

\scriptsize
\begin{verbatim}
nwrite statistics;
vector v;
symbol L,n,SQRTL,SQRT2,factor,[l],cA,cB,cC,cD,cE,eps;
function G,BRA,KET,BRAp,KETp,KETrght,BRAleft;
autodeclare vector l,K,J,Q,R,p,k;
autodeclare symbol blob;
autodeclare index mu=n;
autodeclare function auxline,PI;
autodeclare cfunction V;

*** This program may be executed as
*** $ form42 thisProg.frm | grep "\[l\]"

*** Number of external gluons from the auxiliary quark line and the blob.
#define mm "0"
#define nn "3"

*** The whole numerator. Include a factor (-i_) for one of the two gluon
*** propagators connecting the auxiliary quark lines with the blob. The
*** other will be included with a vertex. The overall factor will be
*** specified at the end.
Local rslt = 
  BRA * G(mu'nn')*(i_/SQRT2) * auxline'mm' * G(mu0)*(i_/SQRT2) * KET * 
  blob'nn' * (-i_) * factor;

*** Definition of the blob for different numbers of external gluons.
id blob0 = d_(mu0,mu'nn');
id blob1 = V1;
id blob2 = V1*V2 + V12;
id blob3 = V1*V2*V3 + V12*V3 + V1*V23;
id blob4 = V1*V2*V3*V4 + V12*V3*V4 + V1*V23*V4 + V1*V2*V34 + V12*V34;
id blob5 = V1*V2*V3*V4*V5 + V12*V3*V4*V5 + V1*V23*V4*V5 + V1*V2*V34*V5
         + V1*V2*V3*V45 + V12*V34*V5 + V12*V3*V45 + V1*V23*V45;

*** Definition auxline for different numbers of external gluons.
id auxline0 = PI0;
id auxline1 = PI0*PI1;
id auxline2 = PI0*PI1*PI2;
id auxline3 = PI0*PI1*PI2*PI3;
id auxline4 = PI0*PI1*PI2*PI3*PI4;
id auxline5 = PI0*PI1*PI2*PI3*PI4*PI5;

*** In the following, external momenta start with K, external currents with J,
*** internal momenta with Q or R, the loop momentum is l, the auxiliary quark line
*** direction is p, and the regulator Lambda is L.

*** Auxiliary quark line numerator factors.
id  PI0 = ( G(l)-L*G(p)+G(Q{0}) )*(i_);
#do i=1,'mm'
id  PI'i' = G(J{'i'})*(i_/SQRT2) * ( G(l)-L*G(p)+G(Q{'i'}) )*(i_);
#enddo

*** Move G(mu0) to the left
repeat;
id G(J?)*G(mu0) = 2*J(mu0) - G(mu0)*G(J);
endrepeat;

*** Gluonic 4-point vertices.
*** Include a factor (-i_) for a propagator. 
*** Include the missing denominator compared to a pair of 3-point vertices
*** by adding it in the numerator.
#do i=1,'nn'-1
id  V'i'{'i'+1} = V( mu{'i'-1} ,J'i' ,J{'i'+1} ,mu{'i'+1} )*(i_/2)*(-i_)*
                  ( l.l + 2*l.R'i' + R'i'.R'i' );
#enddo

*** Gluonic 3-point vertices.
*** Include a factor (-i_) for a propagator. 
#do i=1,'nn'
id  V'i' = V( mu{'i'-1},l+R{'i'-1} ,J'i',K'i' ,mu'i',-l-R'i' )*
          (i_/SQRT2)*(-i_);
#enddo
.sort
id 1/SQRT2/SQRT2 = 1/2;

*** Actual Lorentz form of the vertices.
id V(mu1?,J2?,J3?,mu4?) = 2*J3(mu1)*J2(mu4)-J2(mu1)*J3(mu4)-J2.J3*d_(mu1,mu4);

id V(mu1?,p1?,J2?,p2?,mu3?,p3?) = ( p1(mu3)-p2(mu3))*J2(mu1)
                                + ( p2(mu1)-p3(mu1))*J2(mu3)
                                + ( p3.J2-p1.J2)*d_(mu3,mu1);

*** Feynman gauge for external currents.
#do i=1,'nn'+'mm'+1
id  K'i'.J'i' = 0;
#enddo

*** Symplifications
id G(mu0?)*G(mu0?) = 4-2*eps;
id BRA = BRAp*SQRTL + BRAleft/SQRTL;
id KET = KETp*SQRTL + KETrght/SQRTL;
id SQRTL*SQRTL = L;
id 1/SQRTL/SQRTL = 1/L;
repeat;
id G(J?)*G(l) = 2*l.J - G(l)*G(J);
id G(l)*G(l) = l.l;
id BRAp*G(p) = 0;
endrepeat;
repeat;
id G(J?)*G(p) = 2*p.J - G(p)*G(J);
id G(p)*G(p) = 0;
id BRAp*G(p) = 0;
endrepeat;
id G(p) = KETp*BRAp;
id BRAp*G(J?)*KETp = 2*p.J;
id BRAp*KETp = 0;

*** Counting denominators and powers of the integration momentum
*** in the numerator. The factor has a 1/L^{'mm'+2} coming from
*** all Lambda-denominators and the definition of the amplitude.
id factor = 1/L^{'mm'+2};
id l = l*[l];
id G(l) = G(l)*[l];
id l.l = cA/[l] ;*+ cB/[l]^2;
id p.l = cC/L + cD/[l] ;*+ cE/L/[l];
.sort
bracket L,SQRTL,[l],i_;
print;

.end
\end{verbatim}

\end{appendix}

\end{document}